%% file: ms.tex
\documentclass[lettersize,journal]{IEEEtran}
\usepackage{amsmath,amsfonts}
\usepackage{algorithmic}
\usepackage{amsthm}
\usepackage{algorithm}
\usepackage{array}
\usepackage{textcomp}
\usepackage{multicol}
\usepackage{adjustbox}
\usepackage{makecell}
\usepackage{makecell}
\usepackage{ctable}
\usepackage{booktabs} 
\usepackage{multirow}
\usepackage{stfloats}
\usepackage{subfigure}
\usepackage{url}
\usepackage{verbatim}
\usepackage{graphicx}
\usepackage{cite}

\theoremstyle{definition}
\newtheorem{definition}{Definition}[section]

\begin{document}

\title{Efficient Image Denoising Using Global and Local \\ Circulant Representation}

\author{Zhaoming Kong, Jiahuan Zhang, and Xiaowei Yang
\thanks{Z. Kong and X. Yang are with the School of Software Engineering, South China University of Technology, Guangzhou, 510006, China (e-mail: kong.zm@mail.scut.edu.cn; xwyang@scut.edu.cn).}
\thanks{J. Zhang is with the Department of Clinical Laboratory Medicine, Guangdong Provincial People's Hospital, Southern Medical University, Guangzhou, 510000, China (email: zhangjiahuan@gdph.org.cn).}}

\maketitle

\begin{abstract}
The advancement of imaging devices and countless image data generated everyday impose an increasingly high demand on efficient and effective image denoising. In this paper, we present a computationally simple denoising algorithm, termed Haar-tSVD, aiming to explore the nonlocal self-similarity prior and leverage the connection between principal component analysis (PCA) and the Haar transform under circulant representation. We show that global and local patch correlations can be effectively captured through a unified tensor-singular value decomposition (t-SVD) projection with the Haar transform. This results in a one-step, highly parallelizable filtering method that eliminates the need for learning local bases to represent image patches, striking a balance between denoising speed and performance. Furthermore, we introduce an adaptive noise estimation scheme based on a CNN estimator and eigenvalue analysis to enhance the robustness and adaptability of the proposed method. Experiments on different real-world denoising tasks validate the efficiency and effectiveness of Haar-tSVD for noise removal and detail preservation. Datasets, code and results are publicly available at https://github.com/ZhaomingKong/Haar-tSVD.
\end{abstract}

\vspace{-1.8pt}
\section{Introduction}
\IEEEPARstart{T}he rapid development of modern imaging systems and technologies has significantly enriched the information captured and conveyed by images, delivering a more faithful representation for real scenes. However, images are inevitably corrupted by noise during acquisition and transmission, which can severely degrade the visual quality of acquired image data. Therefore, image denoising plays an important role in numerous applications such as feature extraction, object tracking and medical diagnosis \cite{elad2023image, Collection_denoising_methods}. \\
\indent Many state-of-the-art image denoising techniques weigh highly on deep neural network (DNN) architectures, which are typically trained on datasets containing clean and noisy image pairs. For example, Zhang et al. \cite{zhang2017beyond} incorporated batch normalization (BN) \cite{ioffe2015batch}, rectified linear unit (ReLU) \cite{nair2010rectified} and residual learning \cite{he2016deep} into the convolutional neural network (CNN) model. Zamir et. al \cite{zamir2022restormer} adopted an encoder-decoder Transformer for multi-scale representation. While the data-driven models have shown impressive performance for image restoration, their performance may degrade sharply when the test images do not match the distribution of the current scene. Besides, collecting high-quality data is time-consuming and expensive, and the platform required for training and inference may not be affordable to ordinary users and researchers. \\
\indent Therefore, it is essential and interesting to investigate dataset-free denoising methods at low computational cost. Existing works in this direction fall into two major categories: traditional denoisers and self-supervised models. Briefly, traditional denoisers normally filter out noise based solely on the input noisy observation with different regularization terms and image priors \cite{katkovnik2010local, zoran2011learning}. The representative BM3D framework \cite{dabov2007image} integrated the nonlocal self-similarity (NLSS) characteristic of natural images \cite{buades2005review}, sparse representation \cite{elad2006image} and transform domain techniques \cite{yaroslavsky2001transform} into a subtle paradigm. Zhang et al. \cite{gu2014weighted} replaced the sparsity constraint with the low-rank assumption. Xu et al. \cite{xu2018trilateral} employed the Maximum A-Posterior (MAP) estimation technique \cite{murphy2012machine} and proposed a trilateral weighted sparse coding scheme. As an alternative, self-supervised models get rid of the prerequisite on clean data and leverage the power of DNNs by training on noisy images. For example, Noise2Self \cite{batson2019noise2self} and Noise2Void \cite{krull2019noise2void} introduced blind-spot learning by masking pixels of the noisy input. Self2Self \cite{quan2020self2self} adopted a dropout-based scheme with noisy pairs generated by the Bernoulli sampler. Other methods, such as Noise-As-Clean \cite{xu2020noisy} and R2R \cite{pang2021recorrupted} created noisy pairs by corrupting the input with certain noise distributions. \\
\indent Despite the notable improvements brought by the development of related algorithms, training very deep networks or solving complex optimization problems in the test phase may give rise to high computational overhead. To circumvent these issues and reach a favorable trade-off between denoising performance and computational resources, we propose Haar-tSVD to efficiently capture the local and nonlocal self-similarity of natural images under circulant representation \cite{tee2007eigenvectors, zhang2016exact}. Specifically, the proposed method is built upon the simple and flexible patch-based denoising paradigm, consisting of several key components. First, we adopt the green channel prior (GCP)\cite{guo2021joint, kong2025} to guide patch matching, which can effectively improve the grouping quality and reduce the patch search time by approximately $\frac{2}{3}$. Next, the correlations among stacked image patches are modeled by patch- and group-level circulant structures, which can be reduced to a one-step combination of global t-SVD projection and Haar transform. This design eliminates the need to iteratively learn local bases, resulting in a more efficient and highly parallelizable approach. Moreover, to enhance adaptability and avoid manual parameter setting for each noisy observation, an adaptive noise estimation scheme termed A-Haar-tSVD is introduced by exploiting a CNN-based estimator and eigenvalue analysis. \\
\indent Our main contributions can be summarized as follows: 
\begin{itemize}
  \item We propose to leverage the circulant representation to capture both the inner- and inter-correlations among image patches. Besides, by bridging the gap between PCA and the Haar transform via the eigenvalue decomposition (EVD) of the circulant structure, we show that the circulant similarity can be exploited by a unified t-SVD projection and Haar transform. 
  \item We develop an adaptive scheme to enhance the flexibility and robustness of the proposed method by exploring a CNN-based noise estimator alongside eigenvalue characteristics of circulant structures. To reduce computational overhead, we employ a fast implementation framework by incorporating parallel programming techniques.
  \item We evaluate the applicability and adaptability of the proposed method across various real-world denoising tasks such as images, videos, hyperspectral imaging (HSI) and magnetic resonance imaging (MRI). Experimental results demonstrate the competitive performance of Haar-tSVD and its adaptive variant A-Haar-tSVD in terms of both effectiveness and efficiency.
\end{itemize}

The rest of the paper is structured as follows. Section II summarizes related works. Section III describes the proposed Haar-tSVD denoising method in detail. Section IV presents datasets, experimental settings and results. Besides, discussions of the ablation studies are included in this section. Section V concludes this work.

\section{Related Works} \label{section_background}
This section provides the foundation of the proposed method and a short overview of related works.
\vspace{-1.9pt}
\subsection{Symbols and Notations}
\indent In this paper, we adopt the mathematical notations and preliminaries of tensors from \cite{kolda2009tensor} for image representation. Vectors and matrices are first- and second- order tensors which are denoted by boldface lowercase letters $\mathbf{a}$ and capital letters $\mathbf{A}$, respectively. A higher order tensor (the tensor of order three or above) is denoted by calligraphic letters, e.g., $\mathcal{A}$. An $N$th-order tensor is denoted as $\mathcal{A} \in \mathbb{R}^{I_1\times I_2\times\cdots\times I_N}$. The $n$-mode product of a tensor $\mathcal{A}$ by a matrix $\mathbf{U}\in \mathrm{R}^{P_n\times I_n}$, denoted by $\mathcal{A}\times _n\mathbf{U} \in \mathbb{R}^{I_1 \cdots I_{n-1} P_n I_{n+1} \cdots I_N}$ is also a tensor. The mode-$n$ matricization or unfolding of $\mathcal{A}$ is denoted by $\mathbf{A}_{(n)}$. 
\vspace{-2pt}
\vspace{-3.8pt}
\subsection{Patch-based Denoisers}
\input{Fig_traditional_framework}
Briefly, patch-based denoisers usually accomplish learning and denoising by leveraging the nonlocal self-similarity (NLSS) property. To achieve this goal, the most popular framework mainly follows three consecutive stages: grouping, collaborative filtering and aggregation. The flowchart of this effective three-stage paradigm is illustrated in Fig. \ref{Fig_traditional_framework}. \\
\indent Specifically, for a given reference noisy image patch $\mathcal{P}_{n} \in \mathbb{R}^{ps\times ps\times c}$ with patch size $ps$ and $c$ channels, based on certain patch matching criteria \cite{foi2007pointwise, buades2016patch, Foi2020}, the grouping step stacks $K$ similar (overlapping) patches located within a local window $\Omega_{W}$ into higher dimensional group $\mathcal{G}_n \in \mathbb{R}^{ps \times ps \times c \times K}$. Then collaborative filters operate on $\mathcal{G}_n$ to estimate the corresponding underlying clean group $\mathcal{G}_c$ via
\begin{equation}\label{tensor_collaborative_filtering}
  \hat{\mathcal{G}}_c = \mathop{\arg\min_{\mathcal{G}_c}} \| \mathcal{G}_n - \mathcal{G}_c \|_{F}^2 + \rho\cdot\Psi(\mathcal{G}_c),
\end{equation}
where $\| \mathcal{G}_n - \mathcal{G}_c \|_{F}^2$ measures the difference between the clean and noisy groups, and $\Psi(\cdot)$ denotes a regularization term that encodes prior knowledge. For example, to model the nonlocal redundancies, the low-rank prior is adopted in \cite{dong2012nonlocal, chang2017hyper, xu2017multi}. The dictionary learning model with over-complete representations \cite{elad2006image,mairal2007sparse,xu2018trilateral} is utilized to reconstruct $\mathcal{G}_c$ with a dictionary and sparse coding coefficients. To model the sparsity, a simple and effective approach is to apply thresholding techniques \cite{donoho1994ideal} to shrink the coefficients and attenuate noise.  \\
\indent To further smooth out noise, the estimated clean patches of $\hat{\mathcal{G}}_c$ are averagely written back to their original location. Specifically, at the pixel level, every pixel $\hat{p}_i$ of the denoised image is the (weighted) average of all pixels at the same position of estimated $\hat{\mathcal{G}}_c$, which can be formulated as
 \begin{equation}\label{aggregation}
   \hat{p}_i = \sum_{\hat{p}_{i_k} \in \hat{\mathcal{G}}_c} w_{i_k} \hat{p}_{i_k},
 \end{equation}
where $w_{i_k}$ and $\hat{p}_{i_k}$ are weight and local pixel, respectively.
\vspace{-2.8pt}
\subsection{Circulant Representation and t-SVD for Image Denoising}
\indent Recently, circulant structure and representation \cite{chen2015image, zhang2016exact, kong2019color} has been utilized for image restoration due to its effectiveness of exploring redundancy. The motivation behind the circulant matrices is to encode the inner structure of an image patch by cyclic shift. However, performing circulant operation on a long vector can result in a large matrix and incur high storage and computation cost. To avoid direct vectorization, the block circulant representation (BCR) \cite{zhang2016exact} of an image patch $\mathcal{P} \in \mathbb{R}^{ps \times ps \times c}$ can be exploited to model the patch-level redundancy via
\begin{equation}\label{Equ_block_circulant_matrix}
  bcirc(\mathcal{P}) = \begin{pmatrix}
                    \mathcal{P}^{(1)} & \mathcal{P}^{(c)} & \cdots & \mathcal{P}^{(2)}\\
                    \mathcal{P}^{(2)} & \mathcal{P}^{(1)} & \cdots & \mathcal{P}^{(3)} \\
                    \vdots & \vdots & \ddots & \vdots \\
                    \mathcal{P}^{(c)} & \mathcal{P}^{(c-1)} & \cdots & \mathcal{P}^{(1)} \\
                  \end{pmatrix},
\end{equation}
where $\mathcal{P}^{(i)}$ denotes the $i$-th frontal slice $\mathcal{P}(:,:,i)$ of $\mathcal{P}$, and $bcirc(\mathcal{P}) \in \mathbb{R}^{psc\times psc}$ is a block circulant matrix. In addition to lower storage cost, the BCR preserves the patch structure and recursively models the correlation among different slices, making it suitable for multi-dimensional image data. Another important property of BCR is that its sums, products and inverses are also block circulant. \\
\indent Fortunately, to leverage the circulant structure, we do not need to explicitly construct BCR for a higher-order tensor image patch, since the products between two block circulant matrices can be simplified using tensor t-product \cite{kilmer2011factorization, kilmer2013third}.
\begin{definition}[T-product]
Suppose $\mathcal{A}\in \mathbb{R}^{n_1 \times m \times n_3}$ and $\mathcal{B}\in \mathbb{R}^{m \times n_2 \times n_3}$, then the t-product $\mathcal{C} = \mathcal{A}*\mathcal{B} \in \mathbb{R}^{n_1 \times n_2 \times n_3}$ is defined as 
\begin{equation}\label{Equ_t_product}
  bcirc(\mathcal{C}) = bcirc(\mathcal{A}) bcirc(\mathcal{B}).
\end{equation}
\end{definition}
\noindent Equ. (\ref{Equ_t_product}) can be efficiently computed in the Fourier domain
\begin{equation}\label{Equ_fft_product}
  \mathcal{C}_{F}^{(i)} = \mathcal{A}_{F}^{(i)} \mathcal{B}_{F}^{(i)}, \quad i = 1, 2, \ldots, m,
\end{equation}
where $\mathcal{A}_F$ is obtained by applying the fast Fourier transform (FFT) along the third mode of $\mathcal{A}$ via
\begin{equation}\label{Equ_fft_third_mode}
  \mathcal{A}_F = \mathcal{A} \times _3\mathbf{W}_{FFT},
\end{equation} 
where $\mathbf{W}_{FFT}$ refers to the FFT matrix. Then the popular t-SVD \cite{kilmer2013third} can be defined based on the t-product.
\begin{definition}[T-SVD]
For $\mathcal{A} \in \mathbb{R}^{n_1 \times n_2 \times n_3}$, its t-SVD decomposition is given by 
\begin{equation}\label{Equ_t_SVD}
  \mathcal{A} = \mathcal{U} * \mathcal{S} * \mathcal{V}^T,
\end{equation}
\end{definition}
\noindent where $\mathcal{U} \in \mathbb{R}^{n_1 \times n_1 \times n_3}$ and $\mathcal{V} \in \mathbb{R}^{n_2 \times n_2 \times n_3}$ are orthogonal tensors, and the entries in $\mathcal{S}\in \mathbb{R}^{n_1 \times n_2 \times n_3}$ can be viewed as singular values or coefficients of $\mathcal{A}$. \\
\indent t-SVD has proven effective for image denoising, as it explores the block circulant structure of image patches, which helps to address the unbalance trap of tensor model \cite{bengua2017efficient}, and shares similar properties with its matrix counterpart. For example, Zhang et al. \cite{zhang2016exact} used the tensor nuclear norm regularizer for incomplete and noisy videos. Gong et. al \cite{gong2020low} considered a simultaneously sparse and low-rank tensor representation denoiser with t-SVD and dictionary learning. Kong et. al \cite{kong2019color} combined t-SVD and PCA to leverage NLSS and model sparsity in the transform domain. Although existing t-SVD based methods have certain advantages for noise removal, they are still haunted by the need to iteratively learn local bases and solve optimization problems for collaborative filtering during the test phase. Therefore, it is interesting to investigate the design of a simple and effective denoising scheme with t-SVD and circulant representation.
\subsection{Haar Transform}
As a popular transformation in image filtering \cite{blu2007sure, hou2020nlh}, the Haar transform belongs to the wavelet family, which is derived from the Haar matrix \cite{roeser1982fast}. Specifically, starting from a simple $2 \times 2$ Haar transformation matrix
\begin{equation}\label{Equ_Haar_mtx_example}
    \mathbf{H}_2 = \frac{1}{\sqrt{2}}\begin{bmatrix}
        1 &  1\\
        1 & -1
    \end{bmatrix}.
\end{equation}
We can define the general $2N \times 2N$ Haar matrix by
\begin{equation}\label{Equ_general_Haar_mtx}
  \mathbf{H}_{2N} = \frac{1}{\sqrt{2}} \begin{bmatrix}
        \mathbf{H}_N \otimes [1, 1] \\
        \mathbf{I}_N \otimes [1, -1]
    \end{bmatrix},
\end{equation}
where $\mathbf{I}_N \in \mathbb{R}^{N\times N}$ denotes the identity matrix and $\otimes$ represents the Kronecker product \cite{kolda2009tensor}. The Haar matrix plays an important role in the analysis of the localized features due to its simplicity and orthogonality. In this paper, we study the Haar transform from the perspective of circulant structure, and discuss how it can be efficiently and effectively integrated into the patch-based denoising framework.
\section{Method}
In this section, we provide a detailed description of the proposed Haar-tSVD method, with its flowchart depicted in Fig. \ref{Fig_framework_Haar_tSVD}. We begin by the application of Haar-tSVD to sRGB image, then introduce its adaptive variant and extension to other imaging data.
\input{Fig_framework_Haar_tSVD}
\vspace{-5.8pt}
\subsection{Searching Similar Patches}
Given two image patches $\mathcal{P}_{i}$, $\mathcal{P}_{j}$ $\in \mathbb{R}^{ps\times ps\times 3}$, directly calculating their Euclidean distance is time-consuming and subject to the influence of severe noise. It is noticed that the green channel normally has a higher signal-to-noise (SNR) value \cite{guo2021joint}, thus we can adopt the GCP-based patch search strategy in \cite{kong2025}. Specifically, the distance $d_{ij}$ between two patches can be calculated via
\begin{equation}\label{Equ_GCP_distance_calculation}
  d_{ij}=\left\{
    \begin{aligned}
    &\| \mathcal{P}_{i}^G -  \mathcal{P}_j^G \|, \, \, \, \|\mathcal{P}_{i}^G\| \geq \text{max}(\frac{1}{\gamma}\|\mathcal{P}_{i}^R\|, \frac{1}{\gamma} \|\mathcal{P}_{i}^B\|) \\
    & \| \mathcal{P}_{i}^{avg} -  \mathcal{P}_j^{avg} \|, \, \, \, \text{otherwise},
    \end{aligned}
    \right.
 \end{equation}
where $\mathcal{P}^R$, $\mathcal{P}^G$ and $\mathcal{P}^B$ represents the R, G and B channels of an image patch $\mathcal{P}$, respectively. $\mathcal{P}^{avg}$ is the average value of RGB channels. The weight parameter $\gamma$ is used to measure the importance of the green channel. Similar to \cite{kong2025}, we empirically set $\gamma = 1.2$. \\
\indent The major advantages of adopting GCP for patch-matching are two-fold. First, it avoids directly comparing two image patches and thus reducing the patch matching time by $\frac{2}{3}$. Besides, since the green channel is less noisier, the GCP-guided scheme is able to enhance the conformity between noisy and underlying clean patches, further encouraging group-level sparsity for the following collaborative filtering step.
\subsection{Global and Local Circulant Reperesentation}
In the collaborative filtering step, the choice of algebraic representation plays a crucial role. The proposed method is built upon the BCR in both global and local aspects to model the patch- and group-level correlations, respectively.
\subsubsection{Global Circulant Formulation}
Following \cite{kong2019color}, by integrating the BCR into the nonlocal SVD problem, we can explore the patch-level correlation via 
\begin{equation}\label{Equ_BCR_nonlocal_SVD}
\vspace{-0.69pt}
\begin{aligned}
  & \text{min} \sum_{i = 1}^N \| bcirc(\mathcal{P}_i) -  bcirc(\mathcal{U}) bcirc(\mathcal{S}_i)bcirc(\mathcal{V})^T\|^2, \\
  & \text{s.t} \quad  bcirc(\mathcal{U})^T  bcirc(\mathcal{U}) = \mathbf{I}, \,  bcirc(\mathcal{V})^T  bcirc(\mathcal{V}) = \mathbf{I},
\end{aligned}
\vspace{-0.69pt}
\end{equation}
where $bcirc(\mathcal{S}_i) \in \mathbb{R}^{3ps \times 3ps}$ preserves coefficients, $bcirc(\mathcal{U})$, $bcirc(\mathcal{V})$ $\in \mathbb{R}^{3ps \times 3ps}$ are orthogonal circulant matrices that encode row-wise and column-wise correlation of $N$ image patches. It is noticed that an image patch may share similar features with other patches across the whole image, thus we can learn a pair of global bases $bcirc(\mathcal{U})$ and $bcirc(\mathcal{V})$ to embed the common latent feature space based on randomly sampled patches. In our implementation, we use all reference patches to avoid randomness. The global circulant representation for image patches circumvents repetitive training of local bases and enhances the robustness to severe noise corruption by exploring more patch information. \\
\indent Leveraging the notations of t-SVD and t-product, Equ. (\ref{Equ_BCR_nonlocal_SVD}) can be further simplified as
\begin{equation}\label{Equ_BCR_nonlocal_SVD_simplified}
\vspace{-1.99pt}
\begin{aligned}
  & \text{min} \sum_{i = 1}^N \| \mathcal{P}_i -  \mathcal{U} * \mathcal{S}_i * \mathcal{V}^T\|^2, \\
  & \text{s.t} \quad   \mathcal{U}^T  \mathcal{U}  = \mathcal{I}, \,  \mathcal{V}^T  \mathcal{V}  = \mathcal{I},
\end{aligned}
\vspace{-1.9pt}
\end{equation}
where $\mathcal{S}_i \in \mathbb{R}^{ps\times ps \times 3}$ is a 3D coefficient tensor, $\mathcal{U}$, $\mathcal{V}$ $\in \mathbb{R}^{ps\times ps \times 3}$ are orthogonal 3D tensors, which can be regarded as a pair of pretrained/predefined projections under the gloabl BCR of image patches. 
\subsubsection{Local Circulant Structure}
To further encourage sparsity and suppress noise in the transform domain, an indispensable component is to explore the group-level correlation among similar patches. However, due to the limited search range in the patch-matching step and the interference of noise, it can be challenging to gather sufficient similar patches. Therefore, we propose to extend the circulant structure to the stacked similar patches, enabling recursive modeling of their interdependencies. Specifically, the circulant structure of a group $\mathcal{G}$ with $K$ image patches can be represented via
\begin{equation}\label{Equ_group_circulant_structure}
  circ(\mathcal{G}) = \begin{pmatrix}
                    \mathbf{p}_1^T & \mathbf{p}_K^T & \cdots & \mathbf{p}_2^T\\
                    \mathbf{p}_2^T & \mathbf{p}_1^T & \cdots & \mathbf{p}_{K-1}^T \\
                    \vdots & \vdots & \ddots & \vdots \\
                    \mathbf{p}_K^T & \mathbf{p}_{K-1}^T & \cdots & \mathbf{p}_1^T \\
                  \end{pmatrix} \in \mathbb{R}^{K \times 3Kps^2},
\end{equation}
where $\mathbf{p}_i \in \mathbb{R}^{3ps^2\times 1}$ is the vector form of the $i$-th patch $\mathcal{P}_i$ of $\mathcal{G}$. Equ. (\ref{Equ_group_circulant_structure}) indicates that every patch is modeled repetitively for $K$ times. To capture such group-level redundancy, we can learn an invertible transform matrix $\mathbf{U}_{group}\in \mathbb{R}^{K\times K}$ along the grouping dimension. The most popular approach to obtain $\mathbf{U}_{group}$ is PCA \cite{zhang2010two}, however, it is noticed that the group circulant matrix $circ(\mathcal{G})$ can be very large, thus performing local PCA for each group will lead to high computational burden. To address this concern, we start from the EVD of $circ(\mathcal{G})circ(\mathcal{G})^T$ via
\begin{equation}\label{Equ_eigen_group_circ}
  circ(\mathcal{G})circ(\mathcal{G})^T\mathbf{u} = \lambda\mathbf{u},
\end{equation}
where $circ(\mathcal{G})circ(\mathcal{G})^T$ is also a circulant matrix, $\lambda$ is the eigenvalue corresponding to the eigenvector $\mathbf{u}$ of length $K$. An interesting observation of the circulant pattern $circ(\mathcal{G})$ from Equ. (\ref{Equ_group_circulant_structure}) is that the sum of each row and column of $circ(\mathcal{G})$ is the same and equal to a row vector $\mathbf{r}_{sum}^T$, which can be obtained by
\begin{equation}\label{Equ_sum_row}
  \mathbf{r}_{sum}^T = \sum_{i = 1}^{K} \mathbf{p}_i^T.
\end{equation}
Taking advantage of the circulant property of $circ(\mathcal{G})circ(\mathcal{G})^T$ and $circ(\mathcal{G})$ in Equ. (\ref{Equ_eigen_group_circ}) and Equ. (\ref{Equ_sum_row}), we can derive the largest eigenvalue $\lambda_{max}$ and the corresponding eigenvector $\mathbf{u}_{max}$ of $circ(\mathcal{G})circ(\mathcal{G})^T$ via
\begin{equation}\label{Equ_first_eigen}
\begin{aligned} 
    \lambda_{max} & = \sum_{i = 1}^{K} \mathbf{p}_i^T \sum_{i = 1}^{K} \mathbf{p}_i= \mathbf{r}_{sum}^T \mathbf{r}_{sum}, \\ 
    \mathbf{u}_{max} & = \frac{1}{\sqrt{K}}(1,1,\ldots,1)^T.
\end{aligned}
\end{equation}
Interestingly, we observe that if $K$ is a power of 2, then the dominant vector $\mathbf{u}_{max}^T$ coincides with the first row of the Haar transform matrix defined in Equ. (\ref{Equ_general_Haar_mtx}). This indicates that the Haar transform inherently encodes the first principal component of $circ(\mathcal{G})circ(\mathcal{G})^T$. Since noise can be effectively suppressed by truncating small coefficients in the transform domain, the orthogonal Haar transform can serve as an efficient alternative to PCA under circulant representation. Therefore, the group-level correlation among similar patches can be captured by the Haar transform via
\begin{equation}\label{Equ_group_as_Haar}
  \mathbf{U}_{group} = \mathbf{U}_{Haar}.
\end{equation}
\indent Employing the Haar transform enjoys two computational benefits. First, it enables modeling the group-level redundancy without requiring an explicit circulant structure. Besides, the predefined Haar transform matrix $\mathbf{U}_{Haar}$ eliminates the need to learn a local $\mathbf{U}_{group}$ for each patch group $\mathcal{G}_{noisy}$. \\
\indent With the global and local circulant representation, the collaborative filtering step can be represented by a simple one-step transform-domain scheme \cite{kong2019color}. Specifically, the coefficients $\mathcal{S}_{noisy}$ can be derived via
\begin{equation}\label{Equ_forward_transform_all}
  \mathcal{S}_{noisy} = \mathcal{U}^T*\mathcal{G}_{noisy}*\mathcal{V} \times _4\mathbf{U}_{Haar}^T.
\end{equation}
Then we can apply the hard-thresholding technique \cite{donoho1994ideal} to shrink the coefficients of $\mathcal{S}_{noisy}$ under threshold $\tau$
 \begin{equation}\label{Equ_hard_thresholding}
 \vspace{-1.8pt}
  \mathcal{S}_{truncate}=\left\{
    \begin{aligned}
    {\mathcal{S}}_{noisy}, \quad |{\mathcal{S}}_{noisy}| \geq \tau, \\
    0, \quad |{\mathcal{S}}_{noisy}| < \tau.
    \end{aligned}
    \right.
 \end{equation}
The estimated clean group $\mathcal{G}_{estimate}$ is obtained by taking the inverse of Equ. (\ref{Equ_forward_transform_all}) with $\mathcal{S}_{truncate}$
\begin{equation}\label{Equ_backward_transform_all}
  \mathcal{G}_{estimate} = \mathcal{U}*\mathcal{S}_{truncate}*\mathcal{V}^T \times _4\mathbf{U}_{Haar}^{-1}.
\end{equation}
Finally, the estimated clean patches of $\mathcal{G}_{estimate}$ are averagely written back to their original location according to Equ. (\ref{aggregation}). Algorithm \ref{Algorithm_Haar-tSVD} summarizes the proposed Haar-tSVD denoiser.
\input{Algorithm_Haar-tSVD}
\vspace{-5.998pt}
\subsection{Complexity Analysis}
The computational cost of the proposed Haar-tSVD for a local group comprises three components: (i) the search of $K$ similar patches within a local window $O(KW^2ps)$, (ii) the global t-SVD projection $O(Kps^3)$, and (iii) the Haar transform projection $O(K^3)$. This results in an overall computational complexity of $O([KW^2ps + Kps^3 + K^3])$. Compared to many effective t-SVD based approaches \cite{zhang2016exact, kong2019color, gong2020low, Shi2022tSVD, kong2025}, the proposed method is fully parallelizable, because it gets rid of the need to learn all local transform bases.
\vspace{-3.66pt}
\subsection{Noise Estimation and Adaptive Strategy}
\input{Fig_CNN_noise_est}
Apart from an effective collaborative filtering scheme, noise estimation plays a crucial role for an efficient denoiser. Similar to \cite{kong2025}, we can treat the noise estimation problem as a classification task and train a noise estimator using CNN. Fig. \ref{Fig_CNN_noise_est} shows the CNN-based noise estimation method, the rationale behind this approach lies largely in the robustness of PCA to misclassification of noise level $\sigma$. However, as shown in Fig. \ref{Fig_compare_Haar_PCA_noise_variation}, although adopting the Haar transform can produce almost the same denoising result as PCA-tSVD\cite{kong2025}, its predefined transform bases may be more sensitive to the variation of the input noise level, leading to obvious oversmooth effects.
\input{Fig_compare_Haar_PCA_noise_variation} \\
\indent To enhance the robustness and adaptiveness of the proposed Haar-tSVD, we study the EVD of the circulant matrix $circ(\mathcal{G})circ(\mathcal{G})^T$. Based on the circulant property, one of its eigenvalue-eigenvector pairs $(\hat{\lambda}, \hat{\mathbf{u}})$ can be obtained via
\begin{equation}\label{Equ_other_eigen_value}
  \begin{aligned} 
    \hat{\lambda} & = \sum_{i = 1}^{K} (-1)^i \mathbf{p}_i^T \sum_{i = 1}^{K} (-1)^i\mathbf{p}_i, \\ 
    \hat{\mathbf{u}} & = \frac{1}{\sqrt{K}}(-1,1,\ldots,-1,1)^T.
  \end{aligned}
\end{equation}
Interestingly, Equ. (\ref{Equ_other_eigen_value}) reveals that $\hat{\lambda}$ and $\hat{\mathbf{u}}$ capture the cumulative difference among adjacent similar patches within a local group $\mathcal{G}$. In each group $\mathcal{G}$, all patches are sorted by their Euclidean distance to the reference patch, while the group-level projection on $\hat{\mathbf{u}}$ assigns an equal weight of $\frac{1}{\sqrt{K}}$. \\
\indent Therefore, we can use $\hat{\lambda}$ to model the inner-group similarity. Specifically, in noise-free or low noise images, it is easy to find sufficient similar patches, resulting in minor difference among them. Consequently, $\hat{\lambda}$ tends to appear among the smallest eigenvalues of $circ(\mathcal{G})circ(\mathcal{G})^T$. By contrast, severe noise will lead to low inner-group similarity, and $\hat{\lambda}$ is more likely to appear among the largest eigenvalues. To demonstrate the impact of noise, we denote by $a$ the rank position of $\hat{\lambda}$ among the $K$ eigenvalues sorted in ascending order. Fig. \ref{Fig_change_lambda} illustrates how noise degrades the patch matching process and causes variation in $a$ when $K = 32$.
\input{Fig_change_lambda}\\
\indent From Fig. \ref{Fig_change_lambda}, we observe that $\hat{\lambda}$ is the 10-th smallest eigenvalue in a noise-free group, but becomes the second largest with the presence of noise. This suggests that the rank position $a$ can serve as an indicator of noise severity, thus we can adjust the noise level $\sigma_{est}$ estimated by CNN via
\begin{equation}\label{Equ_adjust_noise_level}
  \hat{\sigma}=
  \left\{
    \begin{aligned}
    \frac{1}{\beta} \sigma_{est}, \quad a \leq \gamma, \\
    \sigma_{est}, \quad a > \gamma,
    \end{aligned}
  \right.
\end{equation}
where $\hat{\sigma}$ is the final estimated noise level, $\beta$ and $\gamma$ are weighting parameters empirically set to 1.2 and 13, respectively. From Equ. (\ref{Equ_adjust_noise_level}), we notice that it is not necessary to perform full EVD to determine $a$. To further reduce computation complexity, we avoid adjusting $\sigma_{est}$ for every local group $\mathcal{G}$. Instead, we randomly sample a few patch groups within a subimage and determine the corresponding $\hat{\sigma}$ through majority voting. We term the proposed Haar-tSVD with the adaptive noise estimation strategy as `A-Haar-tSVD'. Algorithm \ref{Algorithm_A-Haar-tSVD} summarizes the complete procedure of A-Haar-tSVD.
\input{Algorithm_A-Haar-tSVD}

\subsection{Discussion}
\subsubsection{Fast realization of Haar-tSVD} The proposed method consists of two key components: patch-matching and global-local transform. Fortunately, these two steps are inherently parallelizable. Specifically, the patch search step involves the calculation of Euclidean distance, while the transform bases $\mathcal{U}$, $\mathcal{V}$ and $\mathbf{U}_{Haar}$ in Equ. (\ref{Equ_forward_transform_all}) are shared by all patch groups. Concequently, the computational burden lies mainly in matrix-vector multiplications, and thus we are able to exploit parallel computing tools and techniques to achieve at least 10x speedup compared to the baseline MATLAB serial implementation. Besides, in practical applications, the parameter-tuning process can be further accelerated by caching intermediate results such as the similar patch indices and coefficients $\mathcal{S}_{noisy}$.
\subsubsection{Relationship with state-of-the-art patch-based denoisers} Built upon the t-SVD and Haar transforms, the proposed method absorbs the ideas of two representative patch-based approaches: MSt-SVD \cite{kong2019color} and BM3D \cite{dabov2007image}, which can be interpreted by a combination of patch-level projection $\mathcal{T}_{patch}$ and group-level transform $\mathcal{T}_{group}$ via
\begin{equation}\label{Equ_patch_group_transform}
  \mathcal{S} = \mathcal{T}_{group} \circ \mathcal{T}_{patch}(\mathcal{G}_{noisy})
\end{equation}
Compared to MSt-SVD, we alleviate the need to train local PCA basis $\mathcal{T}_{group}$ for each group. Instead, we employ the Haar transform to efficiently capture group-level correlation, and exploit EVD for adaptive noise estimation. Moreover, unlike BM3D that adopts a two-stage filtering scheme with predefined patch size, the proposed method is a one-step algorithm that takes advantage of the t-SVD bases to model patch-level redundancy and thus enjoys more flexibility.
\subsubsection{Extension to other imaging techniques} Since the proposed Haar-tSVD method poses no constraints on the patch size or the number of spectral bands, it can be naturally extended to other imaging techniques such as MRI and HSI. Briefly, an MRI patch can be modeled as a small cube of size $ps \times ps \times ps$, while an HSI patch can be treated as a long tube $ps\times ps \times N_{bands}$. Moreover, it is noticed that human eyes are more sensitive to green color, which corresponds to the medium wavelengths of HSI data. To leverage such spectrum-wise prior \cite{kong2025}, we can utilize the bands of medium wavelengths to guide denoising by applying the proposed adaptive noise estimation scheme to their mean value to avoid manual noise level selection. Therefore, the pre-trained CNN noise estimator can be directly adopted without retraining.

\section{Experiments}
In this section, we present the results of over 40 methods/models across various denoising tasks, including digital images, video sequences and HSI data. For each method, we utilize the authors’ original code or, where available, cite the published results directly. GPU-accelerated methods are executed using Google Colab Pro's computational resources, and all other experiments are conducted on a computer equipped with Core(TM) i7-10700F @2.9 GHz and 16GB RAM.
\subsection{Implementation Details}
The proposed Haar-tSVD involves four primary parameters: the patch size $ps$, the number of similar patches $K$ within a group, the window size $W$ for patch search, and the hard-thresholding parameter $\tau$. For image, video and HSI data, we set $ps = 8$, $K = 32$, $W = 18$ and $\tau = \sigma \sqrt{2log(cKps^2)}$, where $c$ denotes the number of channels or spectral bands. For A-Haar-tSVD, we train the CNN estimator according to \cite{kong2025}. For all comparison methods, parameters and models are carefully chosen to ensure optimal performance.
\subsection{Datasets}
\input{Fig_illus_dataset}
\input{Table_dataset_info_full}
\indent Fig. \ref{Fig_illus_dataset} shows representative samples from datasets used in image, video, and HSI denoising tasks, while Table \ref{Table_dataset_description} summarizes the statistics of popular datasets. In general, datasets for synthetic experiments consist of noise-free (ground-truth) images acquired under ideal conditions with adequate lighting and controlled camera settings, and the noisy images are generated by manually adding noise to the noise-free ones. In many real-world applications, images are inevitably contaminated by noise to various degrees, often decided by the environments and imaging devices. In such cases, the image averaging strategy is often adopted \cite{nam2016holistic, xu2018real, yue2020supervised} to generate the ground-truth data by averaging a series of images captured based on the same, static scene. \\
\indent Specifically, the image averaging strategy is simple and effective, which typically acquires the mean image by averaging over 500 noisy observations of the same scene. Ideally, the quality of the mean image improves with an increasing number of samples. However, collecting a large number of noisy images is time-consuming, and it will cause irreversible damage to the camera devices and requires substantial human effort for preparation and post-processing, which may eventually lead to degraded results due to misalignment and blurry effects. In \cite{kong2023comparison}, Kong et. al have recently shown that in many scenarios, sampling 100 noisy images may be enough to produce high-quality reference image for benchmarking.\\
\indent Compared to the passion for producing real-world image datasets, fewer attempts have been made to sample realistic video sequences. The image averaging strategy can not be directly applied to videos, and the difficulty lies mainly in continuously capturing noisy-clean video pairs for dynamic scenes. Moreover, the strategy of generating clean dynamic videos using low ISO and long exposure time may cause motion blur. Notably, Yue et. al \cite{yue2020supervised} proposed to generate video frames by manually controlling static objects. Kong \cite{kong2023comparison} introduced a video-by-video scheme to reduce human intervention. Nevertheless, producing noisy/clean video pairs is of great difficulty. Similarly, collecting real-world HSI and MRI data is also costly and labor-intensive.
\subsection{Real-world Color Image Denoising}
\input{Table_Color_Image_results}

Table \ref{Table_Color_Image_results} presents the PSNR and SSIM \cite{wang2004image} values of representative traditional patch-based denoisers and supervised/self-supervised DNN models. Overall, the proposed Haar-tSVD method produces competitive results and its adaptive variant A-Haar-tSVD shows steady improvements. Specifically, on the DND and SIDD datasets, A-Haar-tSVD achieves the PSNR gains of at least 0.24dB and 0.84dB over CBM3D and MSt-SVD, respectively. On other benchmarks such as CC, PolyU, HighISO and IOCI datasets, the proposed method is among the most effective denoising algorithms. Interestingly, it is noticed that the performance of many outstanding DNN models deteriorate when training or validation data are not available. Our observations highlight the adaptiveness and generalization capability of the proposed method.\\
\indent To further understand the effectiveness of the proposed method, visual evaluations are presented in Fig. \ref{Fig_compare_Haar_A_Haar} to Fig. \ref{Fig_compare_with_IOCI_CANON5D}. Specifically, from Fig. \ref{Fig_compare_Haar_A_Haar}, we can see that compared to the classic CBM3D, the use of the t-SVD transform better preserves structural information. Besides, the adaptive strategy in A-Haar-tSVD further enhances detail recovery from noisy signals. From Fig. \ref{Fig_compare_with_DND}, it is evident that traditional patch-based denoisers struggle with heavily corrupted images, since the grouping and local transform learning steps are adversely affected by the presence of noise. Nevertheless, the proposed method produces less distortions and color artifacts than methods such as SASL and Bitonic, because it benefits from the GCP-based patch search and the redundancy encoded in the local and global-local circulant structures. Fig. \ref{Fig_compare_with_HighISO} and Fig. \ref{Fig_compare_with_IOCI_CANON5D} show that when confronted with different and unseen noise patterns, the well-trained DNN models may also leave unwanted artifacts and suffer from over-smooth effects to varying degrees. By comparison, the proposed adaptive strategy shows its strengths by exploiting both the CNN estimator and nonlocal information, which render certain robustness and adaptability, therefore achieving a balance between noise suppression and detail preservation. 
\input{Fig_compare_Haar_A_Haar}
\input{Fig_compare_with_DND}
\input{Fig_compare_with_HighISO}
\input{Fig_compare_with_IOCI_CANON5D}\\
\indent From the perspective of denoising speed, as listed in Table \ref{Table_time_complexity_comparison}, supervised DNN models generally achieve fast inference time thanks to the power of modern GPU devices. For instance, Restormer and Condformer can process an image of size $512\times 512 \times 3$ within 1s. The proposed A-Haar-tSVD is among the few methods parallel with the state-of-the-art BM3D in terms of both efficiency and effectiveness, because its grouping step is performed only on the green/opponent channel and it does not need to recursively train local transforms. In addition, compared with other complex network architectures, A-Haar-tSVD is lightweight and enjoys considerably low complexity. Specifically, the training time of the simple CNN noise estimator is about $\frac{1}{100}$ that of the advanced Restormer model. These properties make the proposed scheme a highly practical denoising solution for real-world denoising tasks.
\input{Table_time_complexity_comparison}
\vspace{2.68pt}
\subsection{Real-world Color Video Denoising}
Videos are more informative than images with dynamic objects and temporal continuity. For Haar-tSVD, the patch search is applied to both spatial and temporal dimension. The trained CNN-based estimator and the adaptive noise adjustment of A-Haar-tSVD is adopted in a frame-by-frame fashion. In our experiments, all video sequences of CRVD and IOCV are used for evaluation. The PSNR and SSIM values are computed as the average across all frames \cite{Tassano_2020_CVPR}. Table \ref{Table_Color_Video_Denoising} lists the quantitative results of compared methods. The proposed method achieves very competitive results on both datasets, outperforming baselines by a margin of at least 0.16dB in PSNR. This demonstrates the effectiveness of combining the global t-SVD with Haar transform on capturing the nonlocal characteristics of video data across different frames. 
\input{Fig_compare_with_CRVD_case9}\\
\indent Visual comparisons are provided in Fig. \ref{Fig_compare_with_CRVD_case9} to Fig. \ref{Fig_compare_with_IOCV_case2}. As shown in Fig. \ref{Fig_compare_with_CRVD_case9} and Fig. \ref{Fig_compare_with_IOCV_case2}, even under moderate noise levels, the pretrained DNN models tend to exhibit more obvious over-smooth effects. In particular, Fig. \ref{Fig_compare_with_CRVD_case9} depicts a scene with static background, where the toy girl in blue is dynamic and moves in more than one directions. Consequently, some details and textures appear only in certain frames. Therefore, benefiting from the adaptive noise estimation strategy and the nonlocal characteristics, the proposed A-Haar-tSVD method effectively captures spatiotemporal similarity and preserves more structural information. Besides, Fig. \ref{Fig_compare_with_CRVD_case6} and Fig. \ref{Fig_compare_with_CRVD_case18} demonsgtrate the robustness of the proposed method to severe noise corruptions. By leveraging the patch- and group-level redundancy across different frames, the proposed method is able to suppress noise and mitigate color artifacts.
\input{Table_Color_Video_Denoising}
\input{Fig_compare_with_CRVD_case6}
\input{Fig_compare_with_CRVD_case18}
\input{Fig_compare_with_IOCV_case2}

\subsection{Real-world HSI Denoising}
HSI plays a vital role in a variety of remote sensing applications\cite{song2020unsupervised}. In this subsection, we evaluate the performance of related denoising methods on real-world HSI data with indoor and outdoor scenes \cite{zhang2021hyperspectral}. Alongside the popular spatial-based quality indices PSNR and SSIM, we adopt two widely recognized spectral-based quality indicators for HSI, namely spectral angle mapper (SAM) \cite{yuhas1990determination} and relative dimensionless global error in synthesis (ERGAS) \cite{wald2002data}. Unlike PSNR and SSIM, lower values of SAM and ERGAS indicate better reconstruction quality in the spectral domain.
\input{Table_Real_HSI}

\indent To handle a given HSI data $\mathcal{Y} \in \mathbb{R}^{H\times W \times N_{bands}}$ with the proposed method, we can treat each local patch as a long tube and set $ps = 8\times 8 \times N_{bands}$. Other parameters are kept the same as those used in the image denoising task. To apply the CNN noise estimator without retraining, we estimate the noise of $\mathcal{Y}$ based on the mean value across all spectral bands. Objective results of compared methods on the Real-HSI dataset are given in Table \ref{Table_Real_HSI}. Compared to sRGB images, the HSI data exhibits rich spatial and spectral correlations, which can be effectively modeled with circulant structures. Therefore, from Table \ref{Table_Real_HSI}, we notice that Haar-tSVD is able to produce competitive denoising performance compared with state-of-the-art low rank tensor-based approaches. Meanwhile, by getting rid of the complex iterative filtering strategy and local transform learning step, our method achieves a comparable inference time to advanced DNN models. Notably, since our approach does not require costly training on advanced GPU devices, it is a potentially effective and efficient tool for handling large HSI data in practice. 
\input{Fig_compare_with_Real_HSI_case4}
\input{Fig_compare_with_Real_HSI_case5}

\indent To assess the effectiveness of compared methods, we present their visual performance for the Real-HSI dataset in Fig. \ref{Fig_compare_with_Real_HSI_case4} and Fig. \ref{Fig_compare_with_Real_HSI_case5}. We can see that the proposed method shows pleasing results and balances smoothness and sharp details. In contrast, BM4D tends to produce obvious oversmooth effects, since its predefined patch-level transforms may not fully exploit the correlation among all the spectral bands. Additionally, the state-of-the-art tensor-based method OLRT fail to preserve high-frequency components such as edges and textures, while LTDL introduces residual stripe noise. These observations suggest that increasing the number of iterations and local similar patches may not help preserve fine details and structure of HSI data. Furthermore, the powerful DNN models FlexDLD and RAS2S present impressive noise removal ability with slightly more unwanted artifacts.\\
\indent To further study the denoising capability of the Haar-tSVD transform, we visualize more results on other real-world HSI data in Fig. \ref{Fig_HSI_PaviaU} to Fig. \ref{Fig_HSI_Urban}. Benefiting from the rich nonlocal and redundant information of HSI data, the proposed Haar-tSVD shows competitive performance in term of both noise removal and detail recovery in various challenging cases.
\input{Fig_HSI_PaviaU}
\input{Fig_HSI_EO1}
\input{Fig_HSI_Cuprite}
\input{Fig_HSI_Urban}
\vspace{-5.8pt}
\subsection{Extention To Other Imaging Techniques}
We notice that a large proportion of compared methods adopt the additive white Gaussian noise (AGWN) assumption due to its simplicity. Recently, Romano et al. \cite{romano2017little} have pointed out that the removal of AGWN from an image is largely a solved problem, which may help explain the popularity of such simplified noise modeling. In fact, real-world noise can be more complex and challenging, as it may be multiplicative and signal dependent. Therefore, there are several non i.i.d Gaussian models designed for different applications, such as the strip noise removal \cite{chen2017denoising} and mixture of Gaussian and sparse noise \cite{goossens2009image}. In our experiments, we have demonstrated the effectiveness of the proposed method to handle real-world noise of sRGB images and video data, as well as stripe noise in HSI. To further assess its adaptability, we evaluate its performance on fluorescence microscopy (FM) and MRI data, which are dominated by Possion-Gaussian noise and Rician noise, respectively.
\subsubsection{FM denoising}
Briefly, FM is essential for capturing high-quality images of small specimens such as cells, tissues, and microorganisms. In our experiments, we use the FMDD dataset \cite{zhang2019poisson}, consisting of data captured by confocal, two-photon, and wide-field microscopes. The same parameter setting of Haar-tSVD and the pretrained CNN estimator are exploited. The results presented in Table \ref{Table_FMDD_Haar_tSVD} shows the advantage of the proposed Haar-tSVD and the generalization capability of the corresponding adaptive scheme in terms of both objective evaluations and running time. Visual comparison presented in Fig. \ref{Fig_FMDD_A_Haar_tSVD} shows that the overcomplete circulant representation and the adaptive strategy can also help preserve true color and details for FM data.
\input{Table_FMDD_Haar_tSVD}
\input{Fig_FMDD_A_Haar_tSVD}
\subsubsection{MRI denoising}
MRI is a non-invasive imaging technology that produces three dimensional detailed anatomical images, which has been shown to be helpful in disease detection, diagnosis, and treatment monitoring \cite{vlaardingerbroek2013magnetic}. Different from the AWGN noise modeling, the noise of MRI data is often assumed to follow the Rician distribution \cite{awate2007feature}. To handle Rician noise, the technique of forward and inverse variance stabilizing transforms (VSTs) \cite{foi2011noise} are often adopted. In this context, each patch of MRI data is typically modeled as a 3D cube of size $ps\times ps \times ps$. \\
\indent To evaluate the denoising performance on real-world 3D MRI data, we carry out experiments on T1w MR images from the OASIS dataset \cite{marcus2007open}. The Rician noise levels of two selected T1w data, namely OAS1\_0112 and OAS1\_0092 are estimated to be 3$\%$ and 4.5$\%$ of the maximum intensity, respectively \cite{zhang2015denoising}. We adopt the proposed adaptive approach A-Haar-tSVD with $ps = 4 \times 4\times 4$, $K = 32$ and $W = 5$. Fig. \ref{Fig_MRI_0092} and Fig. \ref{Fig_MRI_0112} present the visual evaluations of competitive nonlocal denoisers, namely BM4D \cite{maggioni2012nonlocal} and MSt-SVD \cite{kong2019color}. By taking advantage of the local adaptive adjustment scheme, the propose method restores more details and anatomical structures, while both BM4D and MSt-SVD exhibit varying degrees of oversmoothing. This interesting observation is another vivid example to show the adaptiveness of the proposed scheme to local structural variations. Furthermore, Table \ref{Table_compare_MRI_OASIS_time} presents a comparison of computational time, and we can see that A-Haar-tSVD achieves a 4x speedup over the classic BM4D and the competitive MSt-SVD, owing to its single-pass filtering strategy and the elimination of local transform learning. Therefore, it is comparatively simple to implement parallelization with multiprocessing libraries.
\input{Table_compare_MRI_OASIS_time}
\input{Fig_MRI_0092}
\input{Fig_MRI_0112}
\subsection{Parameter Analysis}
Following the classic patch-based denoising paradigm, the proposed Haar-tSVD transform relies on several key parameters such as the patch size $ps$, search window range $W$ and number of nonlocal similar patches $K$. Fig. \ref{Fig_parameter_tuning_ps_SR_maxK} evaluates the impact of these parameters on the denoising performance of Haar-tSVD across different datasets. For the patch size $ps$, a comparison between the CC15 and SIDD datasets shows that larger patch size can improve robustness against heavy noise contamination. However, increasing $ps$ may drastically raise computational burden. Similarly, expanding the search range $W$ and choosing more grouped patches $K$ do not guarantee performance enhancement due to the rare patch effect. Based on these observations, we mainly set $ps = 8$, $W = 18$ and $K = 32$ in our experiments to achieve a tradeoff between denoising performance and speed. 
\input{Fig_parameter_tuning_ps_SR_maxK}

\subsection{Ablation Study}
Our experimental results demonstrate the competitive performance of the proposed A-Haar-tSVD method. The effectiveness and robustness of this adaptive strategy can be largely attributed to the adjustment of noise level within a subimage or local group through eigenvalue analysis in Equ. (\ref{Equ_other_eigen_value}). Therefore, it is interesting to investigate the impact of the local noise level adjustment on A-Haar-tSVD. As shown in Table \ref{Table_ablation_study_W_WO_PCA_est}, we can see that the A-Haar-tSVD consistently benefits from the adaptive noise level adjustment scheme in Equ. (\ref{Equ_adjust_noise_level}) across different scenarios, while incurring minimal additional computational cost. This observation justifies the efficiency and applicability of the proposed adaptive mechanism.
\input{Table_ablation_study_W_WO_PCA_est}\\
\indent Apart from the effectiveness of the adaptive noise level adjustment scheme, we notice that it involves two weighting parameters $\beta$ and $\gamma$, which helps to control the local noise level $\sigma$ according to the eigenvalue analysis in Equ. (\ref{Equ_other_eigen_value}). We conduct a sensitivity study to evaluate the impact of $\beta$ and $\gamma$ on the proposed A-Haar-tSVD method, with the results presented in Fig. \ref{Fig_ablation_study_beta_gamma}. From Equ. (\ref{Equ_other_eigen_value}) and Fig. \ref{Fig_ablation_study_beta_gamma}, we notice that a small $\beta(<1)$ will produce a overestimation of $\sigma$ and eventually lead to lower PSNR value due to oversmooth effects. On the other hand, choosing an excessively large $\gamma$ may fail to capture the inner-group similarity under noisy conditions. Hence, a reasonable range for $\gamma$ is between 12 and 16. Based on these findings, $\beta$ and $\gamma$ are empirically set to 1.2 and 13 in our experiments, respectively. 
\input{Fig_ablation_study_beta_gamma}
\vspace{-3.8pt}
\section{Conclusion}
In this paper, we present Haar-tSVD, an efficient and straightforward one-step transform-domain method for image denoising. By establishing the connection between the Haar transform and PCA under circulant structures, we leverage global and local circulant representation to capture similarity and correlation among image patches at both patch and group levels. In addition, to enhance the robustness and effectiveness of Haar-tSVD, we introduce A-Haar-tSVD, an adaptive variant that integrates a flexible noise estimation scheme based on a CNN noise estimator and eigenvalue characteristics of circulant structures. The proposed method reduces computational cost by avoiding the training of local transform bases. To achieve further acceleration, we also design and implement fast, parallelizable strategies. Experiments demonstrate the proposed method's competitive performance in terms of both effectiveness and efficiency across a variety of real-world denoising tasks, such as digital images, video sequences and HSI data, as well as extensions to FM and MRI restoration. The promising results motivate us to further investigate how the proposed method the corresponding adaptive scheme can be better exploited for a further enhancement \cite{xu2024haar} and broader applications in other imaging domains \cite{fu2024weconvene}. 

\bibliographystyle{IEEEtran}
\bibliography{Reference}

\end{document}

%% file: Fig_traditional_framework.tex
\begin{figure}[htbp]
 \vspace{-1.9pt}
  \centering
  \graphicspath{{Figs/Frameworks/}}
  \includegraphics[width=3.46in]{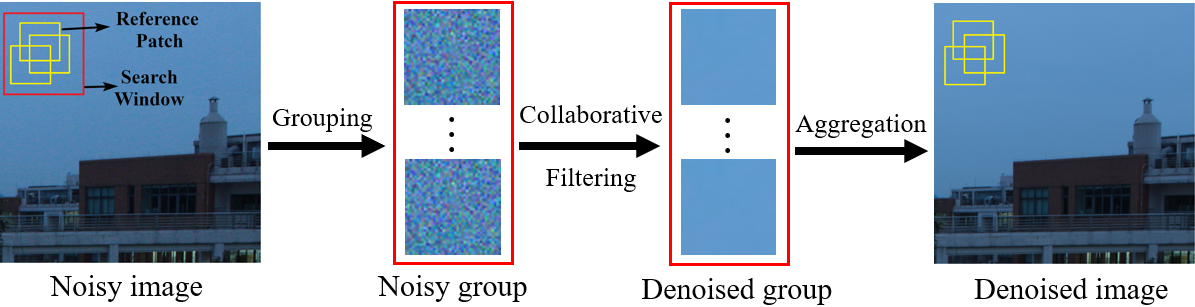}\\
  \caption{Illustration of the patch-based framework for traditional denoisers.}
  \label{Fig_traditional_framework}
  \vspace{-1.6pt}
\end{figure}

%% file: Fig_framework_Haar_tSVD.tex
\begin{figure*}[htbp]
\graphicspath{{Figs/Frameworks/}}
  \centering
  \includegraphics[width=6.998in]{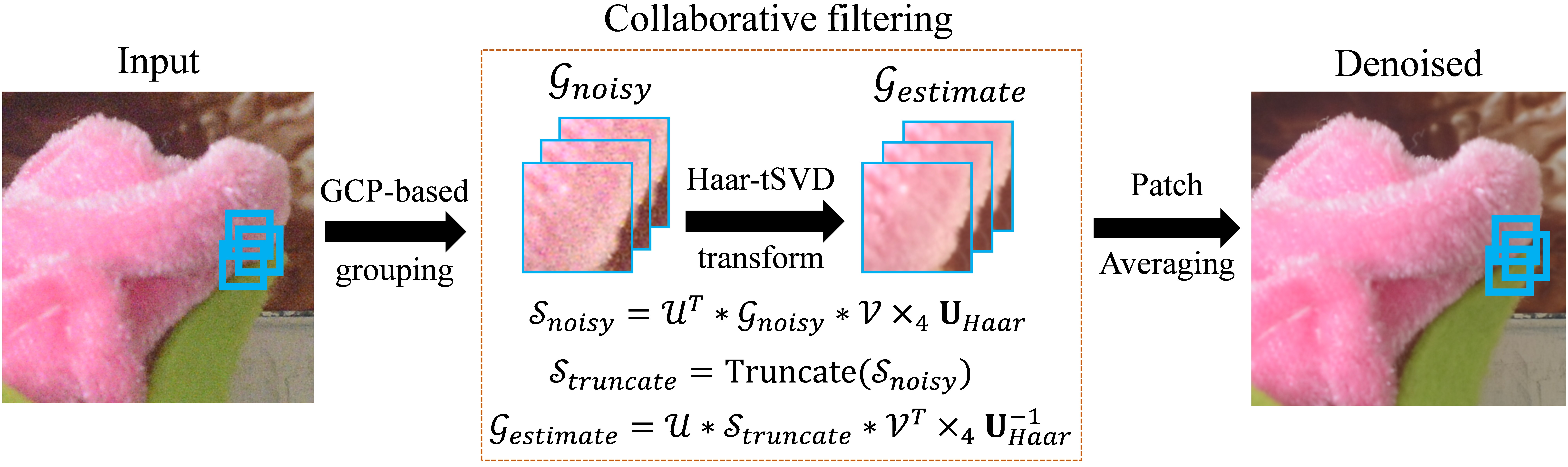}
  \caption{Flowchart of the proposed Haar-tSVD method. It is a one-step filtering algorithm built upon global and local circulant representation.}
  \label{Fig_framework_Haar_tSVD}
  \vspace{-8.18pt}
\end{figure*}

%% file: Algorithm_Haar-tSVD.tex
\setlength{\textfloatsep}{1pt}
\begin{algorithm}[!htbp]
    \caption{Haar-tSVD}
    {\bf Input:} Noisy image data $\mathcal{Y} \in \mathbb{R}^{H\times W \times 3}$, image patch size $ps$, number patches $K$, search window size $W$ and noise level $\sigma$.\\
    {\bf Output:} Estimated clean image $\mathcal{X}$.\\
    {\bf Step 1} (Grouping): For each reference patch $\mathcal{P}_{ref}$, perform similar patch-search according to Equ. (\ref{Equ_GCP_distance_calculation}) and stack $K$ similar patches in a group $\mathcal{G}_{noisy} \in \mathbb{R}^{ps\times ps \times 3 \times K}$.\\
    {\bf Step 2} (Collaborative filtering):\\
     \hspace*{0.18in}(1) Learn the global projection tensors $\mathcal{U}$ and $\mathcal{V}$ in Equ. (\ref{Equ_BCR_nonlocal_SVD}) with all reference patches. \\
     \hspace*{0.18in}(2) Perform the forward Haar transform and t-SVD projection based on Equ. (\ref{Equ_forward_transform_all}) to obtain the coefficient ${\mathcal{S}}_{noisy}$.\\
     \hspace*{0.18in}(3) Shrink ${\mathcal{S}}_{noisy}$ via hard-thresholding in Equ. (\ref{Equ_hard_thresholding}).\\
     \hspace*{0.18in}(4) Apply the inverse Haar and nonlocal t-SVD transform in Equ. (\ref{Equ_backward_transform_all}) to estimate clean group ${\mathcal{G}}_{estimate}$.\\
    {\bf Step 3} (Aggregation): Averagely write all denoised image patches of ${\mathcal{G}}_{estimate}$ to their original locations.
    \label{Algorithm_Haar-tSVD}
\end{algorithm} 

%% file: Fig_CNN_noise_est.tex
\begin{figure}[htbp]
\vspace{-1.8pt}
\graphicspath{{Figs/Method/Noise_estimation/}}
  \centering
  \includegraphics[width=3.26in]{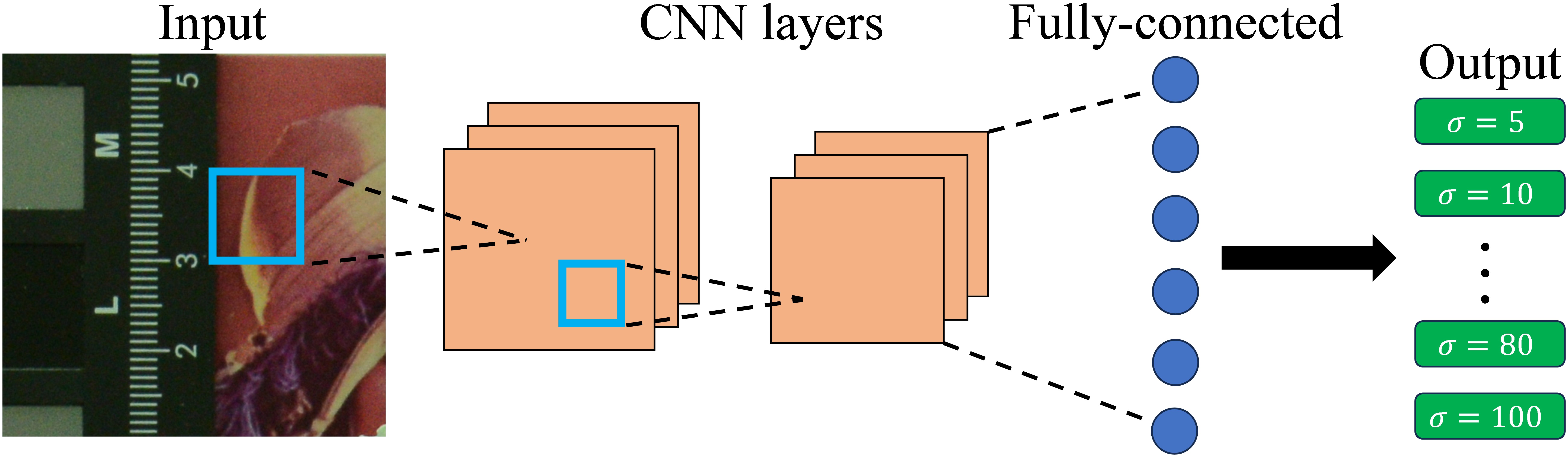}
  \caption{The CNN-based noise estimation approach.}\label{Fig_CNN_noise_est}
  \vspace{-2.8pt}
\end{figure}

%% file: Fig_compare_Haar_PCA_noise_variation.tex
\begin{figure}[htbp]
\graphicspath{{Figs/Method/}}
\centering
\subfigure[Comparison of Haar and PCA]{
\label{Fig4}
\includegraphics[width=1.628in]{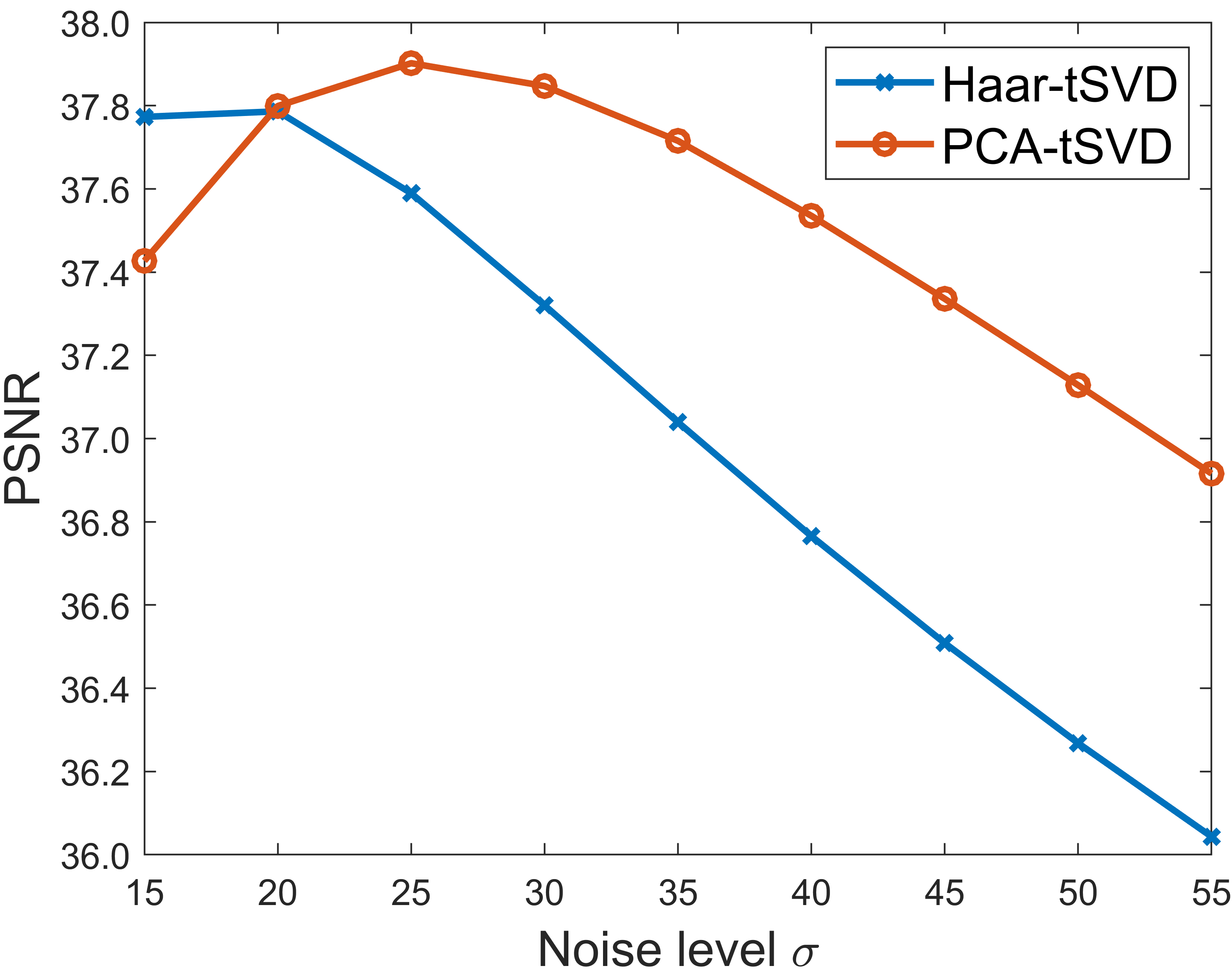}}
\subfigure[Visual effects ($\sigma = 40$)]{
\label{Fig4}
\includegraphics[width=1.66in]{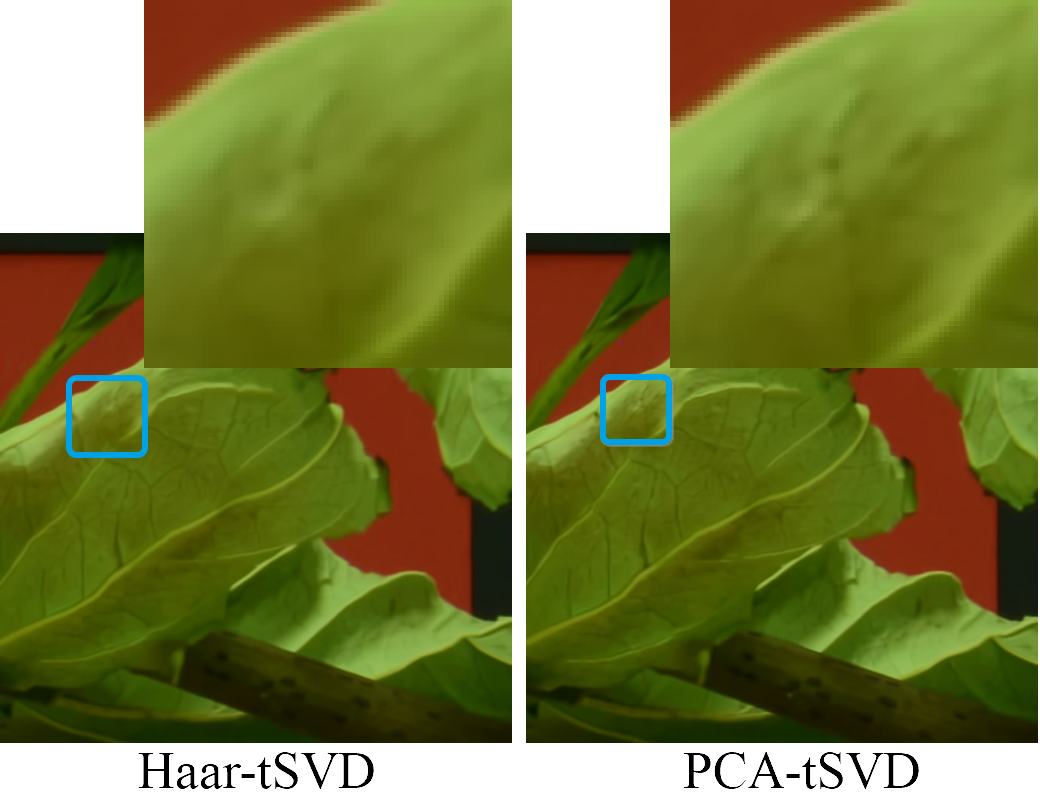}}
\vspace{-2.8pt}
\caption{Denoising effects of the Haar and PCA transforms for $\mathbf{U}_{group}$.}
\label{Fig_compare_Haar_PCA_noise_variation}
\end{figure} 

%% file: Fig_change_lambda.tex
\begin{figure}[htbp]
\vspace{-3.8pt}
\graphicspath{{Figs/Method/Noise_estimation/}}
\centering
\subfigure[Noise-free image]{
\label{Fig4}
\includegraphics[width=1.288in]{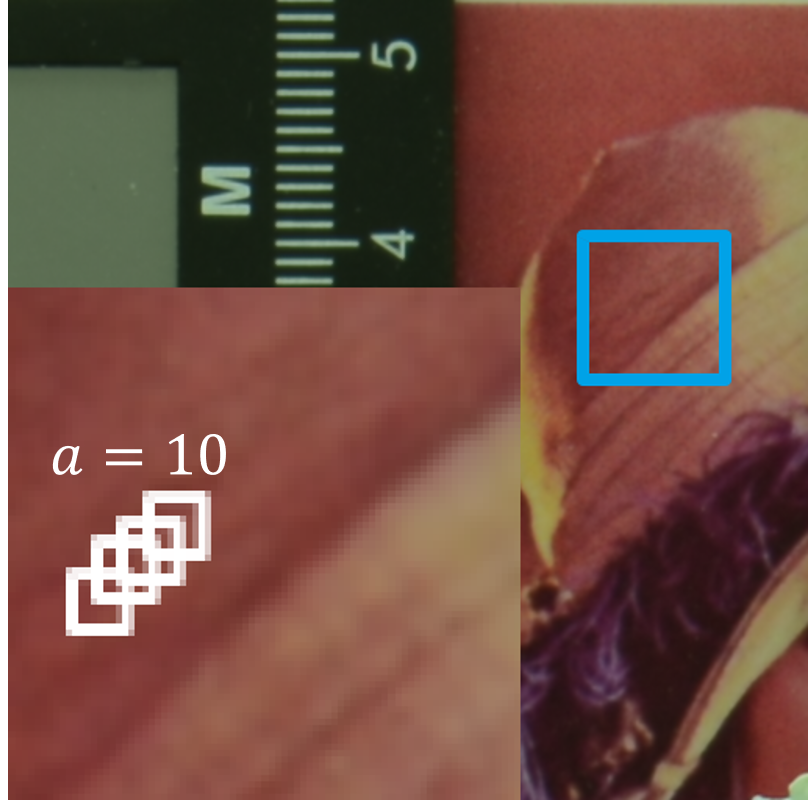}}
\subfigure[Noisy image]{
\label{Fig4}
\includegraphics[width=1.288in]{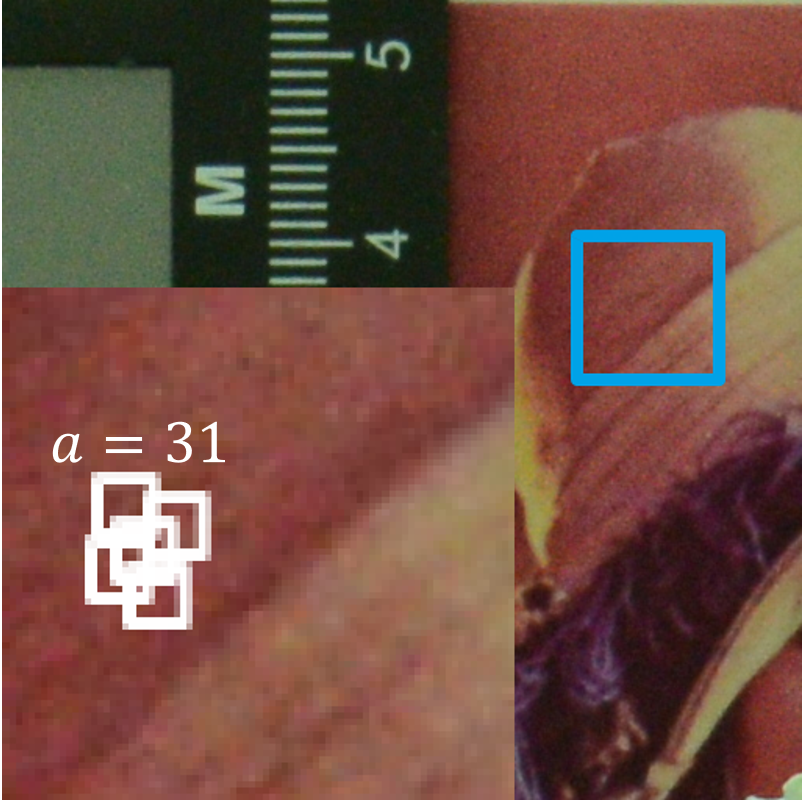}}
\vspace{-2.8pt}
\caption{Influence of real-world noise on patch-search and rank position $a$.}
\vspace{-3.8pt}
\label{Fig_change_lambda}
\end{figure} 

%% file: Algorithm_A-Haar-tSVD.tex
\setlength{\textfloatsep}{1pt}
\begin{algorithm}[!htbp]
    \caption{A-Haar-tSVD}
    {\bf Input:} Noisy image data $\mathcal{Y} \in \mathbb{R}^{H \times W \times 3}$, subimage size $H_{sub}$ and $W_{sub}$.\\
    {\bf Output:} Estimated clean image $\mathcal{X} \in \mathbb{R}^{H \times W \times 3}$.\\
    {\bf Step 1} (Noise estimation): For each subimage $\mathcal{Y}_{sub}\in \mathbb{R}^{H_{sub} \times W_{sub} \times 3}$, determine the noise level $\hat{\sigma}$ according to the CNN estimator and the adaptive strategy in Equ. (\ref{Equ_adjust_noise_level}). \\
    {\bf Step 2} (Image denoising): Based on the estimated noise level $\sigma_{est}$, apply the proposed Haar-tSVD to handle $\mathcal{Y}_{sub}$ and obtain the corresponding denoised subimage $\mathcal{X}_{sub}$.
    \label{Algorithm_A-Haar-tSVD}
\end{algorithm} 
\vspace{-6pt} 

%% file: Fig_illus_dataset.tex
\begin{figure}[htbp]
\vspace{-1.8pt}
\graphicspath{{Figs/Fig_dataset_illus/}}
  \centering
  \includegraphics[width=3.26in]{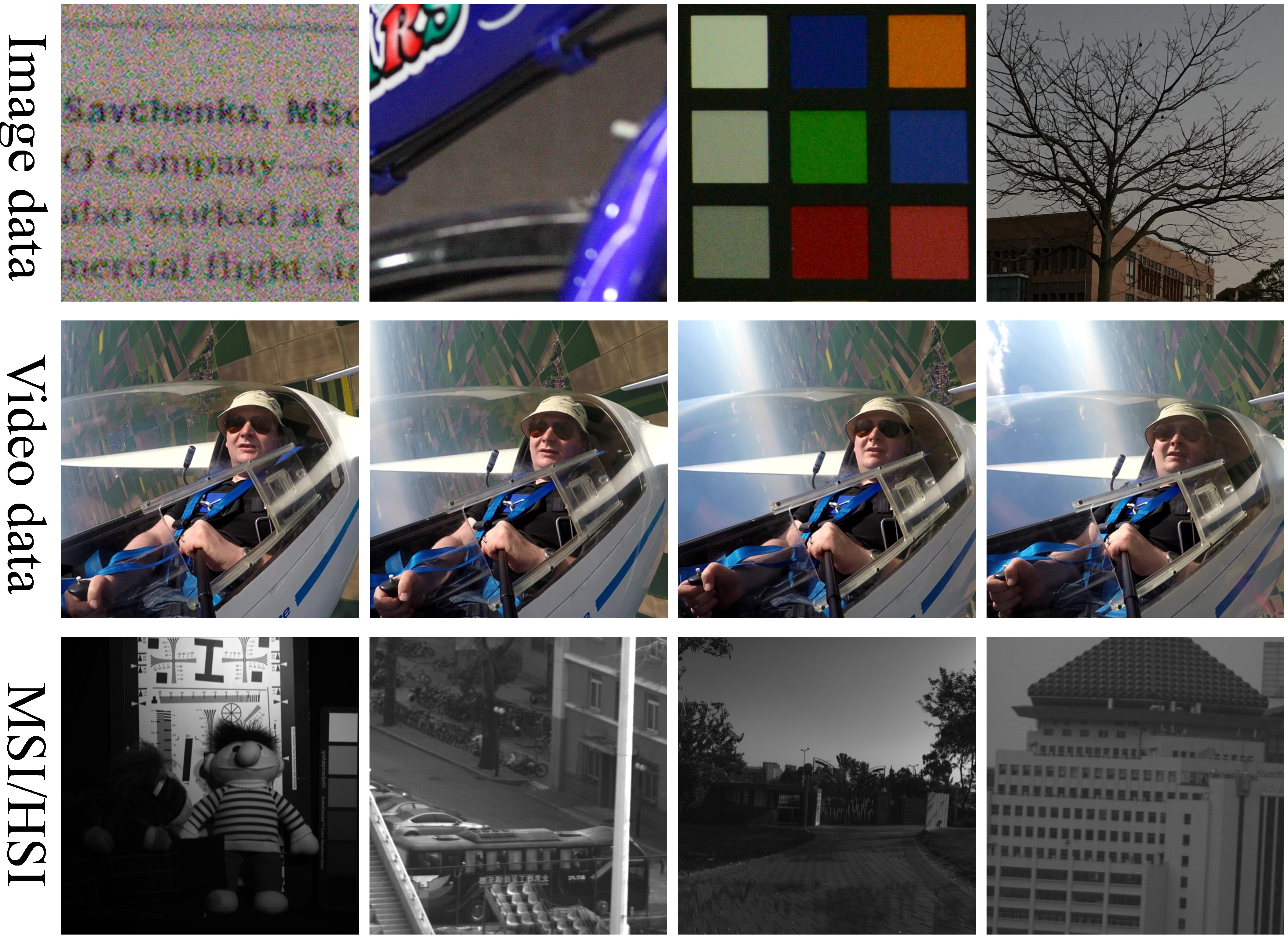}
   \vspace{-1.9pt}
  \caption{Illustrations of popular denoising datasets.}
  \label{Fig_illus_dataset}
  \vspace{-0.6pt}
\end{figure}

%% file: Table_dataset_info_full.tex
\begin{table}[htbp]
\tiny
  \centering
  \caption{Popular datasets for synthetic and real-world experiments. 'GT': ground-truth, '$\surd$': Available, '-': Not available.}
  \renewcommand{\arraystretch}{0.998}
  \scalebox{0.9699}{
    \begin{tabular}{cccccc}
    \toprule
    Applications & Name  & Experiments & GT    & Image size & \# Images \\
    \midrule
    \multicolumn{1}{c}{\multirow{6}[16]{*}{Image}} & Kodak \cite{Kodak} & Synthetic & $\surd$     & 512 $\times$ 512 $\times$ 3 & 24 \\
\cmidrule{2-6}          & Nam-CC \cite{nam2016holistic} & Real-world & $\surd$     & 512 $\times$ 512 $\times$ 3 & 15 \\
\cmidrule{2-6}          & PolyU \cite{xu2018real} & Real-world & $\surd$     & 512 $\times$ 512 $\times$ 3 & 100 \\
\cmidrule{2-6}          & DnD \cite{plotz2017benchmarking} & Real-world & -   & 512 $\times$ 512 $\times$ 3 & 1000 \\
\cmidrule{2-6}          & SIDD \cite{abdelhamed2018high} & Real-world & -   & 256 $\times$ 256 $\times$ 3 & 1280 \\
\cmidrule{2-6}          & HighISO \cite{yue2019high} & Real-world & $\surd$     & 512 $\times$ 512 $\times$ 3 & 100 \\
\cmidrule{2-6}          & IOCI \cite{kong2023comparison} & Real-world & $\surd$     & 1024 $\times$ 1024 $\times$ 3 & 848 \\
    \midrule
    \multirow{3}[6]{*}{Video} & Set8 \cite{Set8} & Synthetic & $\surd$     & 960 $\times$ 540 $\times$ 3 $\times$ frames & 8 \\
\cmidrule{2-6}          & CRVD \cite{yue2020supervised} & Real-world & $\surd$     & 1920 $\times$ 1080 $\times$ 3 $\times$ frames & 55 \\
\cmidrule{2-6}          & IOCV \cite{kong2023comparison} & Real-world & $\surd$     & 512 $\times$ 512 $\times$ 3 $\times$ frames & 39 \\
    \midrule
    \multirow{5}[5]{*}{MSI/HSI} & Real-HSI \cite{zhang2021hyperspectral} & Real-world & $\surd$     & 696 $\times$ 520 $\times$ 34 & 59 \\
\cmidrule{2-6}          & ICVL \cite{arad_and_ben_shahar_2016} & Synthetic & $\surd$     & 1392 $\times$ 1300 $\times$ 31 & 201 \\
\cmidrule{2-6}          & Urban \cite{PURR1947} & Real-world & -   & 307 $\times$ 307 $\times$ 210 & 1 \\
\cmidrule{2-6}          & HHD \cite{chakrabarti2011statistics} & Real-world & -   & 1392 $\times$ 1040 $\times$ 31 & 77 \\    \bottomrule
    \end{tabular}%
    }
  \label{Table_dataset_description}%
\end{table}%

%% file: Table_Color_Image_results.tex
\begin{table*}[htbp]
\scriptsize
  \centering
  \caption{Denoising results of compared methods on real-wolrd sRGB color image datasets.}
  \scalebox{0.6999}{
    \begin{tabular}{cccccccccccccccccccc}
    \toprule
    \multirow{3}[4]{*}{Dataset} & \multicolumn{7}{c}{Traditional denoisers}             & \multicolumn{4}{c}{DNN models (Self-supervised)} & \multicolumn{8}{c}{DNN models (supervised)} \\
\cmidrule{2-20}          & Bitonic & MCWNNM  & NLHCC & CBM3D & MSt-SVD & Haar-tSVD & A-Haar-tSVD & AP-BSN & C2N  & SASL  & TBSN  & CBDNet & ClipDe & CondFormer & DIDN  & FFDNet & DeepSN & NAFNet & Restormer \\
          & \cite{treece2022real} & \cite{xu2017multi} & \cite{hou2020nlh} & \cite{dabov2007image} & \cite{kong2019color} & (Ours) & (Ours) & \cite{lee2022ap} & \cite{jang2021c2n} & \cite{li2023spatially} & \cite{li2024rethinking} & \cite{guo2019toward} & \cite{cheng2024transfer} & \cite{huang2024beyond} & \cite{yu2019deep} & \cite{zhang2018ffdnet} & \cite{deng2025deepsn} & \cite{chen2022simple} & \cite{zamir2022restormer} \\
    \midrule
    \multirow{2}[4]{*}{DnD} & 37.85  & 37.38  & 38.85  & 37.73  & 38.01  & 38.11  & 38.25  & 37.29  & 37.28  & 38.00  & 37.78  & 38.06  & 39.57  & \textbf{40.10} & 39.64  & 37.61  & 39.92  & 38.36  & \textcolor[rgb]{ 0,  .439,  .753}{\textbf{40.03}} \\
\cmidrule{2-20}          & 0.936  & 0.929  & 0.953  & 0.934  & 0.938  & 0.939  & 0.945  & 0.932  & 0.924  & 0.936  & 0.940  & 0.942  & 0.955  & \textbf{0.956} & 0.953  & 0.942  & 0.956  & 0.943  & \textcolor[rgb]{ 0,  .439,  .753}{\textbf{0.956}} \\
    \midrule
    \multirow{2}[4]{*}{SIDD} & 36.67  & 29.54  & -  & 34.74  & 34.38  & 35.05  & 35.58  & 35.97  & -     & -     & 39.01  & 33.26  & 39.42  & \textbf{40.23} & 39.78  & 38.27  & 39.79  & \textcolor[rgb]{ 0,  .439,  .753}{\textbf{40.15}} & 40.02  \\
\cmidrule{2-20}          & 0.933  & 0.888  & -  & 0.922  & 0.901  & 0.914  & 0.925  & 0.925  & -     & -     & 0.945  & 0.869  & 0.956  & -     & 0.958  & 0.948  & 0.958  & \textcolor[rgb]{ 0,  .439,  .753}{\textbf{0.960}} & 0.960  \\
    \midrule
    \multirow{2}[4]{*}{CC15} & 35.22  & 37.02  & \textbf{38.49} & 37.70  & 37.95  & 38.15  & \textcolor[rgb]{ 0,  .439,  .753}{\textbf{38.24}} & 35.44  & 37.02  & 34.93  & 35.73  & 36.20  & 35.40  & 36.34  & 36.06  & 37.67  & 35.89  & 34.39  & 36.33  \\
\cmidrule{2-20}          & 0.925  & 0.950  & \textbf{0.965} & 0.957  & 0.959  & 0.961  & \textcolor[rgb]{ 0,  .439,  .753}{\textbf{0.963}} & 0.936  & 0.945  & 0.936  & 0.931  & 0.919  & 0.916  & 0.922  & 0.946  & 0.956  & 0.936  & 0.923  & 0.941  \\
    \midrule
    \multirow{2}[4]{*}{PolyU} & 36.64  & 38.26  & 38.36  & 38.69  & \textcolor[rgb]{ 0,  .439,  .753}{\textbf{38.85}}  & 38.78 & \textbf{38.87} & 36.99  & 37.69  & 37.13  & 36.51  & 37.81  & 36.87  & 37.27  & 37.36  & 38.76  & 37.25  & 36.38  & 37.66  \\
\cmidrule{2-20}          & 0.940  & 0.965  & 0.965  & 0.970  & \textcolor[rgb]{ 0,  .439,  .753}{\textbf{0.971}}  & 0.969 & \textbf{0.971} & 0.956  & 0.958  & 0.954  & 0.954  & 0.956  & 0.939  & 0.948  & 0.953  & 0.970  & 0.957  & 0.947  & 0.956  \\
    \midrule
    \multirow{2}[4]{*}{HighISO} & 37.37  & 39.89  & 40.29  & 40.35  & 40.49  & \textcolor[rgb]{ 0,  .439,  .753}{\textbf{40.51}} & \textbf{40.63} & 38.26  & 38.86  & 38.24  & 38.84  & 38.18  & 37.61  & 37.79  & 38.24  & 40.28  & 38.12  & 37.88  & 38.29  \\
\cmidrule{2-20}          & 0.943  & 0.970  & 0.971  & 0.974  & \textcolor[rgb]{ 0,  .439,  .753}{\textbf{0.974}}  & 0.973 & \textbf{0.974} & 0.965  & 0.960  & 0.964  & 0.966  & 0.942  & 0.955  & 0.928  & 0.950  & 0.973  & 0.951  & 0.954  & 0.948  \\
    \midrule
    \multirow{2}[4]{*}{IOCI} & 39.10  & 41.04  & 41.22  & 41.46  & 41.48  & 41.45  & \textbf{41.52} & 39.71  & 40.23  & 39.44  & 39.95  & 40.36  & 40.03  & 39.86  & 39.86  & \textcolor[rgb]{ 0,  .439,  .753}{\textbf{41.49}} & 39.04  & 38.27  & 40.10  \\
\cmidrule{2-20}          & 0.954  & 0.972  & 0.974  & 0.976  & 0.977  & 0.976  & \textbf{0.978} & 0.966  & 0.967  & 0.964  & 0.966  & 0.966  & 0.970  & 0.957  & 0.964  & \textcolor[rgb]{ 0,  .439,  .753}{\textbf{0.976}} & 0.966  & 0.939  & 0.966  \\
    \bottomrule
    \end{tabular}}%
  \label{Table_Color_Image_results}%
  \vspace{-6.18pt}
\end{table*}%

%% file: Fig_compare_Haar_A_Haar.tex
\begin{figure}[htbp]
\vspace{-8.18pt}
\graphicspath{{Figs/Selected_color_images/}}
\centering
\includegraphics[width=3.368in]{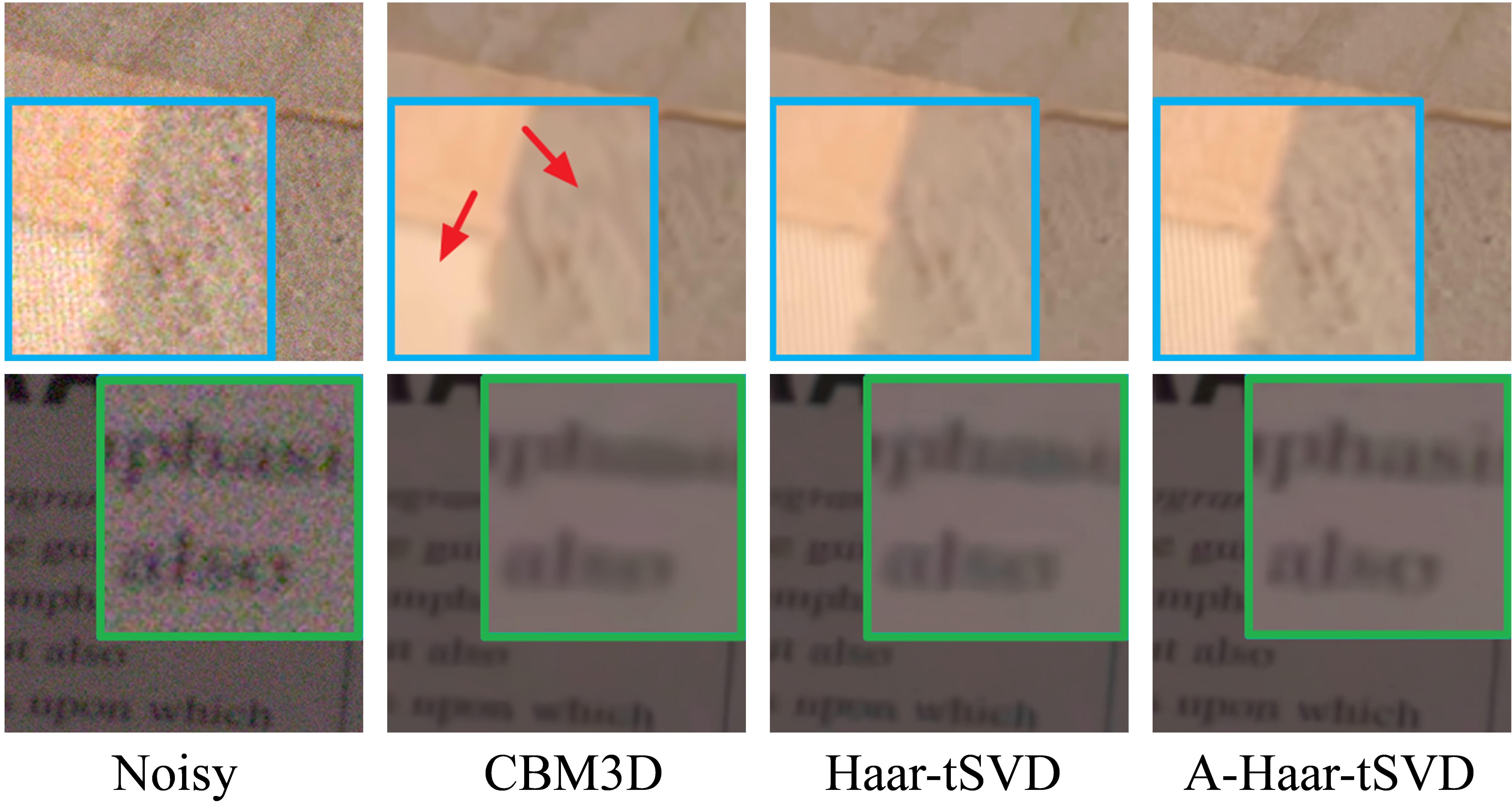}
\vspace{-2.8pt}
\caption{Denoising comparison of CBM3D, the proposed Haar-tSVD method and its adaptive variant A-Haar-tSVD.}
\vspace{-9.98pt}
\label{Fig_compare_Haar_A_Haar}
\end{figure} 

%% file: Fig_compare_with_DND.tex
\begin{figure}[htbp]
\vspace{-6.8pt}
\graphicspath{{Figs/Selected_color_images/DND/Case1/Combined/}}
\centering
\subfigure[Noisy]{
\label{Fig4}
\includegraphics[width=0.8116in]{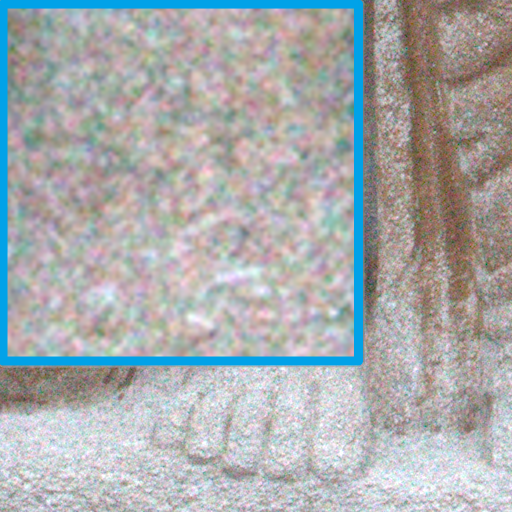}} \hfill
\subfigure[Bitonic]{
\label{Fig4}
\includegraphics[width=0.8116in]{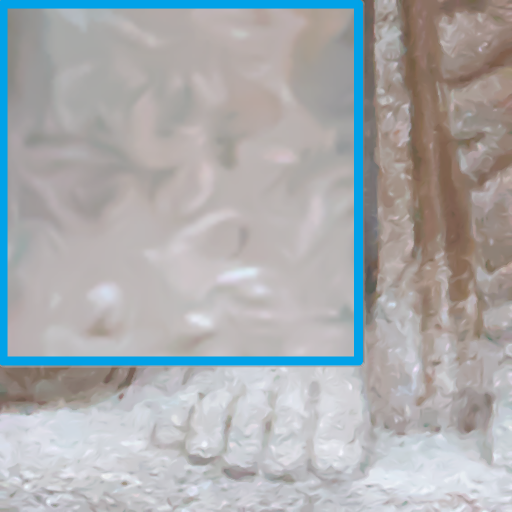}}\hfill
\subfigure[NLHCC]{
\label{Fig4}
\includegraphics[width=0.8116in]{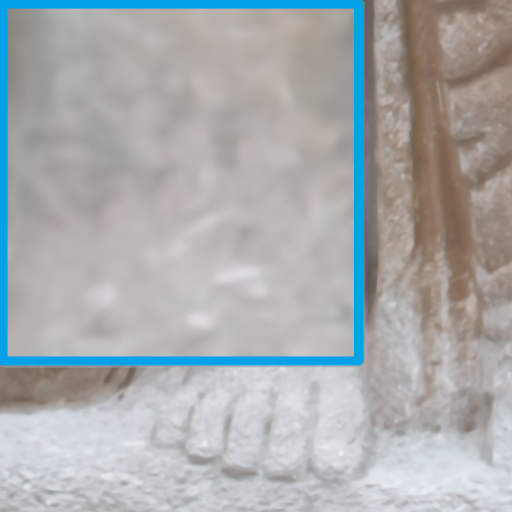}}\hfill
\subfigure[SASL]{
\label{Fig4}
\includegraphics[width=0.8116in]{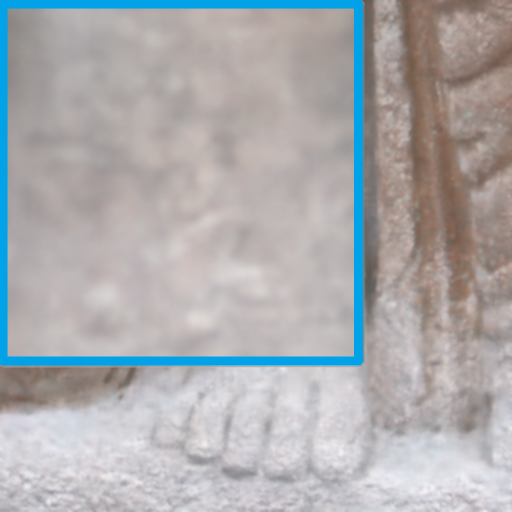}}\hfill
\vspace{-6.8pt}
\subfigure[DeepSN]{
\label{Fig4}
\includegraphics[width=0.8116in]{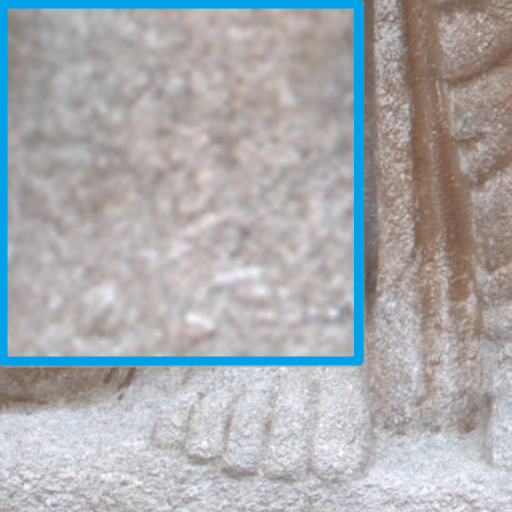}}\hfill
\subfigure[Restormer]{
\label{Fig4}
\includegraphics[width=0.8116in]{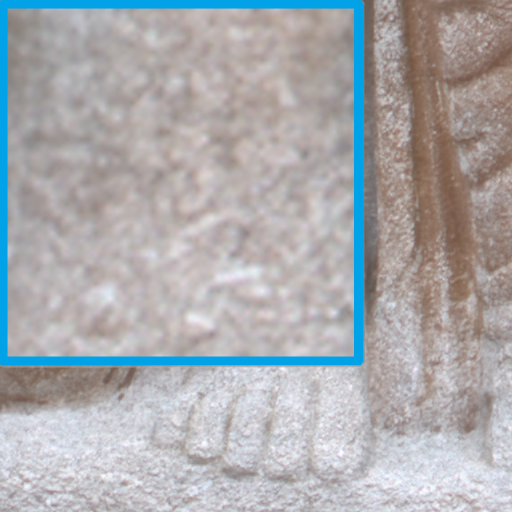}}\hfill
\subfigure[NAFNet]{
\label{Fig4}
\includegraphics[width=0.8116in]{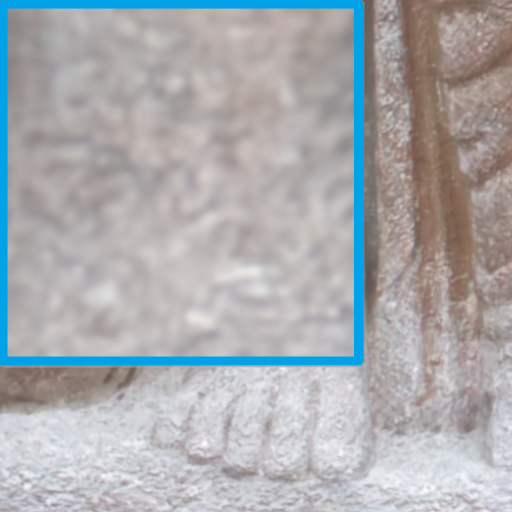}}\hfill
\subfigure[A-Haar-tSVD]{
\label{Fig4}
\includegraphics[width=0.8116in]{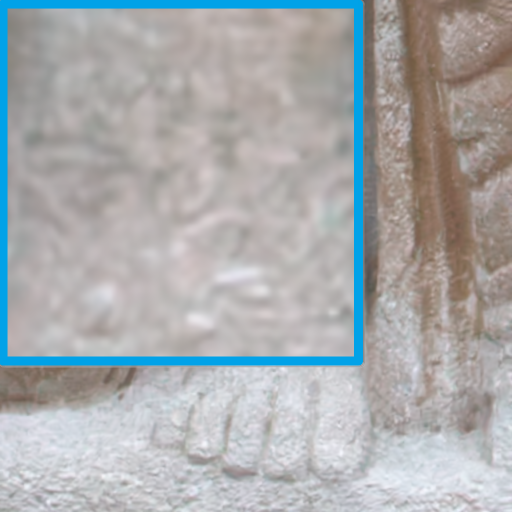}}
\vspace{-2.8pt}
\caption{Denoising results of compared methods on the DND datset.}
\label{Fig_compare_with_DND}
\vspace{-3.8pt}
\end{figure}

%% file: Fig_compare_with_HighISO.tex
\begin{figure}[htbp]
\vspace{-9.8pt}
\graphicspath{{Figs/Selected_color_images/HighISO/Case1/Combined/}}
\centering
\subfigure[Clean]{
\label{Fig4}
\includegraphics[width=0.8116in]{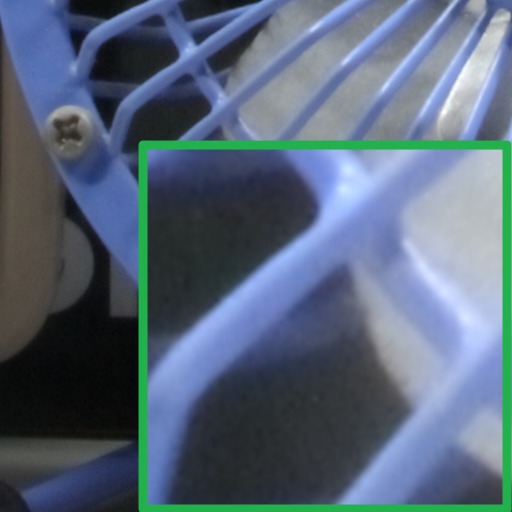}} \hfill
\subfigure[Noisy]{
\label{Fig4}
\includegraphics[width=0.8116in]{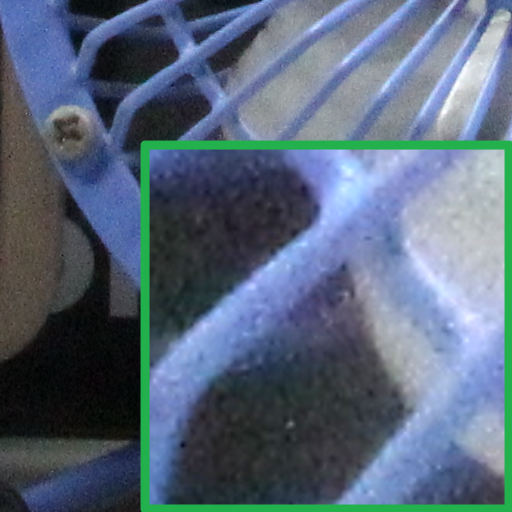}}\hfill
\subfigure[Bitonic]{
\label{Fig4}
\includegraphics[width=0.8116in]{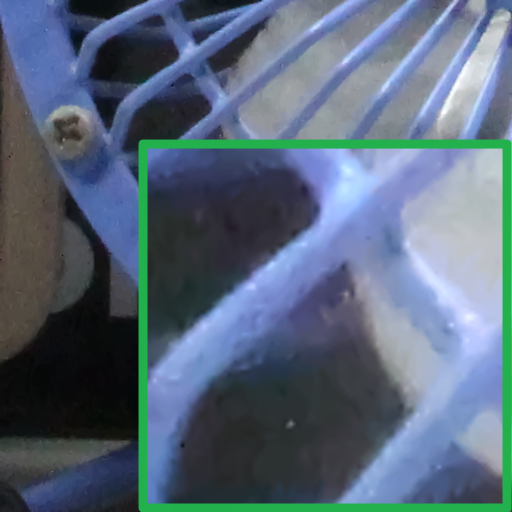}}\hfill
\subfigure[Condformer]{
\label{Fig4}
\includegraphics[width=0.8116in]{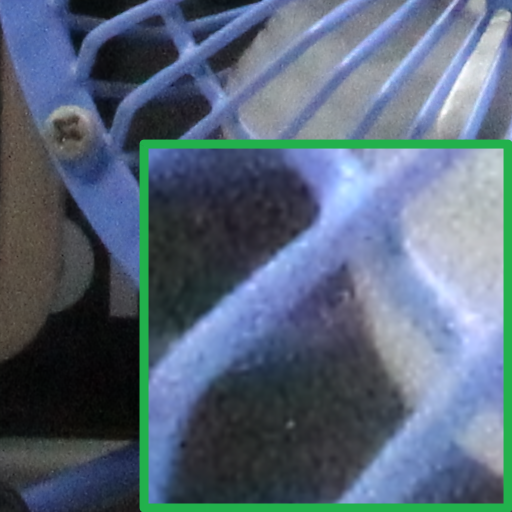}}\hfill
\vspace{-6.8pt}
\subfigure[DeepSN]{
\label{Fig4}
\includegraphics[width=0.8116in]{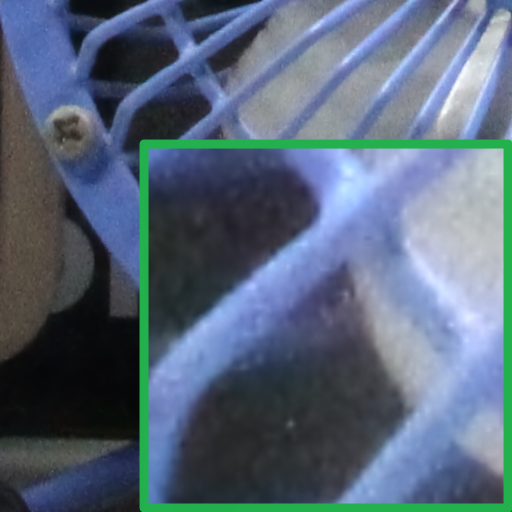}}\hfill
\subfigure[DIDN]{
\label{Fig4}
\includegraphics[width=0.8116in]{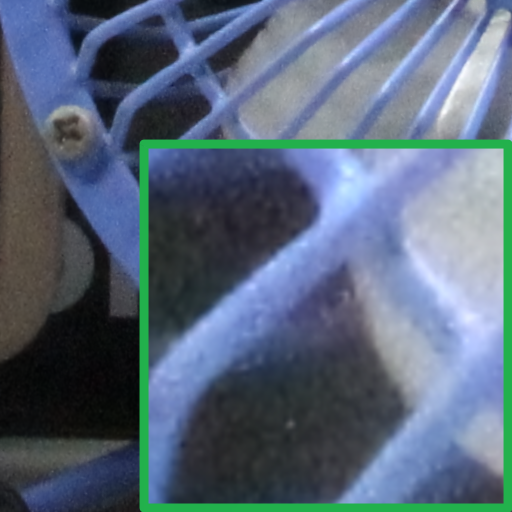}}\hfill
\subfigure[Restormer]{
\label{Fig4}
\includegraphics[width=0.8116in]{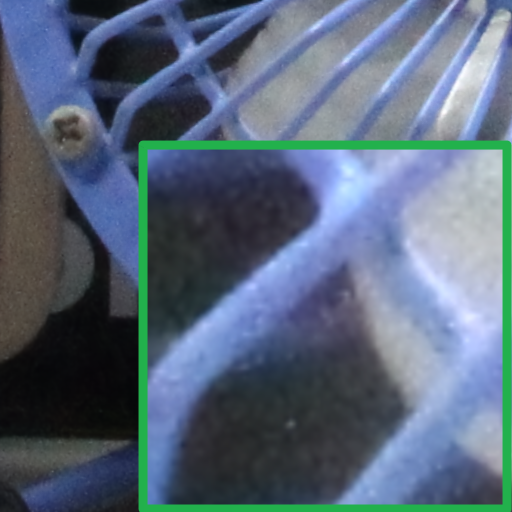}}\hfill
\subfigure[A-Haar-tSVD]{
\label{Fig4}
\includegraphics[width=0.8116in]{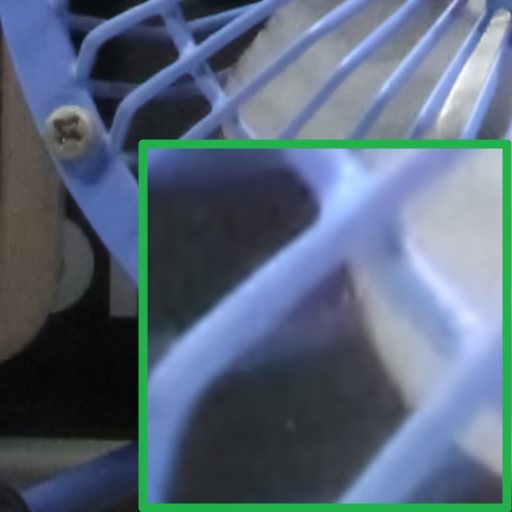}}
\vspace{-5.18pt}
\caption{Denoising results of compared methods on the HighISO datset.}
\label{Fig_compare_with_HighISO}
\vspace{-5.8pt}
\end{figure}

%% file: Fig_compare_with_IOCI_CANON5D.tex
\begin{figure}[htbp]
\vspace{-13.8pt}
\graphicspath{{Figs/Selected_color_images/IOCI/Case3/Combined/}}
\centering
\subfigure[Clean]{
\label{Fig4}
\includegraphics[width=0.8116in]{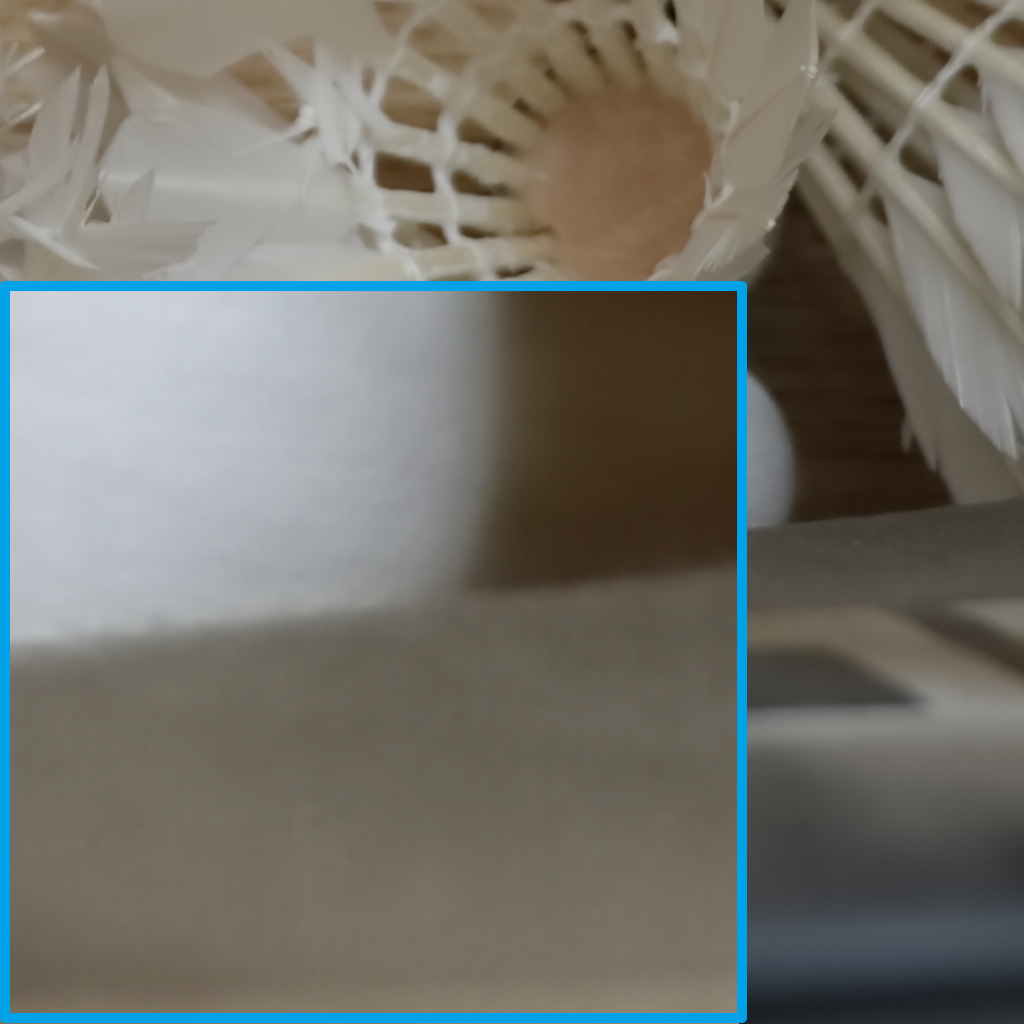}} \hfill
\subfigure[Noisy]{
\label{Fig4}
\includegraphics[width=0.8116in]{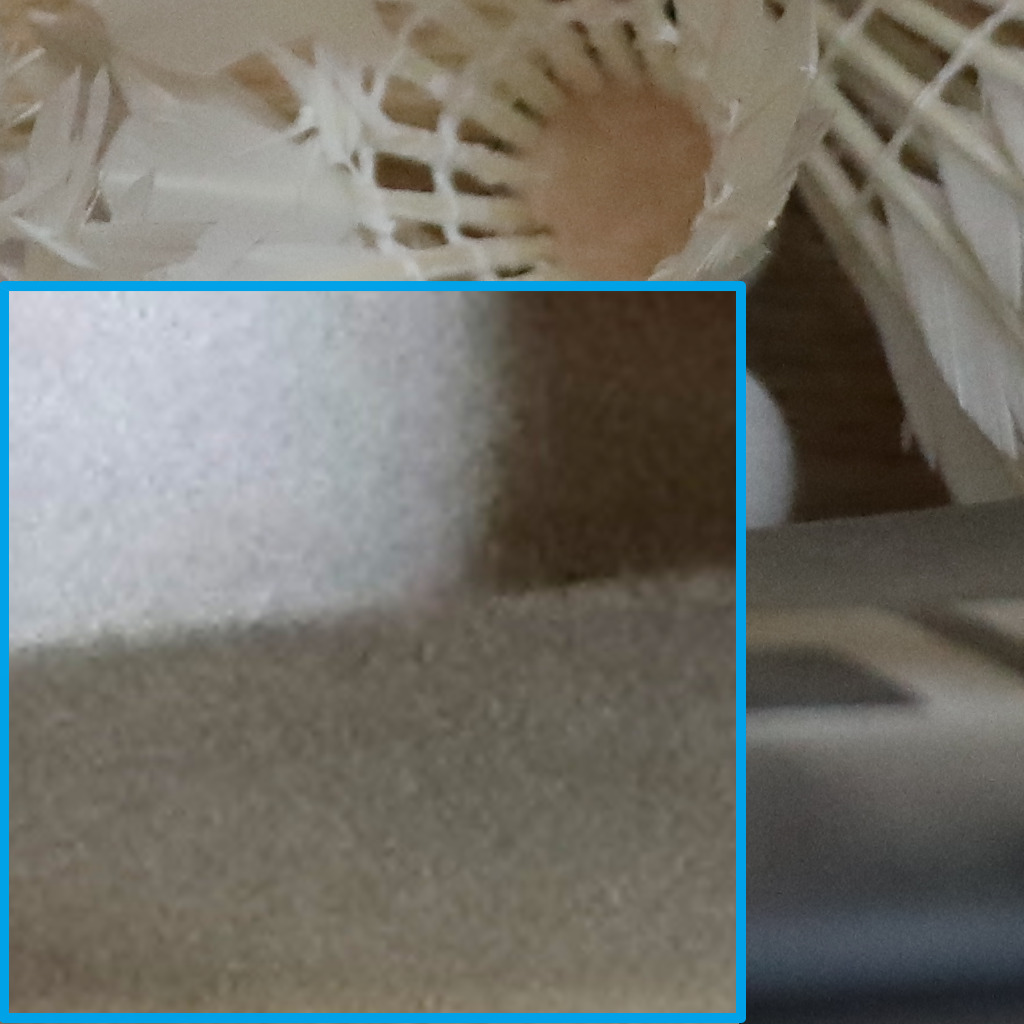}}\hfill
\subfigure[Condformer]{
\label{Fig4}
\includegraphics[width=0.8116in]{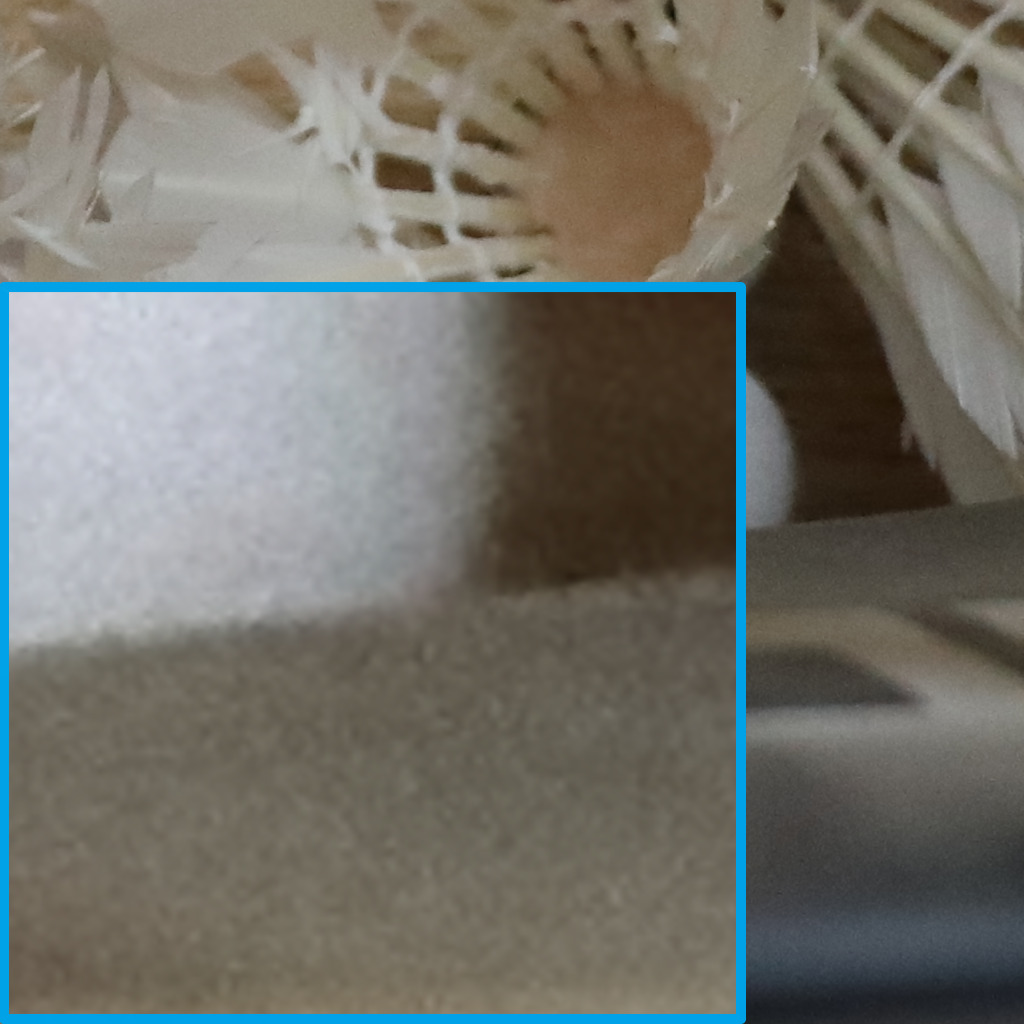}}\hfill
\subfigure[DeepSN]{
\label{Fig4}
\includegraphics[width=0.8116in]{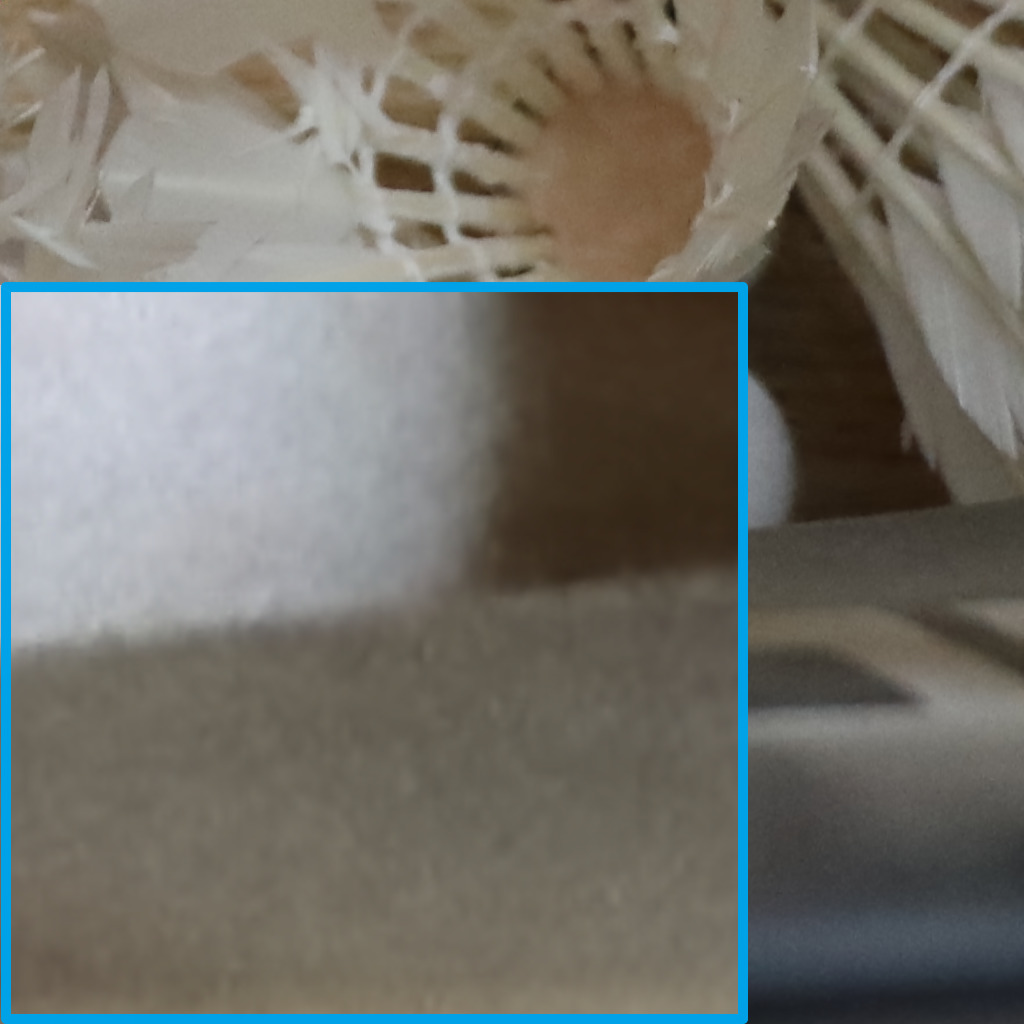}}\hfill
\vspace{-6.8pt}
\subfigure[DIDN]{
\label{Fig4}
\includegraphics[width=0.8116in]{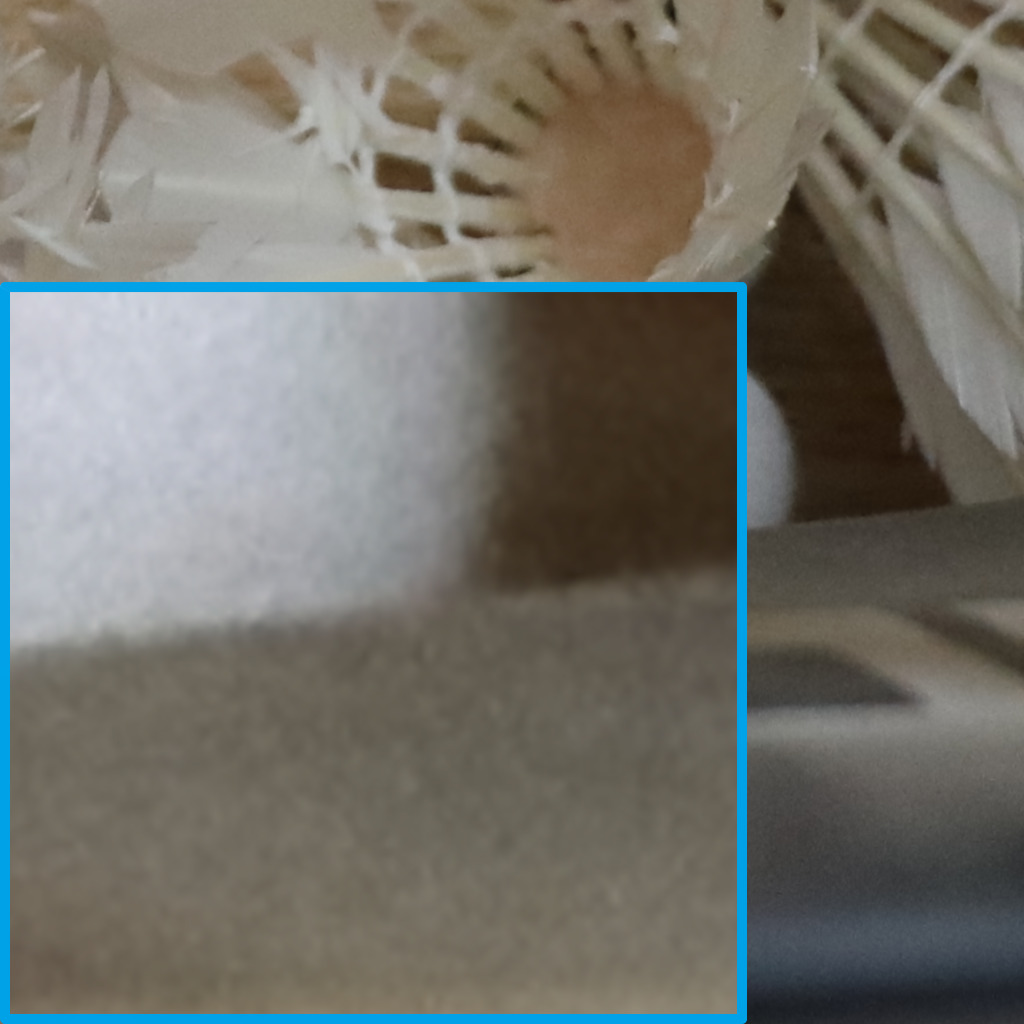}}\hfill
\subfigure[NAFNet]{
\label{Fig4}
\includegraphics[width=0.8116in]{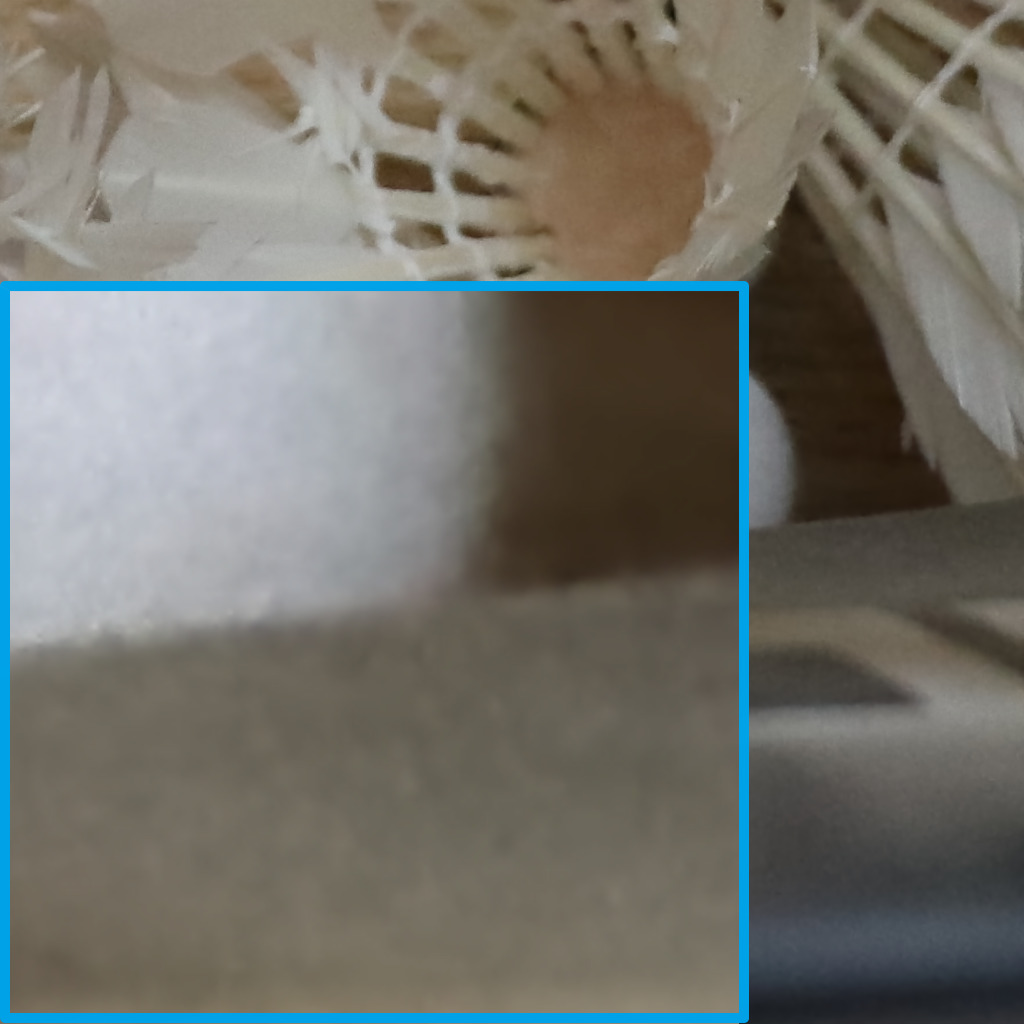}}\hfill
\subfigure[Restormer]{
\label{Fig4}
\includegraphics[width=0.8116in]{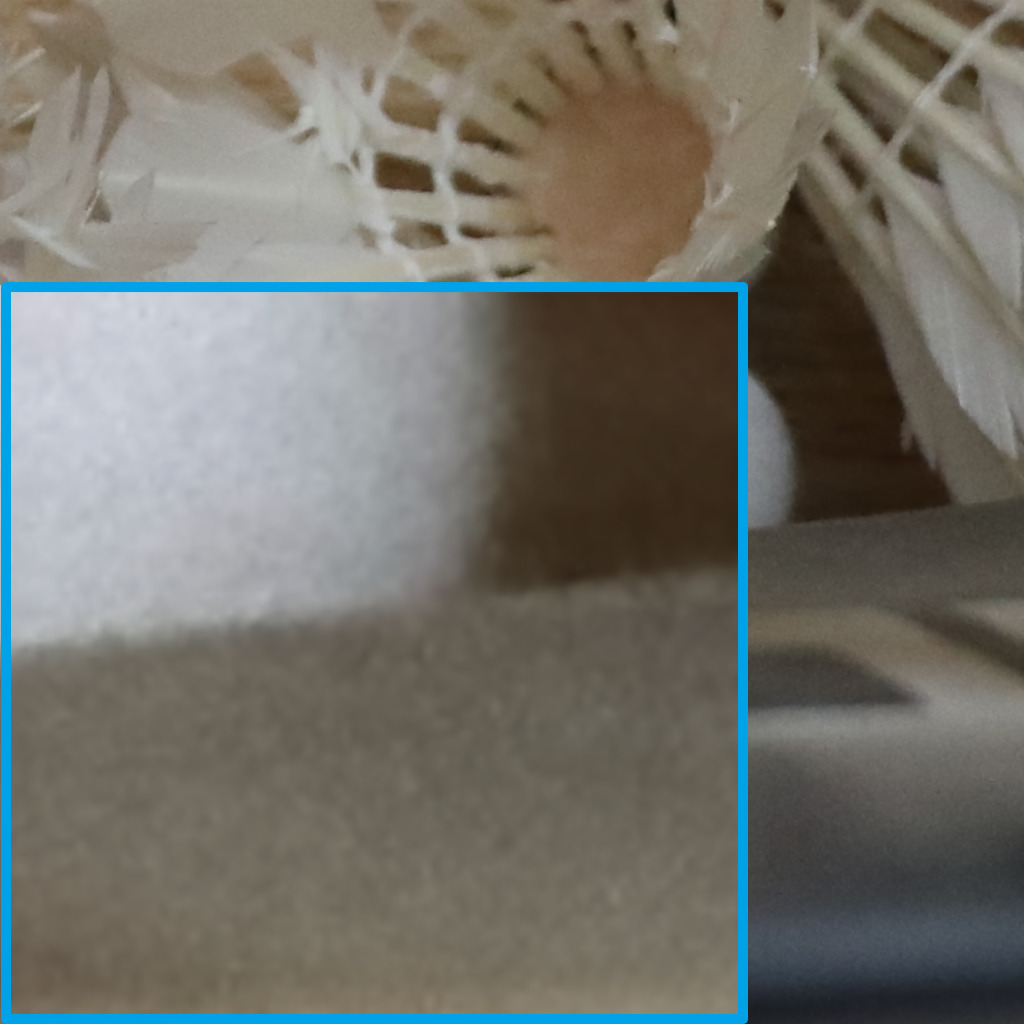}}\hfill
\subfigure[A-Haar-tSVD]{
\label{Fig4}
\includegraphics[width=0.8116in]{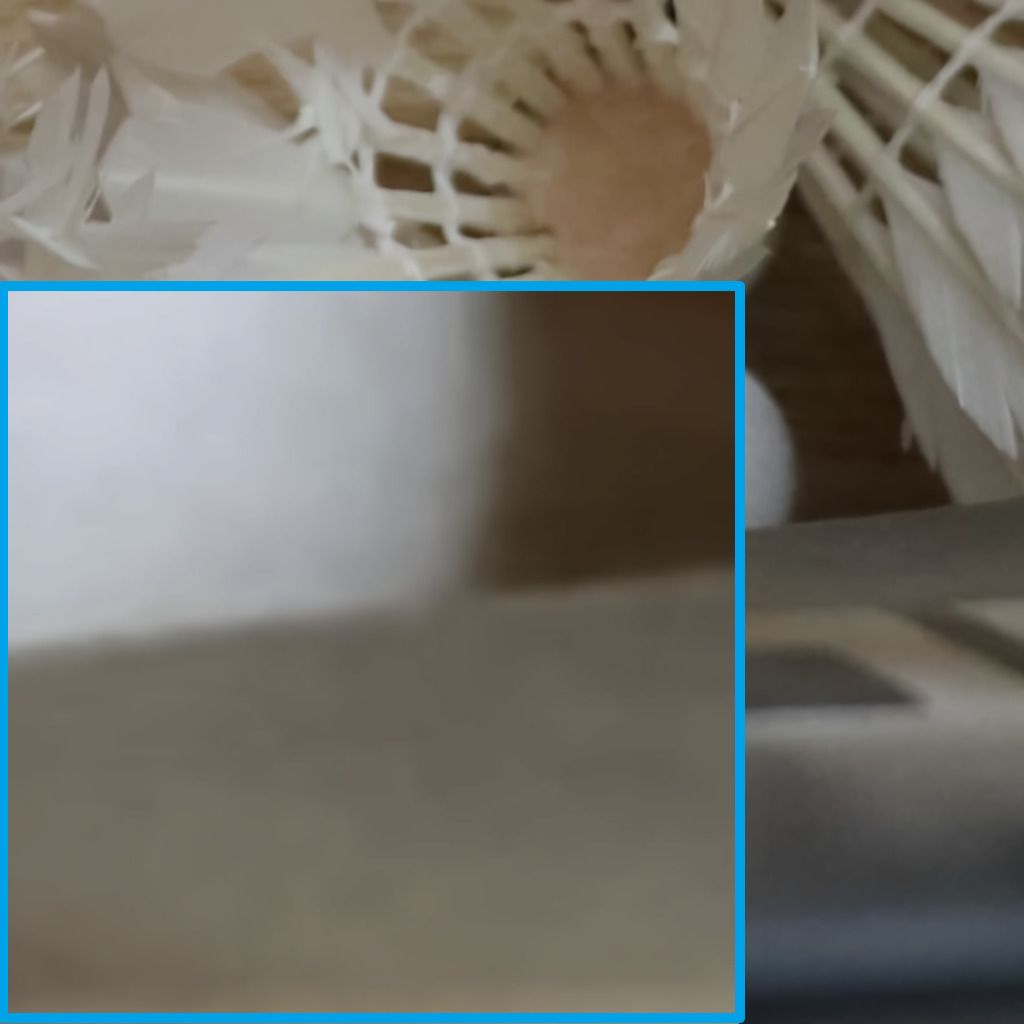}}
\vspace{-3.8pt}
\caption{Denoising results of compared methods on the IOCI datset.}
\label{Fig_compare_with_IOCI_CANON5D}
\vspace{0pt}
\end{figure}

%% file: Table_time_complexity_comparison.tex
\begin{table}[htbp]
\vspace{-3.98pt}
\scriptsize
  \centering
  \vspace{-1.98pt}
  \caption{Computational complexity comparisons of different denoising methods when processing $512 \times 512 \times 3$ sRGB images.}
  \scalebox{0.8398}{
    \begin{tabular}{ccccccc}
    \toprule
    Method & CBM3D & NLHCC & A-Haar-tSVD & DIDN  & Condformer & Restormer \\
    \midrule
    Test time (s) & 3.6   & 41.2  & 3.9   & 7.3   & 0.9   & \textbf{0.8} \\
    \midrule
    Train time (m) & -    & -     & \textbf{15.6} & -     & -     & $>$1500 \\
    \midrule
    Params (M) & -     & -     & \textbf{6.6} & 218.1   & 26.4 & 25.3 \\
    \bottomrule
    \end{tabular}}%
  \label{Table_time_complexity_comparison}%
\end{table}%

%% file: Fig_compare_with_CRVD_case9.tex
\begin{figure}[htbp]
\vspace{-9.8pt}
\graphicspath{{Figs/Selected_color_images/CRVD/Case9/Combined/}}
\centering
\subfigure[Mean]{
\label{Fig4}
\includegraphics[width=0.8116in]{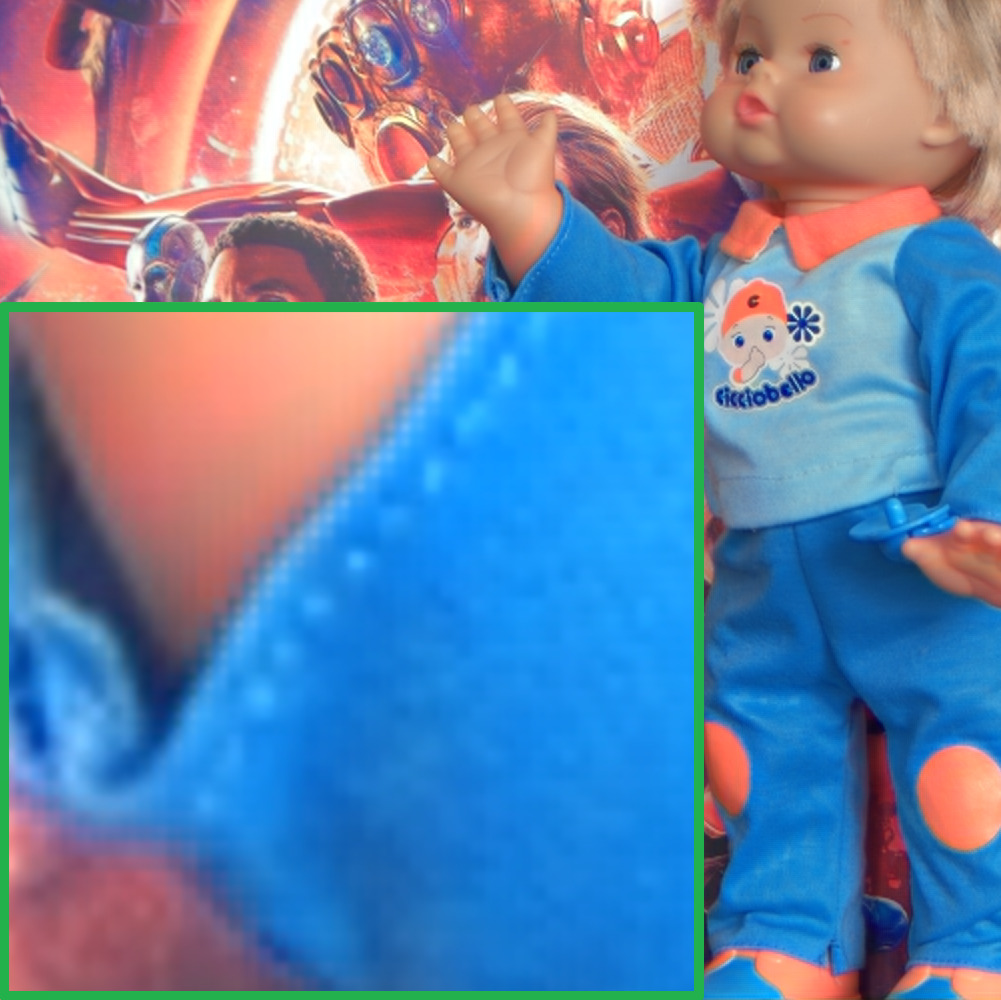}} \hfill
\subfigure[Noisy]{
\label{Fig4}
\includegraphics[width=0.8116in]{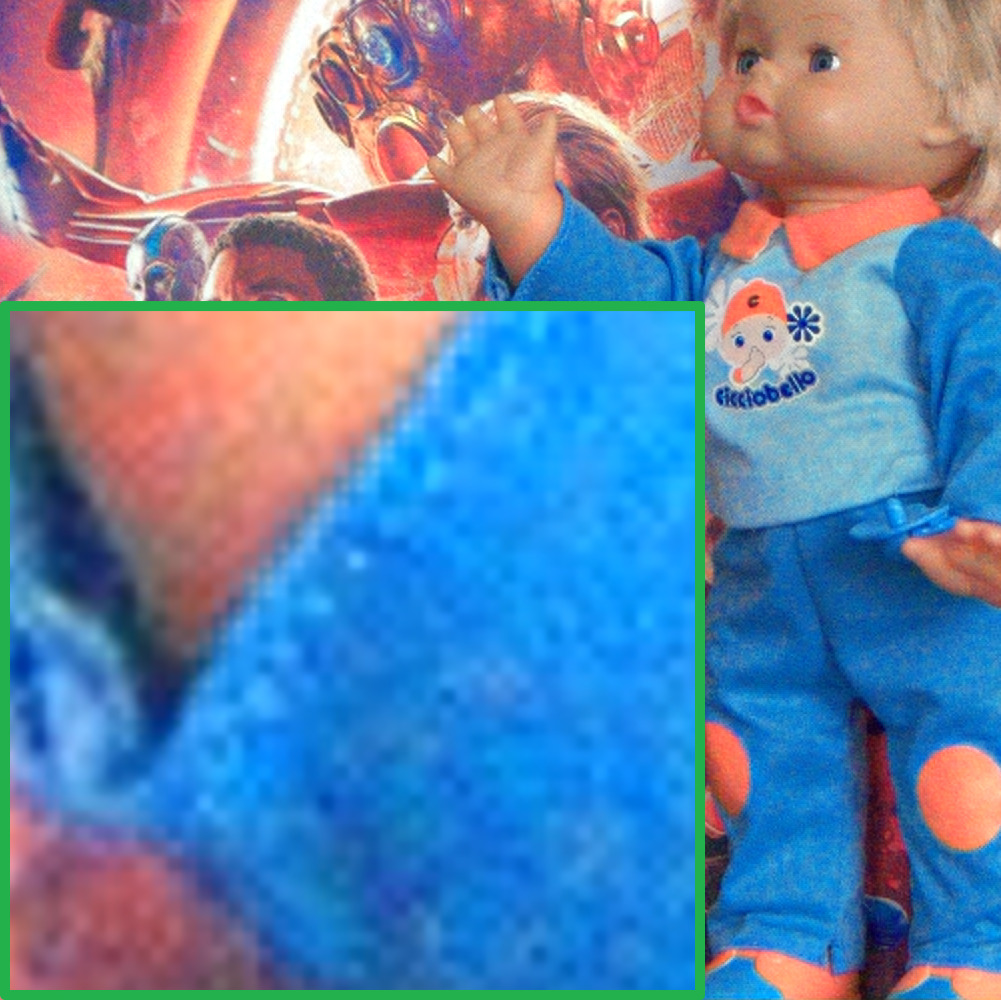}}\hfill
\subfigure[VBM4D]{
\label{Fig4}
\includegraphics[width=0.8116in]{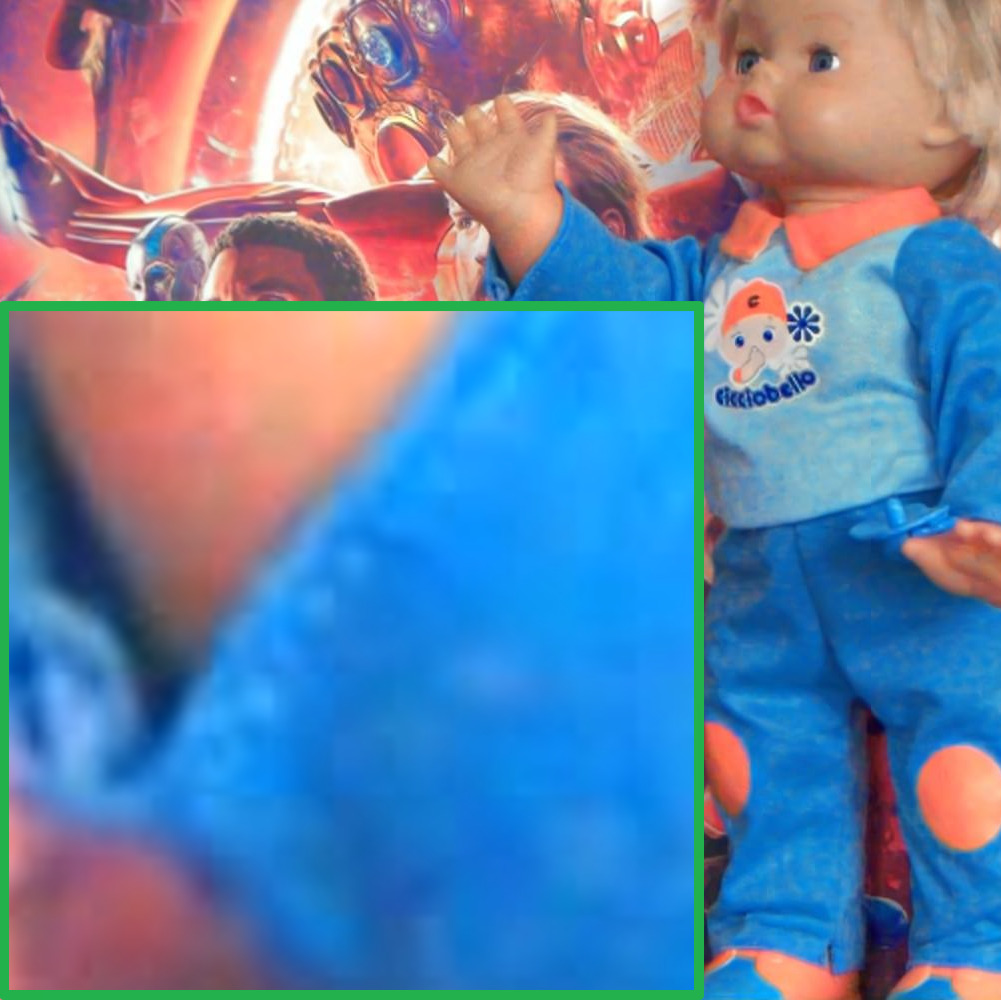}}\hfill
\subfigure[FastDVDNet]{
\label{Fig4}
\includegraphics[width=0.8116in]{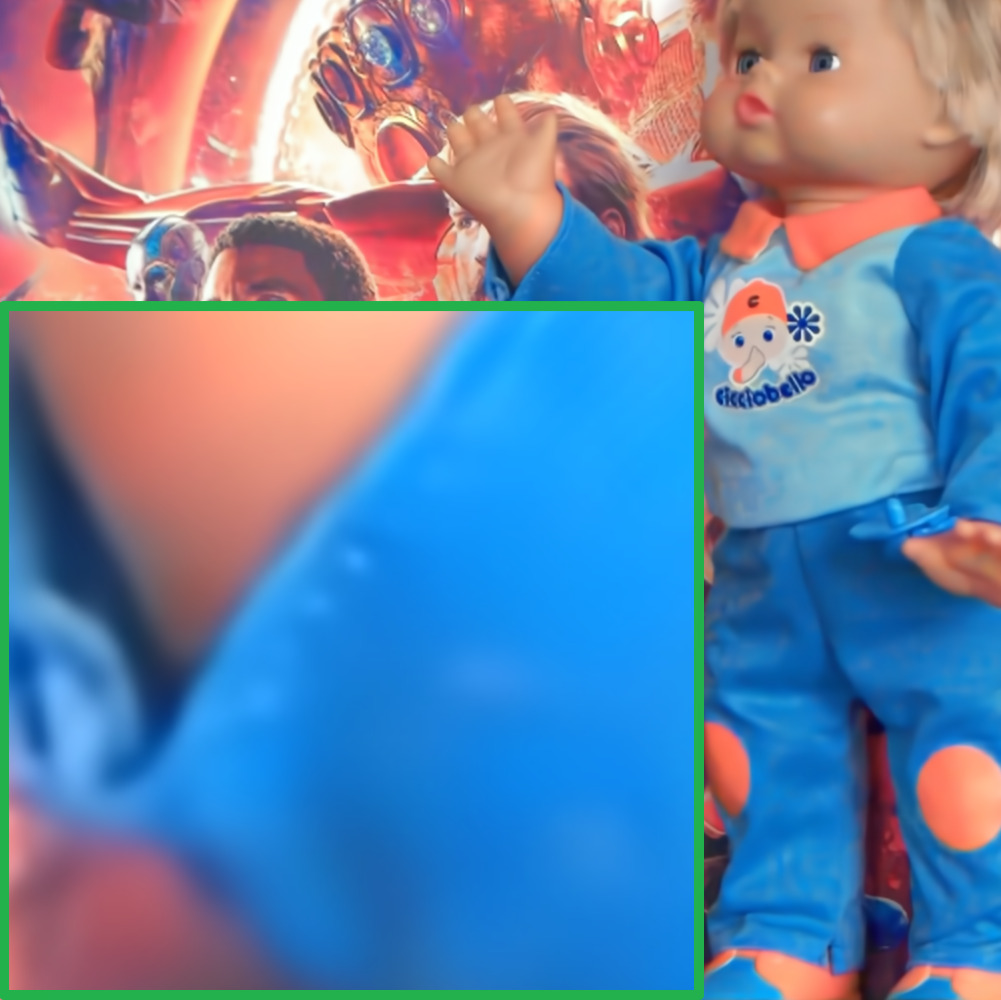}}\hfill
\vspace{-6.8pt}
\subfigure[FloRNN]{
\label{Fig4}
\includegraphics[width=0.8116in]{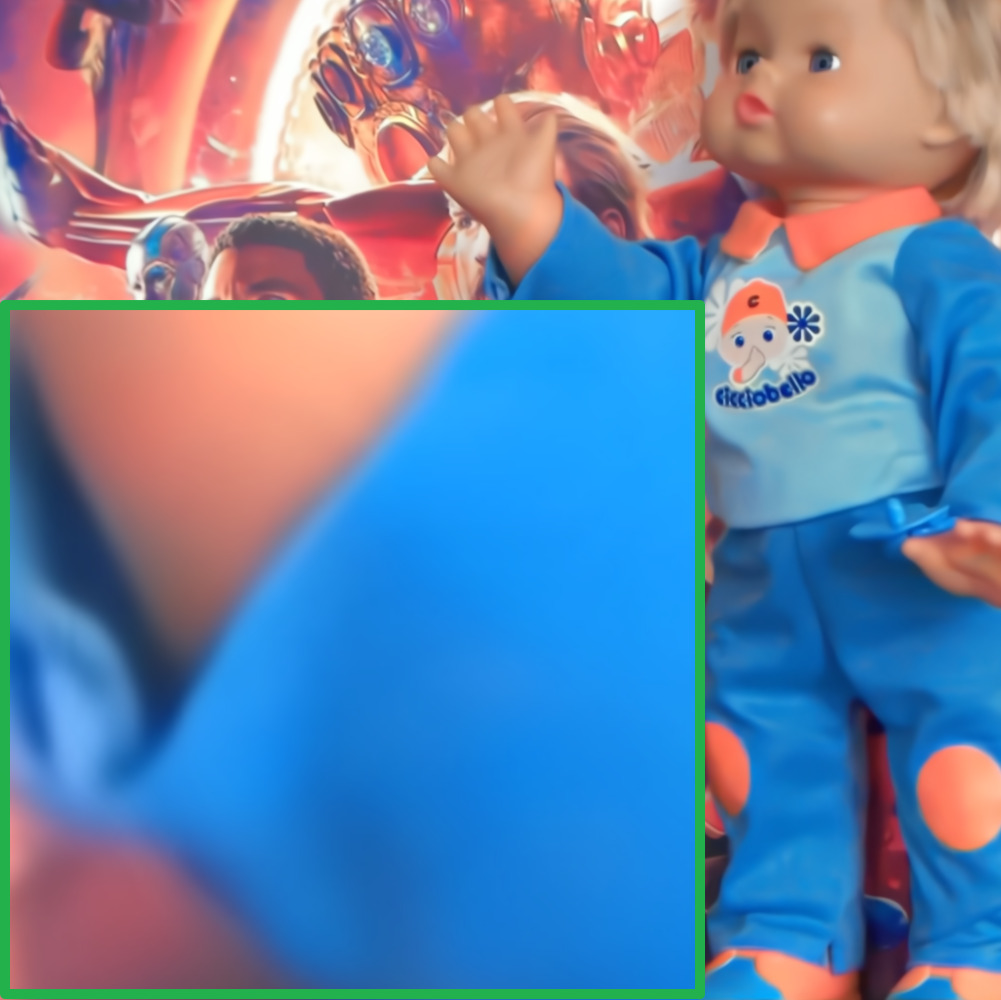}}\hfill
\subfigure[VRT]{
\label{Fig4}
\includegraphics[width=0.8116in]{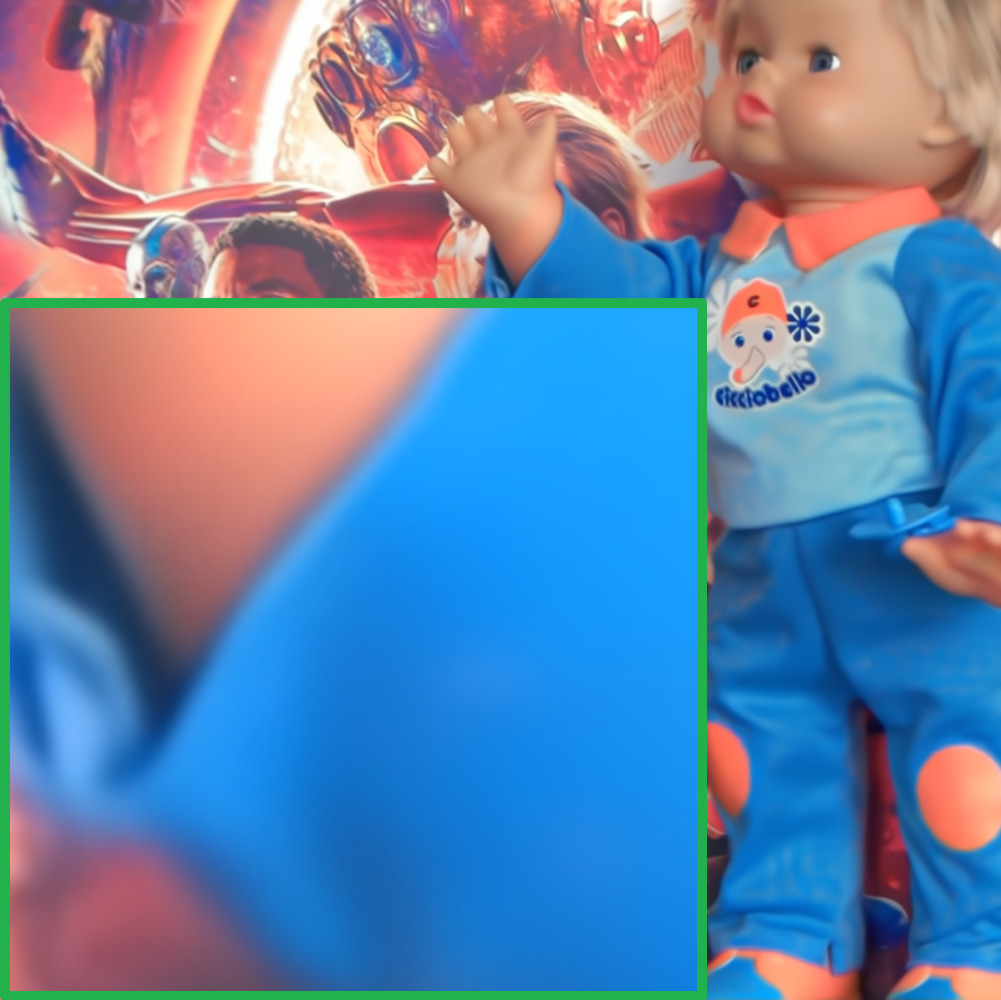}}\hfill
\subfigure[VNLNet]{
\label{Fig4}
\includegraphics[width=0.8116in]{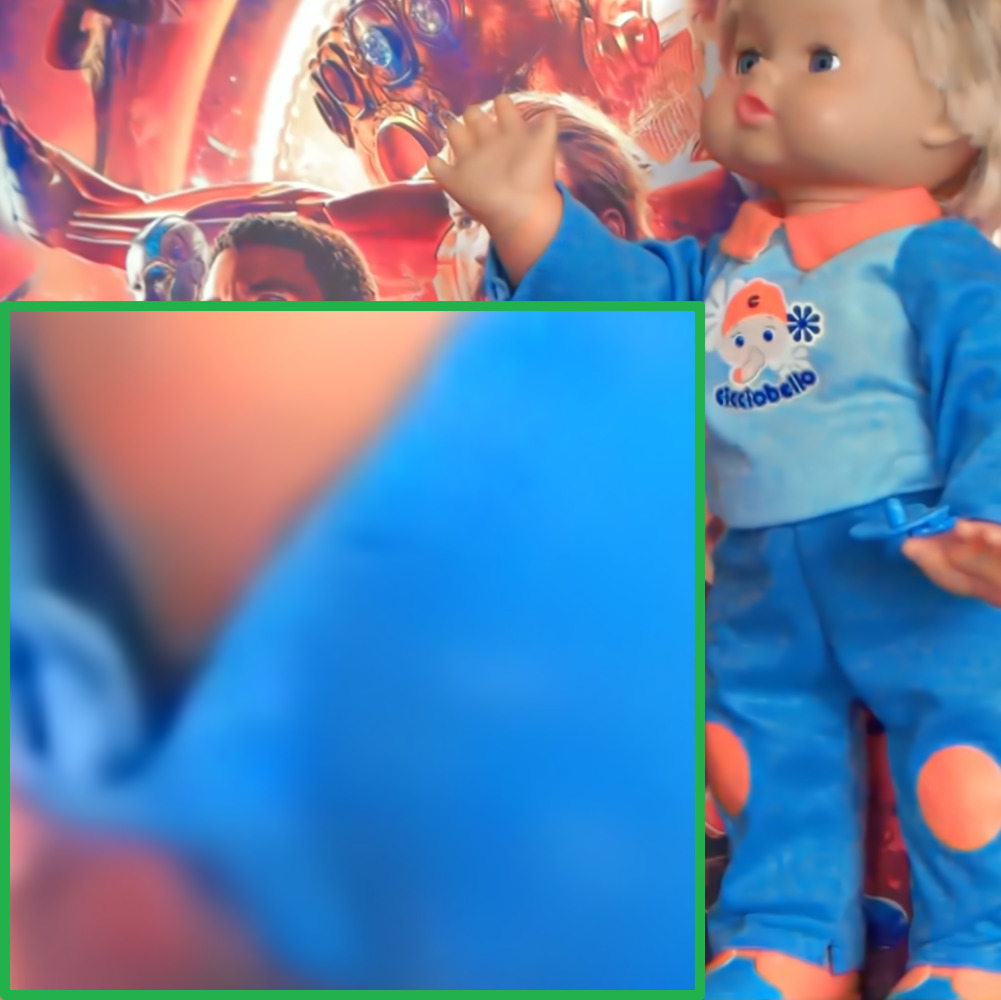}}\hfill
\subfigure[A-Haar-tSVD]{
\label{Fig4}
\includegraphics[width=0.8116in]{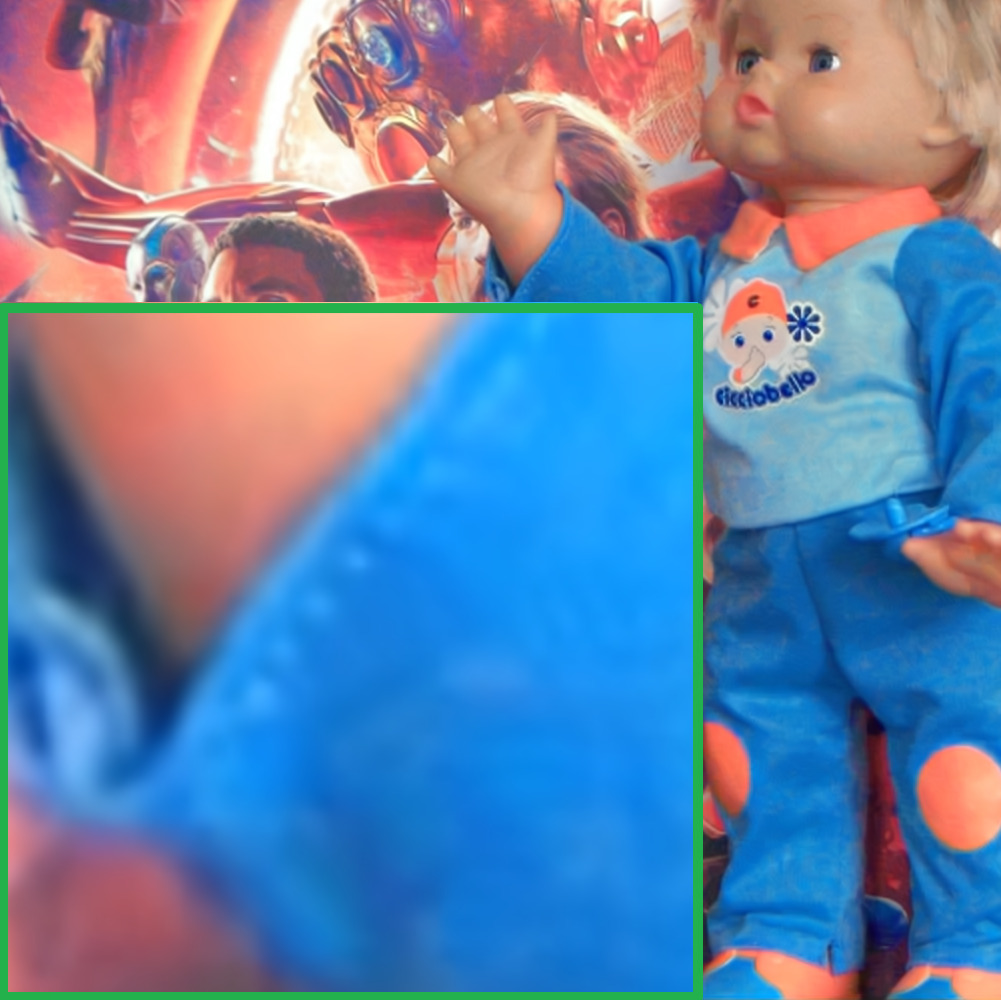}} \hfill
\vspace{-2.8pt}
\caption{Denoising results comparison on the CRVD dataset (ISO = 6400).}
\label{Fig_compare_with_CRVD_case9}
\vspace{-8.8pt}
\end{figure}

%% file: Table_Color_Video_Denoising.tex
\begin{table*}[htbp]
\scriptsize
  \centering
  \caption{Denoising results of compared methods on real-wolrd sRGB color video datasets..}
  \scalebox{0.906}{
    \begin{tabular}{cccccccccccccc}
    \toprule
    \multirow{3}[4]{*}{Dataset} & \multicolumn{4}{c}{Traditional denoisers} &       & \multicolumn{8}{c}{DNN methods} \\
\cmidrule{2-14}          & VMSt-SVD & VBM4D & VIDOSAT & Haar-tSVD & A-Haar-tSVD & DVDNet & FastDVDNet & FloRNN & MAP-VDNet & VRT  & UDVD  & ViDeNN & VNLNet \\
          &  \cite{kong2019color}     &  \cite{maggioni2012video}     &  \cite{wen2018vidosat}      &   (Ours)     &   (Ours)   & \cite{tassano2019dvdnet}  &  \cite{Tassano_2020_CVPR}     &   \cite{li2022unidirectional}    &   \cite{sun2021deep}    &   \cite{liang2024vrt}    &   \cite{sheth2021unsupervised}    &  \cite{claus2019videnn}     & \cite{davy2019non} \\
    \midrule
    \multirow{2}[2]{*}{CRVD} & 36.66  & 34.14  & 34.16  & 37.12  & \textbf{37.16} & 34.50  & 35.84  & 36.66  & -     & 36.94  & -     & 32.31  & 36.11  \\
\cmidrule{2-14}          & 0.946  & 0.908  & 0.938  & \textbf{0.964}  & \textcolor[rgb]{0,  .439,  .753}{\textbf{0.963}} & 0.949  & 0.931  & 0.961  & -     & 0.956  & -     & 0.845  & 0.945  \\
    \midrule
    \multirow{2}[2]{*}{IOCV} & 38.22  & 38.76  & -     & \textcolor[rgb]{0,  .439,  .753}{\textbf{38.92}}  & \textbf{38.97} & 38.53  & 37.57  & 38.64  & 35.52  & 38.50  & 35.02  & 36.13  & 38.76  \\
\cmidrule{2-14}          & 0.974  & 0.977  & -     & \textcolor[rgb]{0,  .439,  .753}{\textbf{0.978}}  & \textbf{0.978} & 0.975  & 0.970  & 0.974  & 0.931  & 0.968  & 0.966  & 0.951  & 0.977  \\
    \bottomrule
    \end{tabular}}%
  \label{Table_Color_Video_Denoising}%
  \vspace{-12.98pt}
\end{table*}%

%% file: Fig_compare_with_CRVD_case6.tex
\begin{figure}[htbp]
\vspace{-11.88pt}
\graphicspath{{Figs/Selected_color_images/CRVD/Case6/Combined/}}
\centering
\subfigure[Mean]{
\label{Fig4}
\includegraphics[width=0.8116in]{frame6_clean_and_slightly_denoised_combined_marked}} \hfill
\subfigure[Noisy]{
\label{Fig4}
\includegraphics[width=0.8116in]{frame6_noisy6_combined_marked}}\hfill
\subfigure[VBM4D]{
\label{Fig4}
\includegraphics[width=0.8116in]{6_VBM4D2_combined_marked}}\hfill
\subfigure[FastDVDNet]{
\label{Fig4}
\includegraphics[width=0.8116in]{flip_6_fastdvdnet_sig20_combined_marked}}\hfill
\vspace{-6.8pt}
\subfigure[FloRNN]{
\label{Fig4}
\includegraphics[width=0.8116in]{flip_6_FloRNN_sig30_combined_marked}}\hfill
\subfigure[VRT]{
\label{Fig4}
\includegraphics[width=0.8116in]{flip_6_RVRT_sig30_combined_marked}}\hfill
\subfigure[VNLNet]{
\label{Fig4}
\includegraphics[width=0.8116in]{6_vnlnet_sig20_combined_marked}}\hfill
\subfigure[A-Haar-tSVD]{
\label{Fig4}
\includegraphics[width=0.8116in]{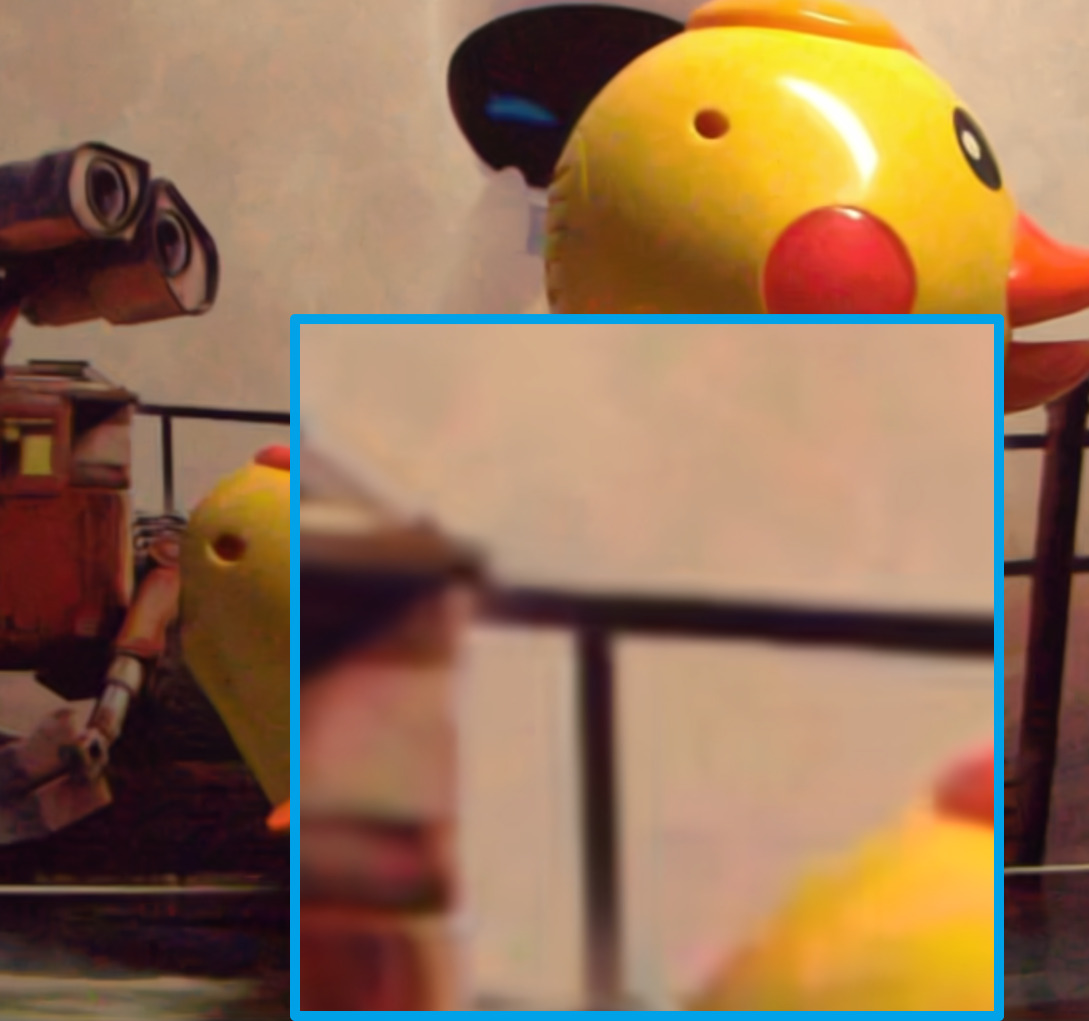}} \hfill
\vspace{-2.8pt}
\caption{Denoising results comparison on the CRVD dataset (ISO = 25600).}
\label{Fig_compare_with_CRVD_case6}
\vspace{-9.18pt}
\end{figure} 

%% file: Fig_compare_with_CRVD_case18.tex
\begin{figure}[htbp]
\vspace{-12.8pt}
\graphicspath{{Figs/Selected_color_images/CRVD/Case7/Combined/}}
\centering
\subfigure[Mean]{
\label{Fig4}
\includegraphics[width=0.8116in]{frame6_clean_and_slightly_denoised_combined_marked}} \hfill
\subfigure[Noisy]{
\label{Fig4}
\includegraphics[width=0.8116in]{frame6_noisy6_combined_marked}}\hfill
\subfigure[VBM4D]{
\label{Fig4}
\includegraphics[width=0.8116in]{6_VBM4D2_combined_marked}}\hfill
\subfigure[FastDVDNet]{
\label{Fig4}
\includegraphics[width=0.8116in]{flip_6_fastdvdnet_sig20_combined_marked}}\hfill
\vspace{-6.8pt}
\subfigure[FloRNN]{
\label{Fig4}
\includegraphics[width=0.8116in]{flip_6_FloRNN_sig30_combined_marked}}\hfill
\subfigure[VRT]{
\label{Fig4}
\includegraphics[width=0.8116in]{flip_6_RVRT_sig30_combined_marked}}\hfill
\subfigure[VNLNet]{
\label{Fig4}
\includegraphics[width=0.8116in]{6_vnlnet_sig20_combined_marked}}\hfill
\subfigure[A-Haar-tSVD]{
\label{Fig4}
\includegraphics[width=0.8116in]{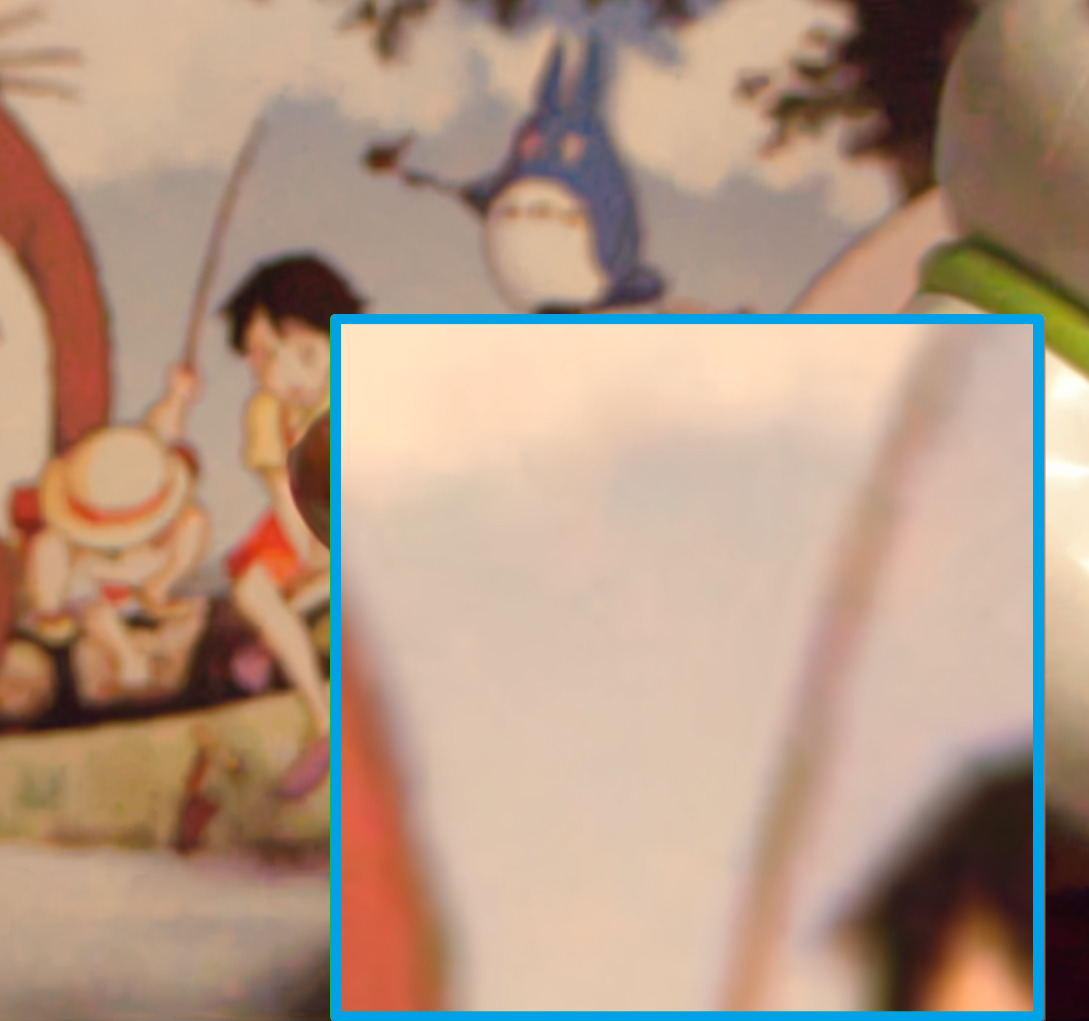}}
\vspace{-3.98pt}
\caption{Denoising results comparison on the CRVD dataset (ISO = 25600).}
\label{Fig_compare_with_CRVD_case18}
\vspace{-12.98pt}
\end{figure}

%% file: Fig_compare_with_IOCV_case2.tex
\begin{figure}[htbp]
\vspace{-3.98pt}
\graphicspath{{Figs/Selected_color_images/IOCV/Case2/Combined/}}
\centering
\subfigure[Mean]{
\label{Fig4}
\includegraphics[width=0.8116in]{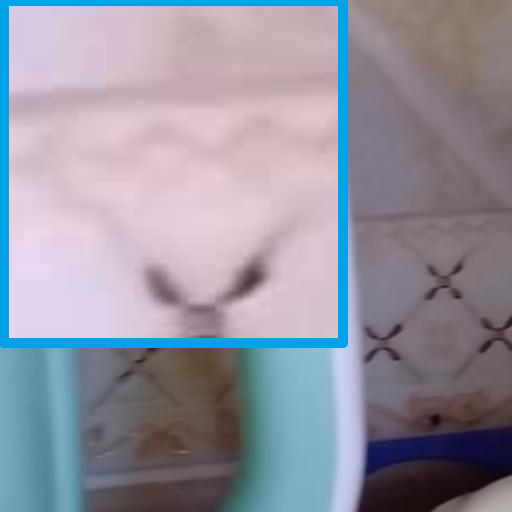}} \hfill
\subfigure[Noisy]{
\label{Fig4}
\includegraphics[width=0.8116in]{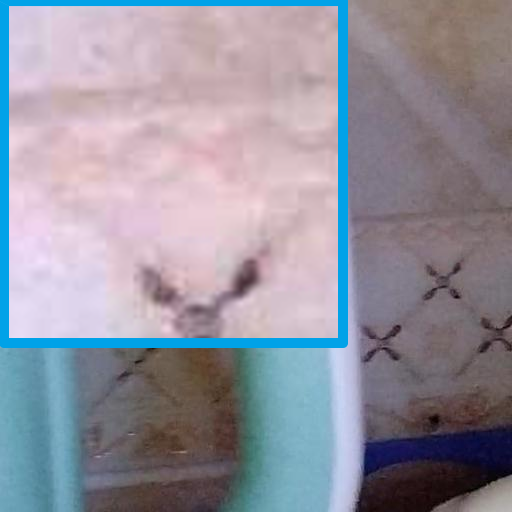}}\hfill
\subfigure[VBM4D]{
\label{Fig4}
\includegraphics[width=0.8116in]{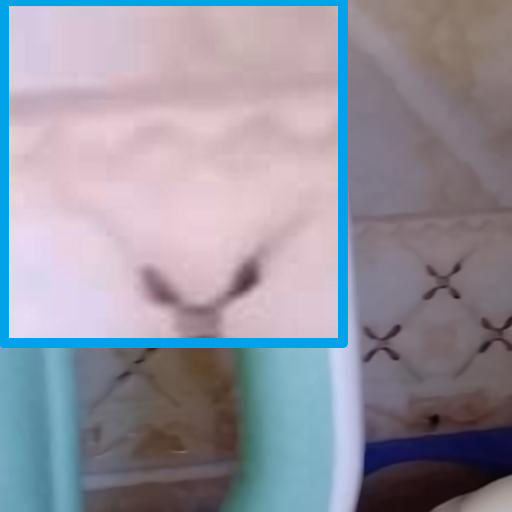}}\hfill
\subfigure[FastDVDNet]{
\label{Fig4}
\includegraphics[width=0.8116in]{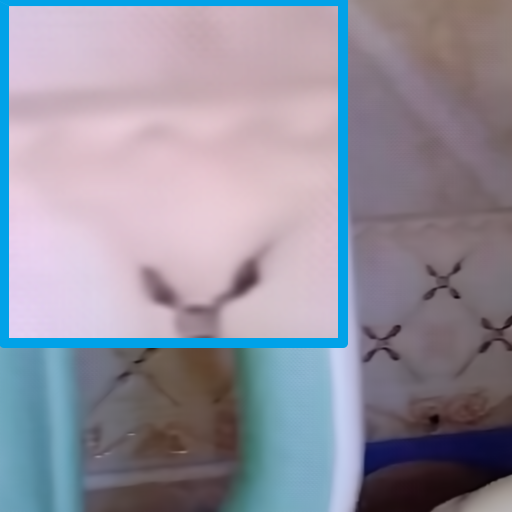}}\hfill
\vspace{-6.8pt}
\subfigure[FloRNN]{
\label{Fig4}
\includegraphics[width=0.8116in]{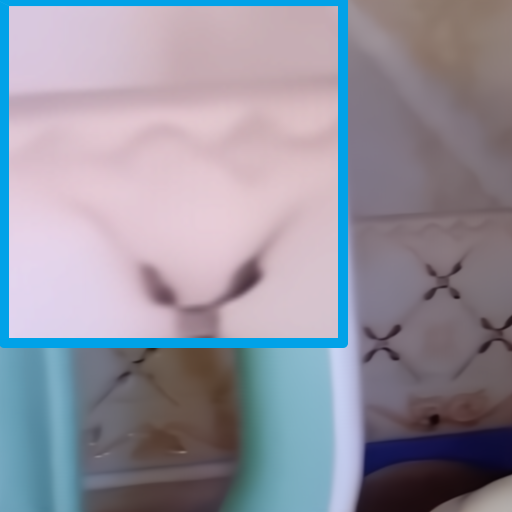}}\hfill
\subfigure[VRT]{
\label{Fig4}
\includegraphics[width=0.8116in]{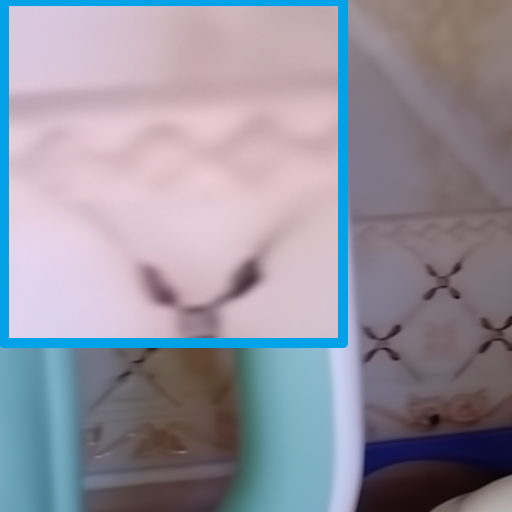}}\hfill
\subfigure[VNLNet]{
\label{Fig4}
\includegraphics[width=0.8116in]{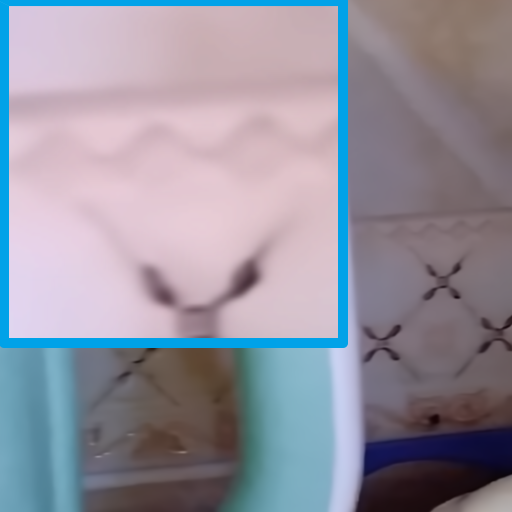}}\hfill
\subfigure[A-Haar-tSVD]{
\label{Fig4}
\includegraphics[width=0.8116in]{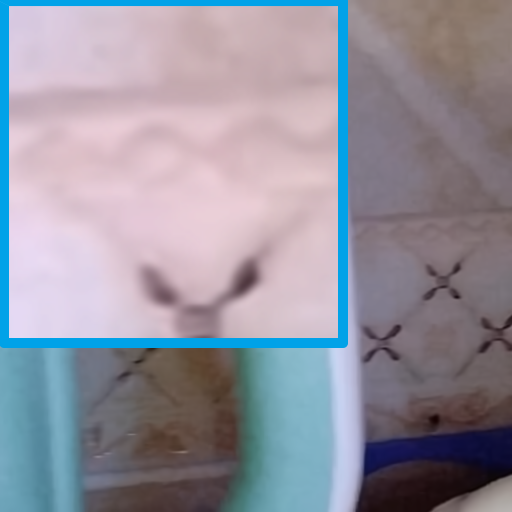}}
\vspace{-1.98pt}
\caption{Denoising results comparison on the IOCV dataset.}
\label{Fig_compare_with_IOCV_case2}
\end{figure} 

%% file: Table_Real_HSI.tex
\begin{table*}[htbp]
  \centering
  \caption{Denoising results of compared methods on the Real-HSI dataset.}
  \scalebox{0.838}{
    \begin{tabular}{cccccccccccccc}
    \toprule
    \multirow{3}[3]{*}{Datasets} & \multirow{3}[3]{*}{Metrics} & \multicolumn{7}{c}{Traditonal denoisers}              & \multicolumn{5}{c}{DNN methods} \\
\cmidrule{3-14}          &       & BM4D  & LLRT  & LTDL  & MSt-SVD & NGMeet & Haar-tSVD & A-Haar-tSVD & FlexDID & HSI-DeNet & QRNN3D & RAS2S & sDeCNN \\
          &       & \cite{maggioni2012nonlocal} & \cite{chang2017hyper} & \cite{gong2020low} & \cite{kong2019color} & \cite{he2019non} & (Ours) & (Ours) & \cite{chen2024flex}& \cite{chang2018hsi} & \cite{wei2020physics} & \cite{xiao2024region} & \cite{maffei2019single} \\
    \midrule
    \multirow{4}[8]{*}{Real-HSI} & PSNR  & 25.88  & \textbf{25.90} & 25.80  & 25.86  & 25.87  & 25.82  & 25.81  & 25.31  & 25.63  & 25.82  & 25.87  & 25.70  \\
\cmidrule{2-14}          & SSIM  & 0.865  & 0.861  & 0.841  & 0.866  & 0.866  & 0.865  & 0.866  & 0.820  & 0.853  & \textbf{0.869} & 0.868  & 0.860  \\
\cmidrule{2-14}          & SAM   & 0.066  & 0.060  & 0.074  & 0.063  & \textbf{0.051} & 0.063  & 0.064  & 0.075  & 0.092  & 0.064  & 0.054  & 0.093  \\
\cmidrule{2-14}          & EGRAS & 222.68  & 224.05  & 223.32  & \textbf{222.64} & 222.69  & 222.96  & 223.12  & 232.12  & 232.73  & 225.32  & 222.83  & 227.88  \\
    \midrule
    Platform & -     & CPU   & CPU   & CPU   & CPU   & CPU   & CPU   & CPU   & GPU (T4) & GPU (T4) & GPU (T4) & GPU (T4) & GPU (T4) \\
    \midrule
    Time  & minutes & 4.1   & 16.5  & 35.0  & 2.8   & 4.1   & \textbf{0.1} & \textbf{0.1} & 6.1   & 0.8   & \textbf{0.1} & \textbf{0.1} & 0.7  \\
    \bottomrule
    \end{tabular}}%
  \label{Table_Real_HSI}%
  \vspace{-6.8pt}
\end{table*}%

%% file: Fig_compare_with_Real_HSI_case4.tex
\begin{figure}[htbp]
\vspace{-8.8pt}
\graphicspath{{Figs/Selected_color_images/Real-HSI/Case4/Combined/}}
\centering
\subfigure[Mean]{
\label{Fig4}
\includegraphics[width=0.816in]{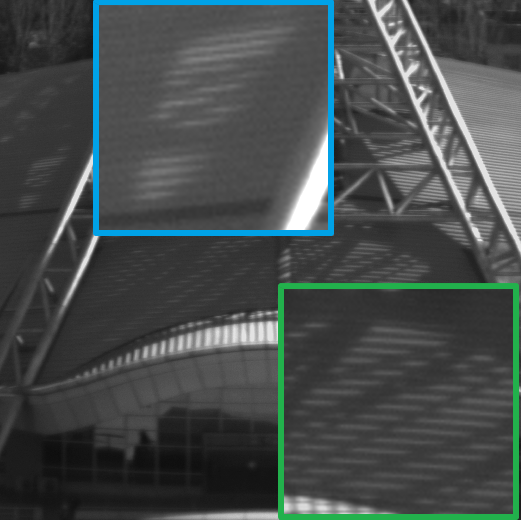}}\hfill
\subfigure[Noisy]{
\label{Fig4}
\includegraphics[width=0.816in]{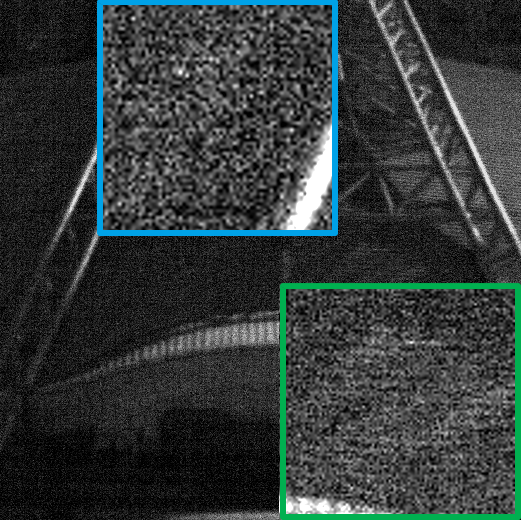}}\hfill
\subfigure[BM4D]{
\label{Fig4}
\includegraphics[width=0.816in]{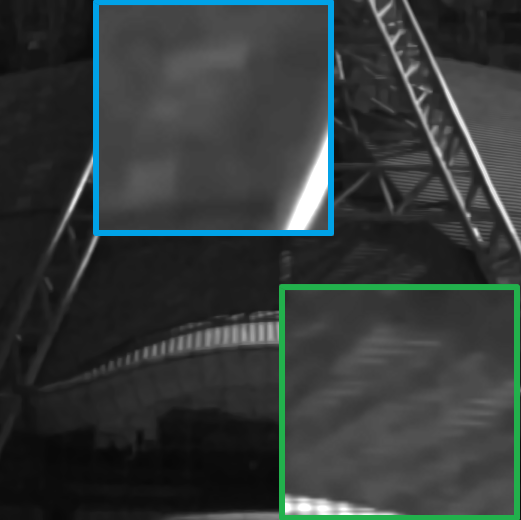}}\hfill
\subfigure[OLRT]{
\label{Fig4}
\includegraphics[width=0.816in]{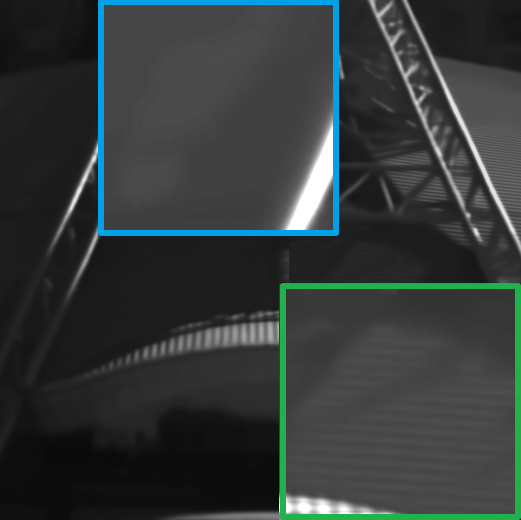}}\hfill \\
\vspace{-6.98pt}
\subfigure[FlexDLD]{
\label{Fig4}
\includegraphics[width=0.816in]{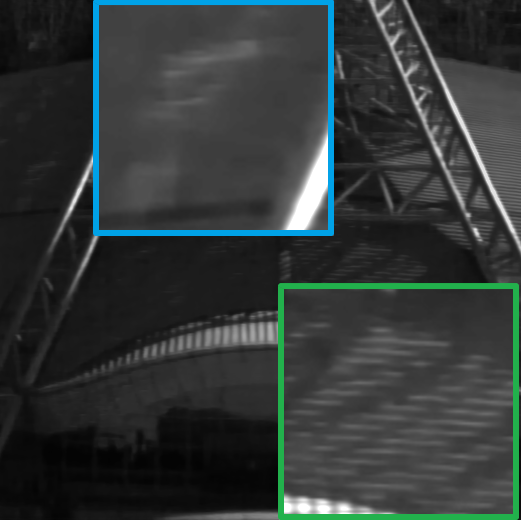}}\hfill
\subfigure[QRNN3D]{
\label{Fig4}
\includegraphics[width=0.816in]{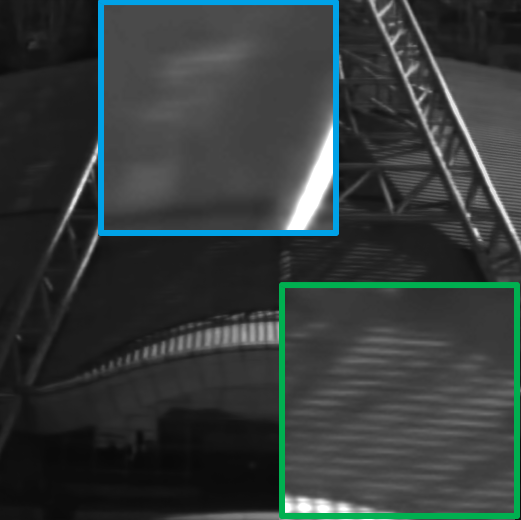}}\hfill
\subfigure[RAS2S]{
\label{Fig4}
\includegraphics[width=0.816in]{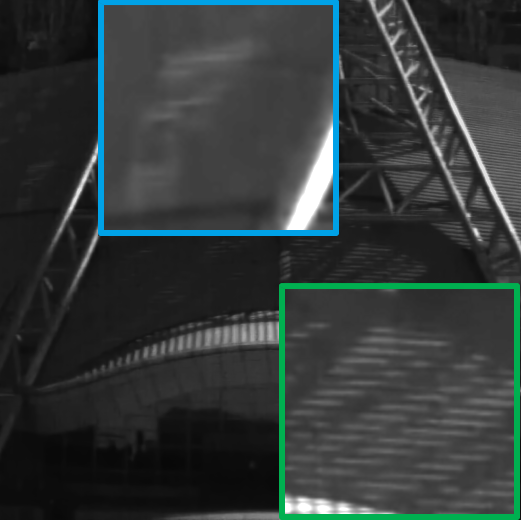}}\hfill
\subfigure[A-Haar-tSVD]{
\label{Fig4}
\includegraphics[width=0.816in]{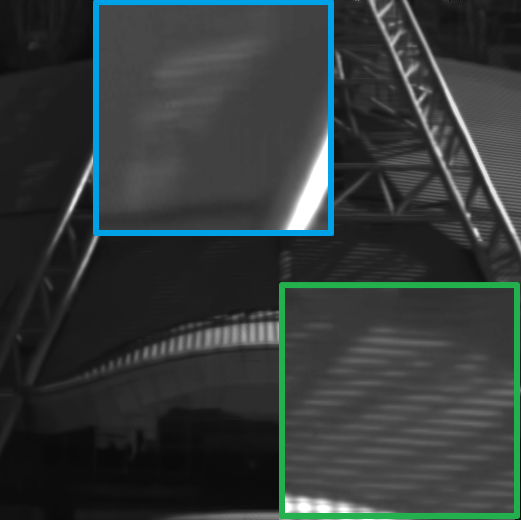}}
\vspace{-6.8pt}
\caption{Denoising comparison on the Real-HSI dataset.}
\label{Fig_compare_with_Real_HSI_case4}
\end{figure}

%% file: Fig_compare_with_Real_HSI_case5.tex
\begin{figure}[htbp]
\vspace{-6.98pt}
\graphicspath{{Figs/Selected_color_images/Real-HSI/Case5/Combined/}}
\centering
\subfigure[Mean]{
\label{Fig4}
\includegraphics[width=0.816in]{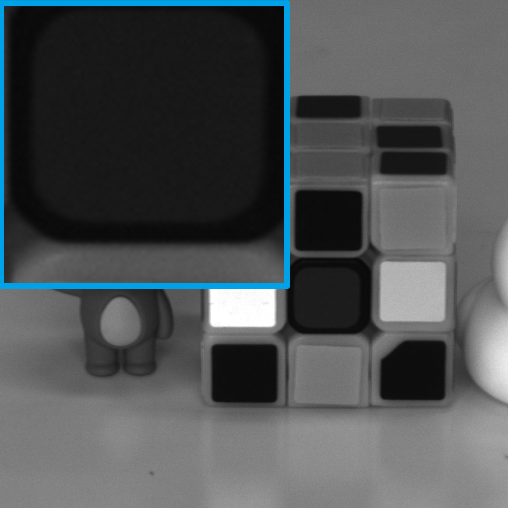}}\hfill
\subfigure[LTDL]{
\label{Fig4}
\includegraphics[width=0.816in]{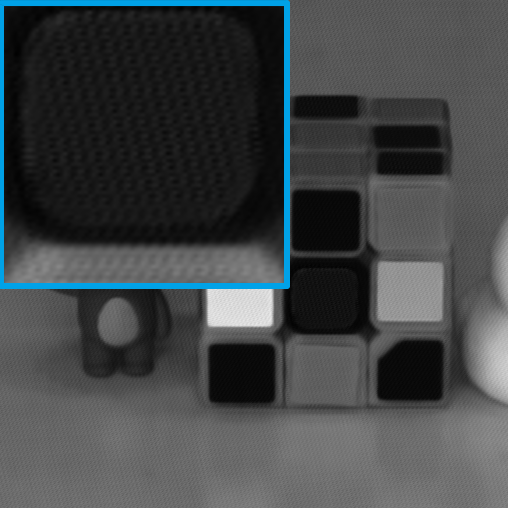}}\hfill
\subfigure[RAS2S]{
\label{Fig4}
\includegraphics[width=0.816in]{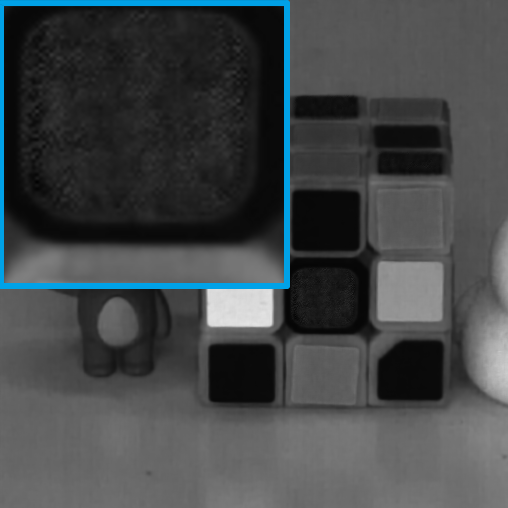}}\hfill
\subfigure[A-Haar-tSVD]{
\label{Fig4}
\includegraphics[width=0.816in]{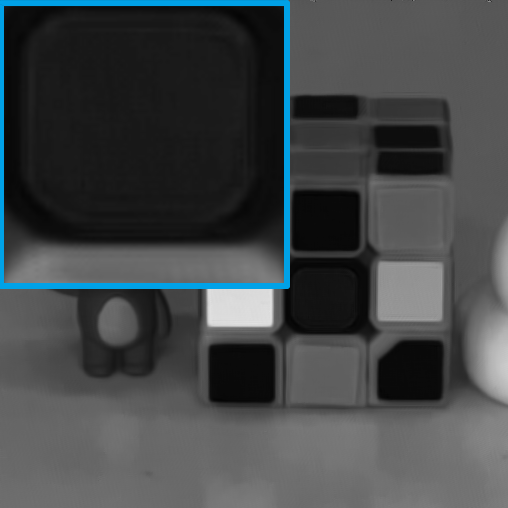}}
\vspace{-6.98pt}
\caption{Denoising comparison on the Real-HSI dataset.}
\vspace{-2.8pt}
\label{Fig_compare_with_Real_HSI_case5}
\end{figure}

%% file: Fig_HSI_PaviaU.tex
\begin{figure}[htbp]
\vspace{-6.98pt}
\graphicspath{{Figs/Real_HSI_witout_ref_selected/PaviaU/Combined/}}
\centering
\subfigure[Noisy]{
\label{Fig4}
\includegraphics[width=0.8116in]{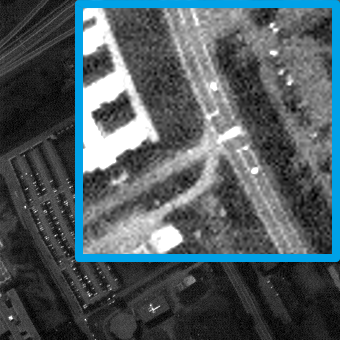}}\hfill
\subfigure[BM4D]{
\label{Fig4}
\includegraphics[width=0.8116in]{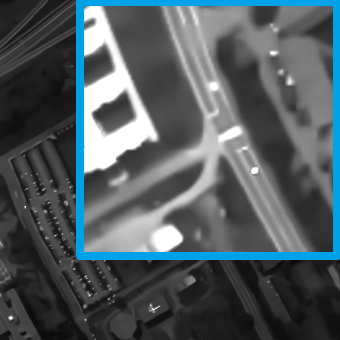}}\hfill
\subfigure[RAS2S]{
\label{Fig4}
\includegraphics[width=0.8116in]{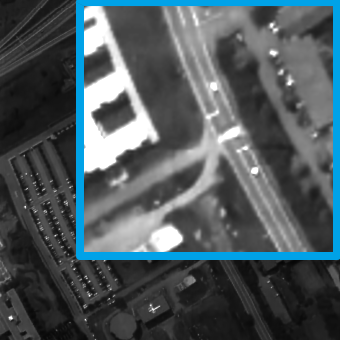}}\hfill
\subfigure[Haar-tSVD]{
\label{Fig4}
\includegraphics[width=0.8116in]{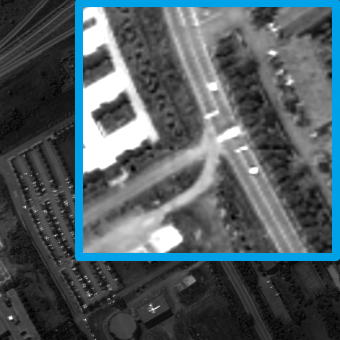}}
\vspace{-2.8pt}
\caption{Denoising comparison on the PaviaU data \cite{CupriteHSI}.}
\label{Fig_HSI_PaviaU}
\end{figure}

%% file: Fig_HSI_EO1.tex
\begin{figure}[H]
\vspace{-8.8pt}
\graphicspath{{Figs/Real_HSI_witout_ref_selected/EO1/combined/}}
\centering
\subfigure[Noisy]{
\label{Fig4}
\includegraphics[width=0.8116in]{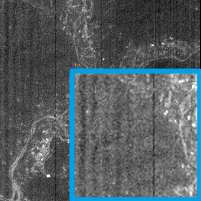}}\hfill
\subfigure[BM4D]{
\label{Fig4}
\includegraphics[width=0.8116in]{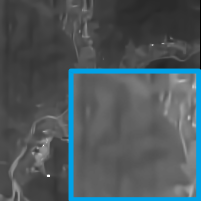}}\hfill
\subfigure[RAS2S]{
\label{Fig4}
\includegraphics[width=0.8116in]{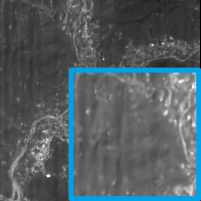}}\hfill
\subfigure[Haar-tSVD]{
\label{Fig4}
\includegraphics[width=0.8116in]{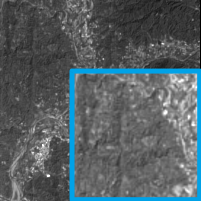}}
\vspace{-5.8pt}
\caption{Denoising comparison on the EO1 data \cite{middleton2013earth}.}
\label{Fig_HSI_EO1}
\end{figure}

%% file: Fig_HSI_Cuprite.tex
\begin{figure}[htbp]
\vspace{-8.8pt}
\graphicspath{{Figs/Real_HSI_witout_ref_selected/Cuprite/Combined/}}
\centering
\subfigure[Noisy]{
\label{Fig4}
\includegraphics[width=0.816in]{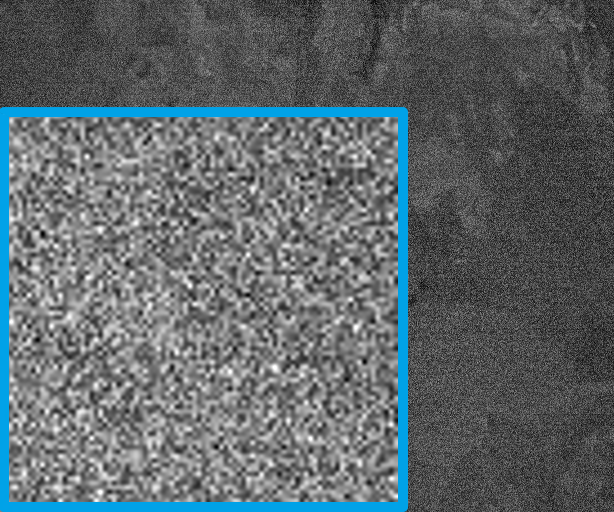}}\hfill
\subfigure[BM4D]{
\label{Fig4}
\includegraphics[width=0.816in]{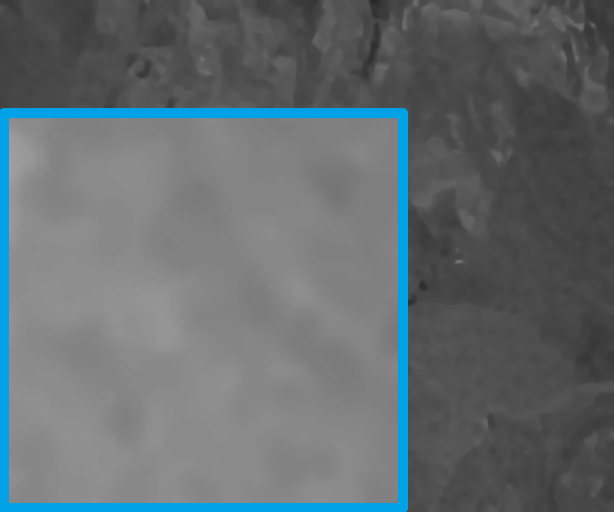}}\hfill
\subfigure[RAS2S]{
\label{Fig4}
\includegraphics[width=0.816in]{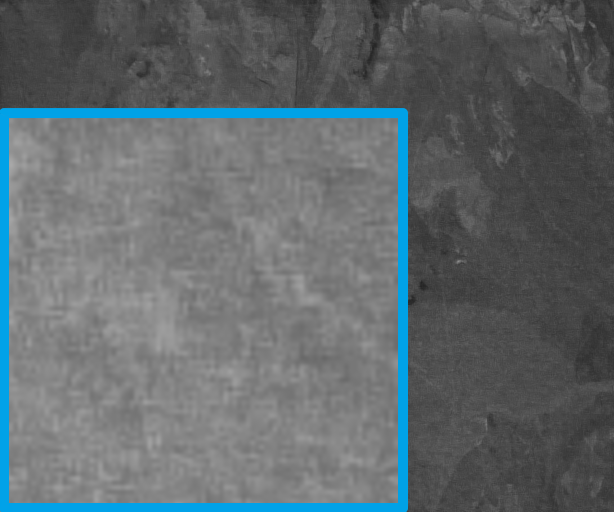}}\hfill
\subfigure[Haar-tSVD]{
\label{Fig4}
\includegraphics[width=0.816in]{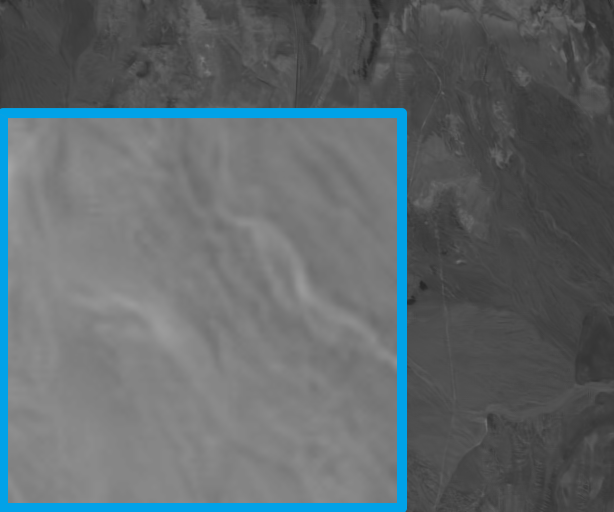}}
\vspace{-6.998pt}
\caption{Denoising comparison on the Cuprite data \cite{middleton2013earth}.}
\label{Fig_HSI_Cuprite}
\vspace{-9.8pt}
\end{figure}

%% file: Fig_HSI_Urban.tex
\begin{figure}[htbp]
\vspace{-2.8pt}
\graphicspath{{Figs/Real_HSI_witout_ref_selected/Urban/Combined/}}
\centering
\subfigure[Noisy]{
\label{Fig4}
\includegraphics[width=0.8196in]{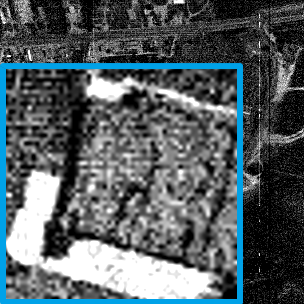}}\hfill
\subfigure[BM4D]{
\label{Fig4}
\includegraphics[width=0.8196in]{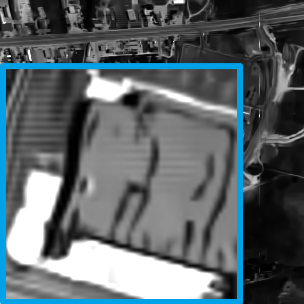}}\hfill
\subfigure[RAS2S]{
\label{Fig4}
\includegraphics[width=0.8196in]{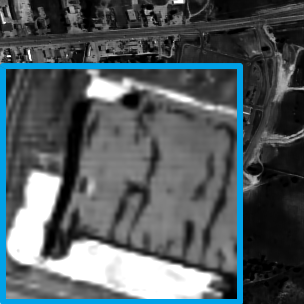}}\hfill
\subfigure[Haar-tSVD]{
\label{Fig4}
\includegraphics[width=0.8196in]{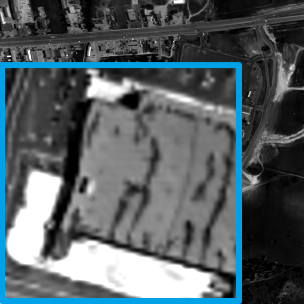}}
\vspace{-8.18pt}
\caption{Denoising comparison on the Urban data \cite{kalman1997classification}.}
\label{Fig_HSI_Urban}
\end{figure}

%% file: Table_FMDD_Haar_tSVD.tex
\begin{table}[htbp]
\vspace{-3.6pt}
\scriptsize
  \centering
  \vspace{-2.6pt}
  \caption{Denoising results of compared methods on the FMDD dataset.}
  \scalebox{0.826}{
    \begin{tabular}{cccccccc}
    \toprule
    \multirow{2}[2]{*}{Metric} & Bitonic & Haar-tSVD & A-Haar-tSVD & DDT   & DMID  & Pixel2Pixel & TBSN \\
          &  \cite{treece2022real}     &  (Ours)      &  (Ours)      &  \cite{liu2023ddt}     &  \cite{li2024stimulating}    &  \cite{ma2025pixel2pixel}     &  \cite{li2024rethinking}  \\
    \midrule
    PSNR  & 31.91  & \textbf{34.15} & \textcolor[rgb]{ .267,  .447,  .769}{\textbf{33.81}} & 31.12  & 33.09  & 33.78  & 30.68  \\
    \midrule
    SSIM  & 0.800  & \textbf{0.883} & \textcolor[rgb]{ .267,  .447,  .769}{\textbf{0.876}} & 0.789  & 0.862  & 0.850  & 0.719  \\
    \midrule
    Time (s) & 3.57  & \textcolor[rgb]{ .267,  .447,  .769}{\textbf{2.83}} & 3.36  & \textbf{1.37} & -     & 303.51  & 14.22 \\
    \bottomrule
    \end{tabular}}%
  \label{Table_FMDD_Haar_tSVD}%
  \vspace{-3.6pt}
\end{table}%

%% file: Fig_FMDD_A_Haar_tSVD.tex
\begin{figure}[htbp]
\graphicspath{{Figs/Selected_color_images/FMDD/Combined/}}
\vspace{-3.98pt}
\centering
\subfigure[Mean]{
\label{Fig4}
\includegraphics[width=0.8006in]{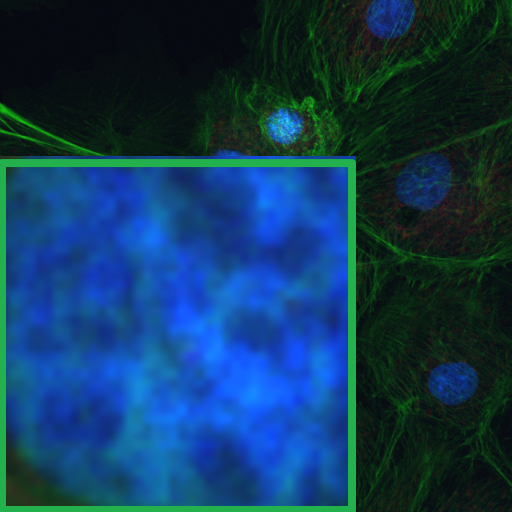}}\hfill
\subfigure[Noisy]{
\label{Fig4}
\includegraphics[width=0.8006in]{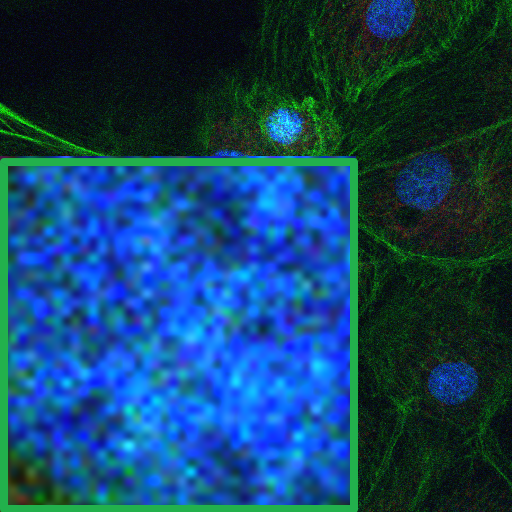}}\hfill
\subfigure[Pixel2Pixel]{
\label{Fig4}
\includegraphics[width=0.8006in]{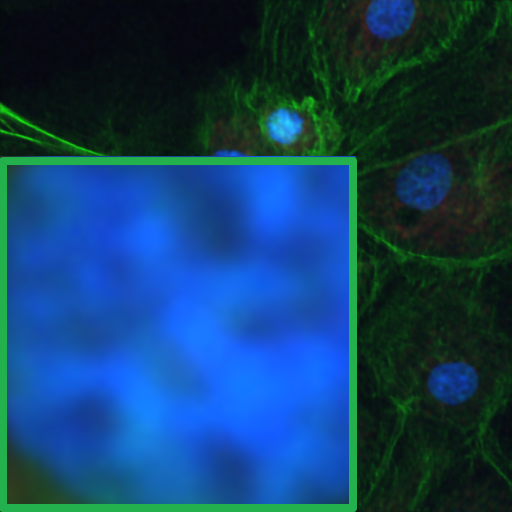}}\hfill
\subfigure[A-Haar-tSVD]{
\label{Fig4}
\includegraphics[width=0.8006in]{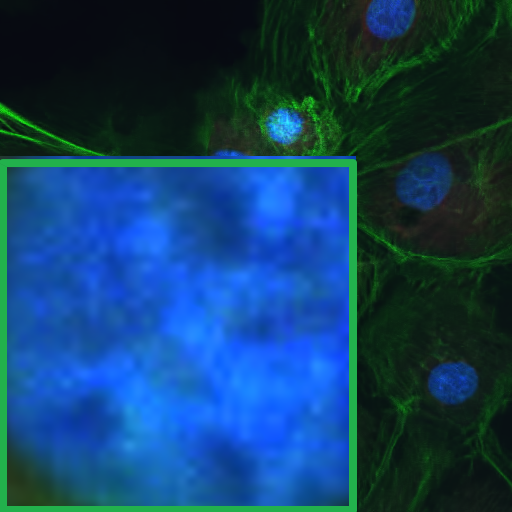}}\hfill
\vspace{-2.6pt}
\caption{Denoising results of compared methods on the FMDD dataset.}
\label{Fig_FMDD_A_Haar_tSVD}
\end{figure}

%% file: Table_compare_MRI_OASIS_time.tex
\begin{table}[htbp]
\scriptsize
  \centering
  \vspace{-2.6pt}
  \caption{Average running time comparison on the OASIS dataset.}
    \begin{tabular}{cccc}
    \toprule
    Method & BM4D  & MSt-SVD & A-Haar-tSVD \\
    \midrule
    Implementation & Matlab + Mex   & Matlab + Mex   & Matlab + Mex \\
    \midrule
    Time (minutes) & 6.6   & 6.8   & \textbf{1.5} \\
    \bottomrule
    \end{tabular}%
  \label{Table_compare_MRI_OASIS_time}%
\end{table}%

%% file: Fig_MRI_0092.tex
\begin{figure}[htbp]
\vspace{-2.8pt}
\graphicspath{{Figs/Selected_color_images/MRI_OASIS/}}
\centering
\includegraphics[width=3.58in]{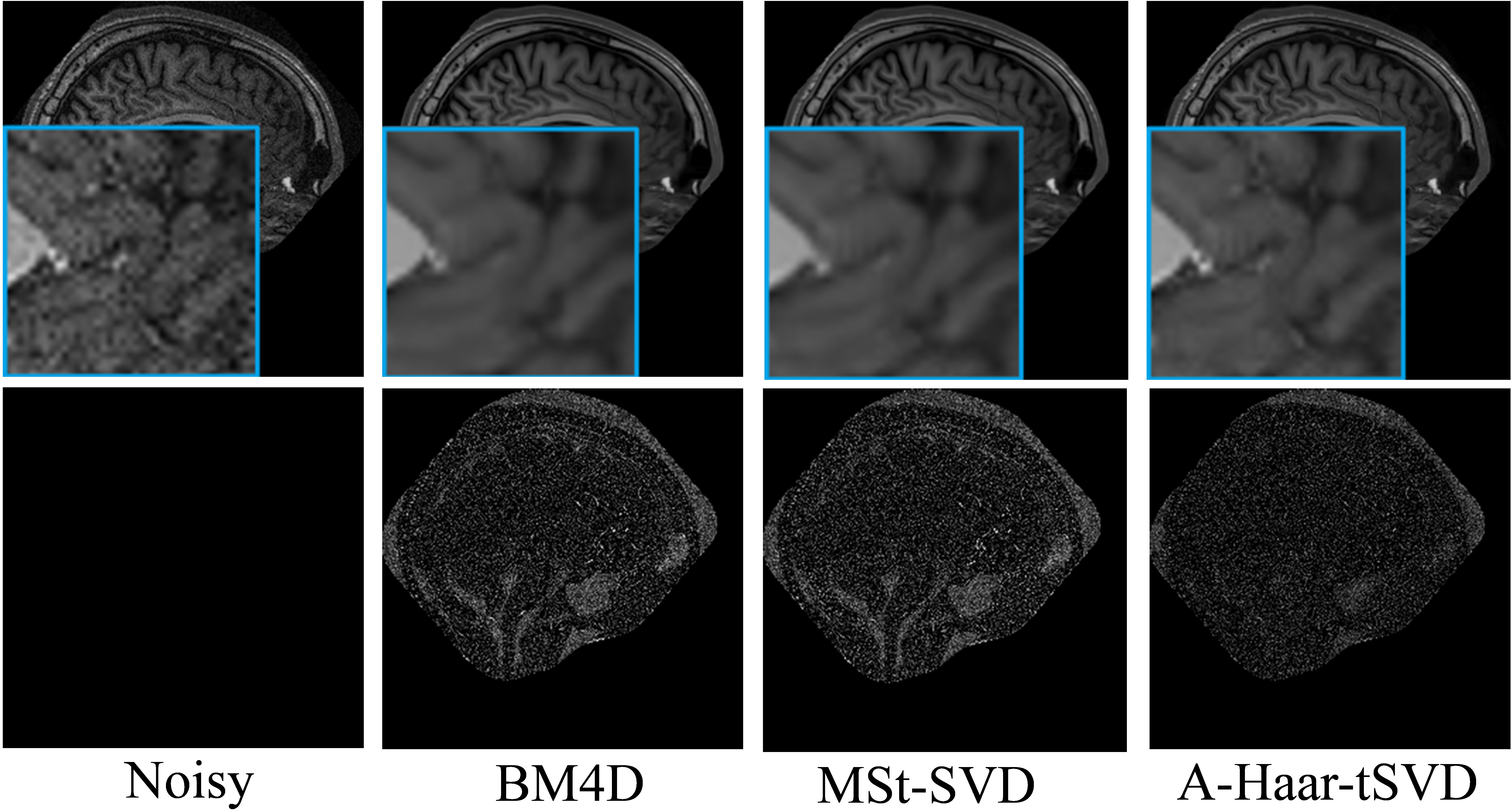}
\caption{Denoising comparison with OAS1\_0092 data. The first and second rows denote the denoised results and difference images, respectively.}
\label{Fig_MRI_0092}
\vspace{-5.98pt}
\end{figure}

%% file: Fig_MRI_0112.tex
\begin{figure}[htbp]
\vspace{-1.8pt}
\graphicspath{{Figs/Selected_color_images/MRI_OASIS/}}
\centering
\includegraphics[width=3.5018in]{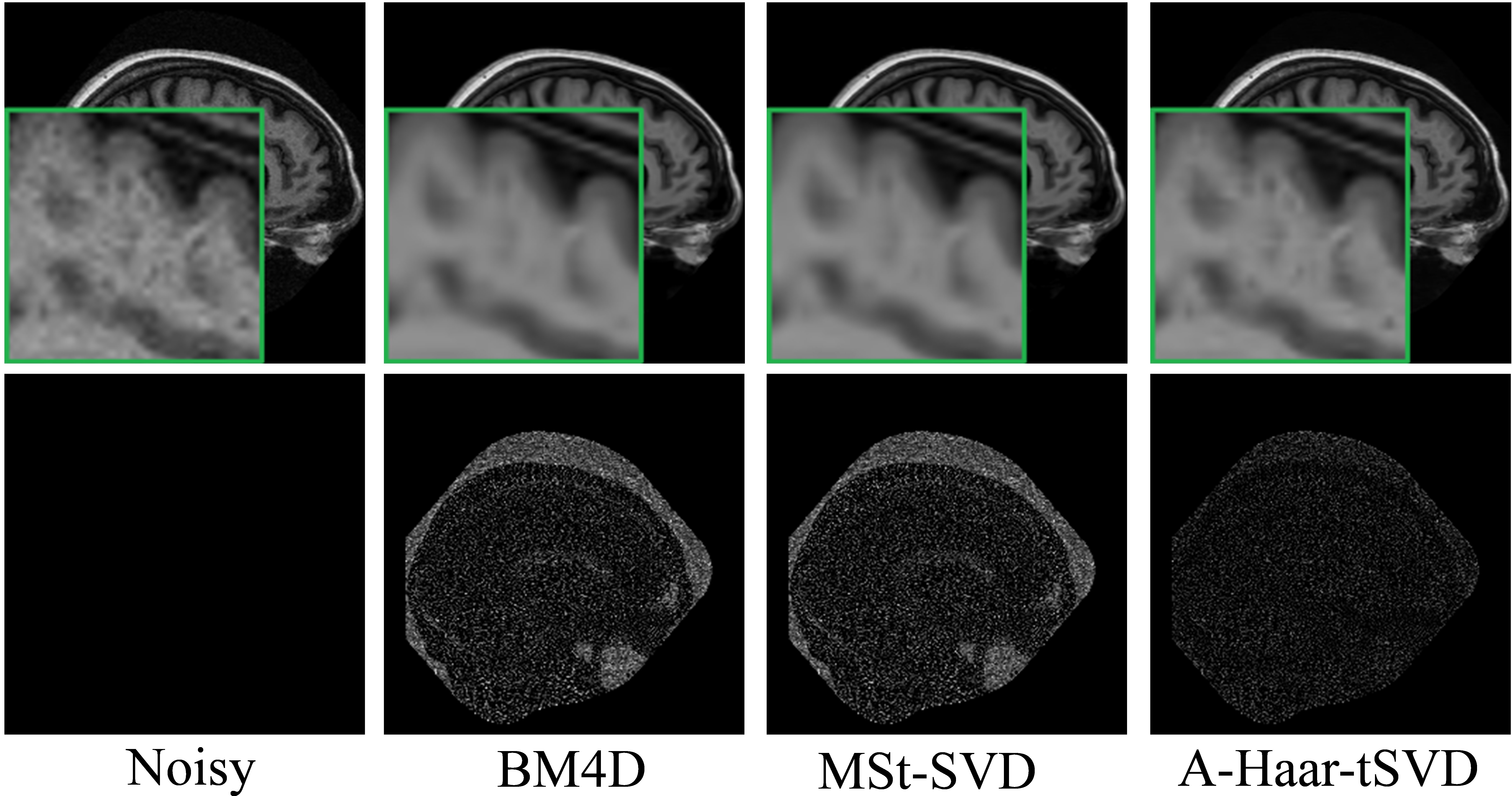}
\vspace{-12.18pt}
\caption{Denoising comparison with OAS1\_0112 data. The first and second rows denote the denoised results and difference images, respectively.}
\label{Fig_MRI_0112}
\vspace{3.98pt}
\end{figure}

%% file: Fig_parameter_tuning_ps_SR_maxK.tex
\begin{figure}[htbp]
\vspace{-6.88pt}
\graphicspath{{Figs/Parameter_tuning/}}
\centering
\subfigure[Patch size $ps$]{
\label{Fig4}
\includegraphics[width=1.1129in]{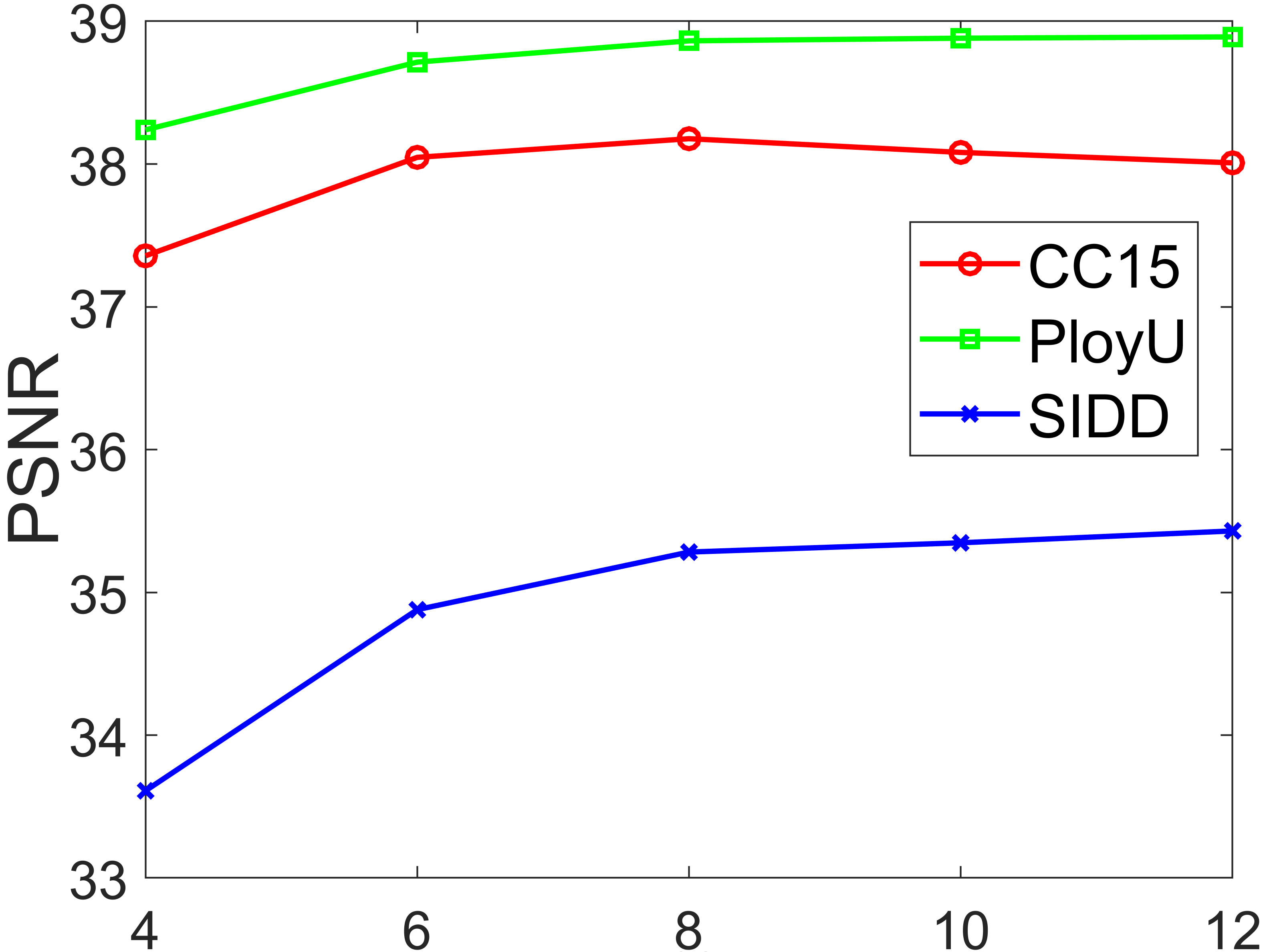}}\hfill
\subfigure[Search range $W$]{
\label{Fig4}
\includegraphics[width=1.1129in]{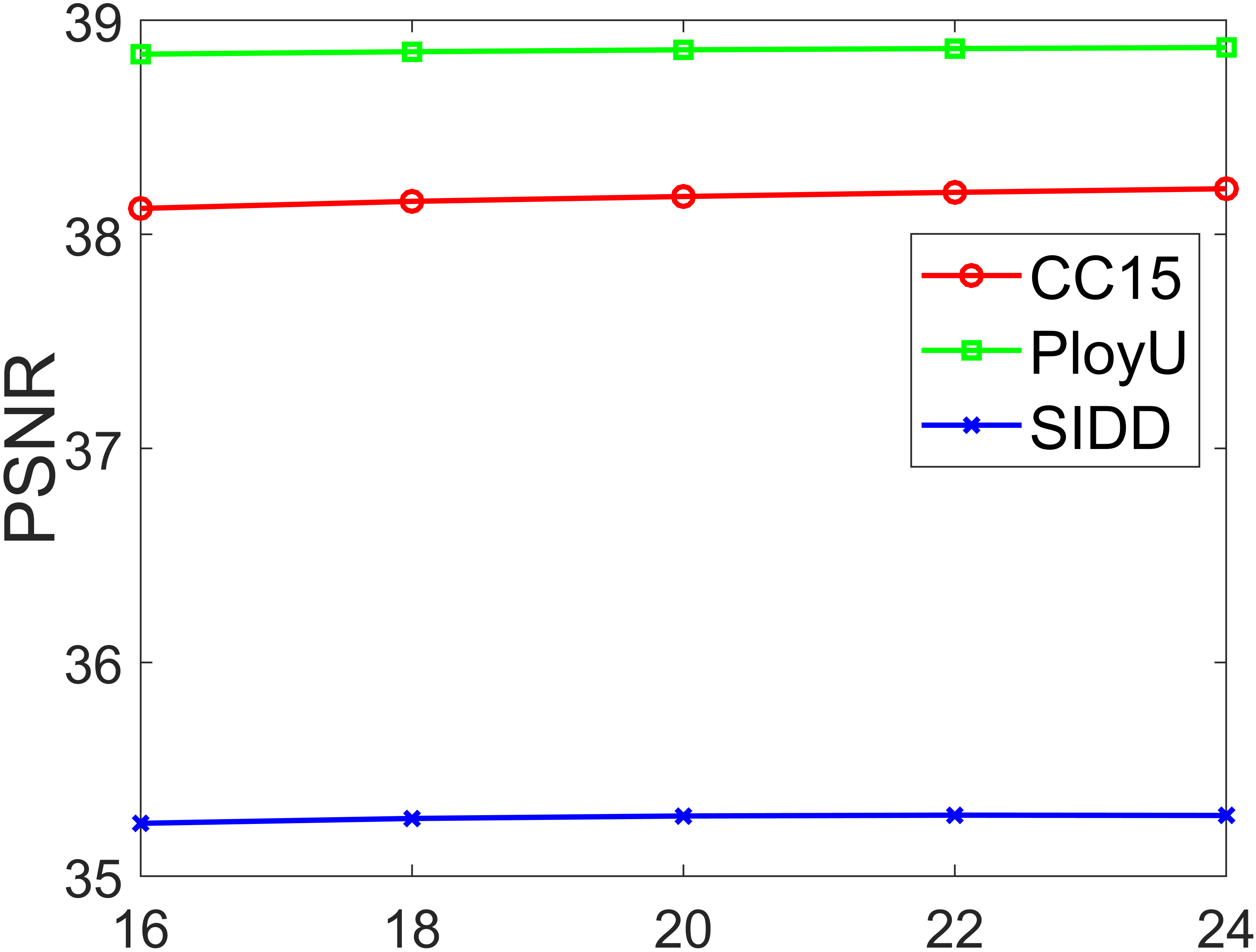}}\hfill
\subfigure[Similar patches $K$]{
\label{Fig4}
\includegraphics[width=1.1129in]{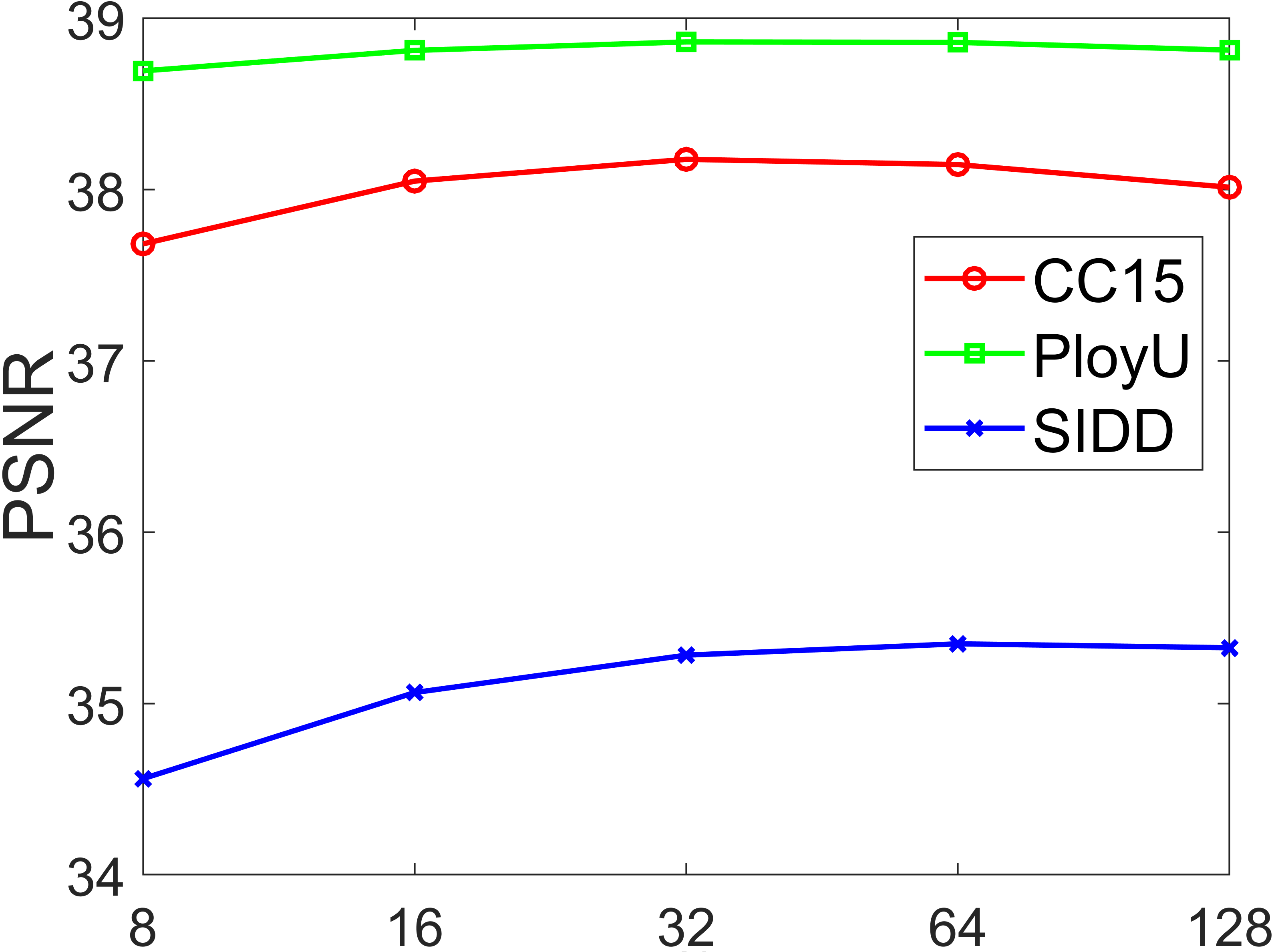}}\hfill
\vspace{-2.8pt}
\caption{Influence of different parameters on the Haar-tSVD transform.}
\label{Fig_parameter_tuning_ps_SR_maxK}
\vspace{-3.98pt}
\end{figure}

%% file: Table_ablation_study_W_WO_PCA_est.tex
\begin{table}[htbp]
\vspace{-3.98pt}
  \centering
  \caption{Comparison of the proposed A-Haar-tSVD with and without local noise level adjustment. Computational time (s) is evaluated based on images of size $512 \times 512 \times 3$.}
  \scalebox{0.936}{
    \begin{tabular}{ccc}
    \toprule
    Dataset & Without local adjustment & With local adjustment \\
    \midrule
    CC15  & 38.10/0.961 & \textbf{38.24/0.963} \\
    \midrule
    PolyU & 38.77/0.969 & \textbf{38.87/0.971} \\
    \midrule
    HighISO & 40.53/0.974 & \textbf{40.63/0.974} \\
    \midrule
    IOCI  & 41.39/0.977 & \textbf{41.52/0.978} \\
    \midrule
    SIDD-val & 35.20/0.893 & \textbf{35.28/0.894} \\
    \midrule
    Time (s) & \textbf{3.62} & 3.86 \\
    \bottomrule
    \end{tabular}}%
  \label{Table_ablation_study_W_WO_PCA_est}%
  \vspace{-2.8pt}
\end{table}%

%% file: Fig_ablation_study_beta_gamma.tex
\begin{figure}[htbp]
\vspace{-5.18pt}
\graphicspath{{Figs/Ablation_study/}}
\centering
\subfigure[$\beta$]{
\label{Fig4}
\includegraphics[width=1.6519in]{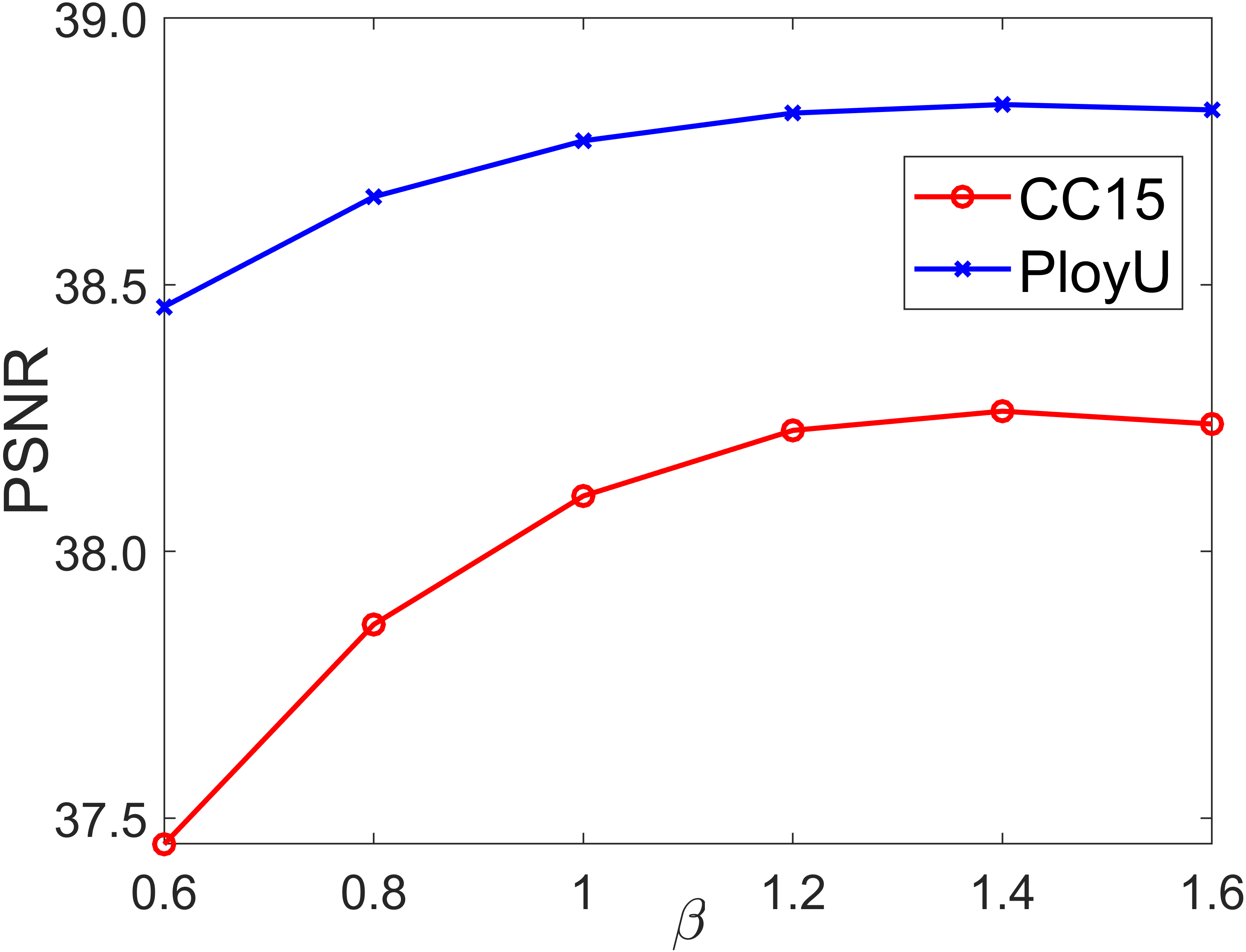}}
\subfigure[$\gamma$]{
\label{Fig4}
\includegraphics[width=1.6519in]{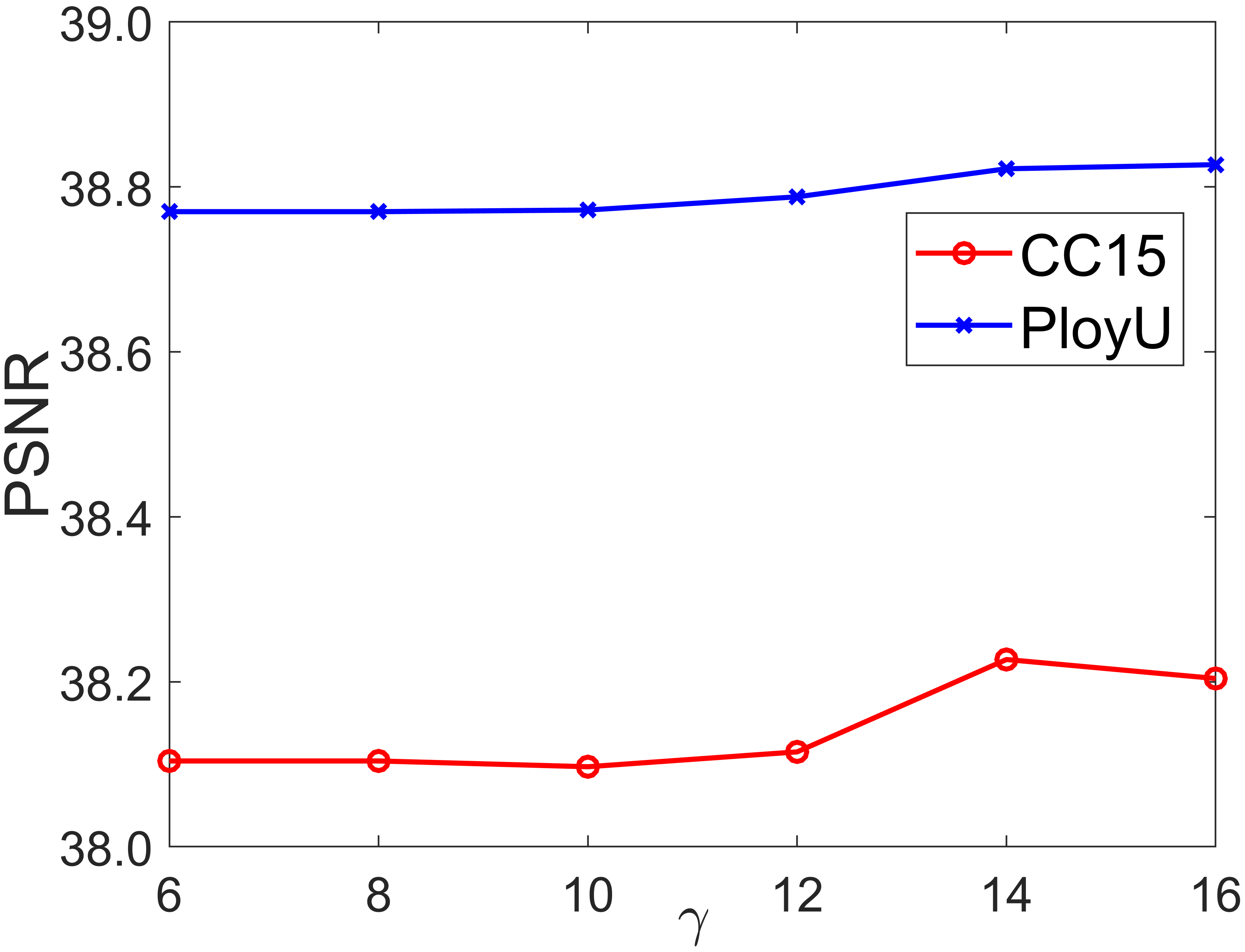}}
\vspace{-3.8pt}
\caption{Impact of weighting parameters $\beta$ and $\gamma$ on A-Haar-tSVD.}
\label{Fig_ablation_study_beta_gamma}
\vspace{-0.8pt}
\end{figure}